\documentclass[%
 reprint,
 prd,
 superscriptaddress,
 nofootinbib,
 amsmath,amssymb,
 aps,
]{revtex4-2}

\usepackage{journal_abbrevs}
\usepackage{physics}
\usepackage{graphicx}
\usepackage[caption=false]{subfig}
\usepackage{dcolumn}
\usepackage{bm}
\usepackage{hyperref}

\DeclareMathOperator*{\argmax}{arg\,max}
\newcommand{\fstat}{$\mathcal{F}$-statistic}
\newcommand{\bstat}{$\mathcal{B}$-statistic}

\begin{document}
\title{Search for continuous gravitational waves from neutron stars in five globular clusters with a phase-tracking hidden Markov model in the third LIGO observing run}

\author{L. Dunn}
\affiliation{School of Physics, University of Melbourne, Parkville, Victoria 3010, Australia}%
\affiliation{OzGrav-Melbourne, Australian Research Council Centre of Excellence for Gravitational Wave
Discovery, Parkville, Victoria 3010, Australia}
\author{A. Melatos}
\affiliation{School of Physics, University of Melbourne, Parkville, Victoria 3010, Australia}%
\affiliation{OzGrav-Melbourne, Australian Research Council Centre of Excellence for Gravitational Wave
Discovery, Parkville, Victoria 3010, Australia}
\author{P. Clearwater}
\affiliation{School of Physics, University of Melbourne, Parkville, Victoria 3010, Australia}%
\affiliation{OzGrav-Melbourne, Australian Research Council Centre of Excellence for Gravitational Wave
Discovery, Parkville, Victoria 3010, Australia}
\author{S. Suvorova}
\affiliation{School of Physics, University of Melbourne, Parkville, Victoria 3010, Australia}%
\affiliation{OzGrav-Melbourne, Australian Research Council Centre of Excellence for Gravitational Wave
Discovery, Parkville, Victoria 3010, Australia}
\affiliation{Department of Electrical and Electronic Engineering, University of Melbourne, Parkville, Victoria 3010, Australia}
\author{L. Sun}
\affiliation{OzGrav-ANU, Centre for Gravitational Astrophysics, College of Science, The Australian National University, Canberra ACT 2601, Australia}
\author{W. Moran}
\affiliation{Department of Electrical and Electronic Engineering, University of Melbourne, Parkville, Victoria 3010, Australia}
\author{R. J. Evans}
\affiliation{OzGrav-Melbourne, Australian Research Council Centre of Excellence for Gravitational Wave
Discovery, Parkville, Victoria 3010, Australia}
\affiliation{Department of Electrical and Electronic Engineering, University of Melbourne, Parkville, Victoria 3010, Australia}

\date{\today}

\begin{abstract}
A search is performed for continuous gravitational waves emitted by unknown neutron stars in five nearby globular clusters using data from the third Laser Interferometer Gravitational-Wave Observatory (LIGO) observing run, over the frequency range $100$--$800\,\mathrm{Hz}$.
The search uses a hidden Markov model to track both the frequency and phase of the continuous wave signal from one coherent segment to the next.
It represents the first time that a phase-tracking hidden Markov model has been used in a LIGO search.
After applying vetoes to reject candidates consistent with non-Gaussian artifacts, no significant candidates are detected.
Estimates of the strain sensitivity at 95\% confidence $h_{0,\mathrm{eff}}^{95\%}$ and corresponding neutron star ellipticity $\epsilon^{95\%}$ are presented.
The best strain sensitivity, $h_{0,\mathrm{eff}}^{95\%} = 2.7 \times 10^{-26}$ at $211\,\mathrm{Hz}$, is achieved for the cluster NGC6544.
\end{abstract}

\maketitle

\section{Introduction}
Continuous gravitational waves (CWs) are long-lasting and nearly monochromatic signals which may be emitted by a variety of sources, but which have so far remained undetected \citep{Riles2023,Wette2023}.
In the frequency band covered by existing terrestrial interferometers, one promising class of CW sources is rapidly rotating neutron stars located within the Milky Way.
There are many possible emission mechanisms through which neutron stars can emit gravitational waves, involving both the crust and the interior.
These include deformations created by thermal \cite{Bildsten1998,UshomirskyCutler2000, OsborneJones2020, HutchinsJones2023}, magnetic \cite{BonazzolaGourgoulhon1996, Cutler2002, PayneMelatos2004,MelatosPayne2005,PayneMelatos2006, RossettoFrauendiener2023}, and tectonic \cite{CaplanHorowitz2017, GittinsAndersson2021b, KerinMelatos2022} effects, as well as fluid oscillations (particularly $r$-modes) \cite{AnderssonKokkotas1999, ArrasFlanagan2003, BondarescuTeukolsky2007}.
CW searches thus serve to probe the physics involved in these mechanisms.

Globular clusters are an interesting target for CW searches, as their dense cores provide a natural formation site for millisecond pulsars (MSPs), fast-spinning neutron stars which have been ``recycled'' by accretion from a binary companion \cite{Katz1975, Clark1975, AlparCheng1982, PooleyLewin2003}.
The gravitational wave strain scales as the square of the spin frequency for several important emission mechanisms \cite{Riles2023}.
Once recycled, MSPs may also experience further accretion episodes later in their life, as the stellar density in the cluster core enhances the likelihood of stellar encounters which disrupt the orbits of debris disks \cite{WangChakrabarty2006, AbbottAbbott2017b, JenningsCordes2020} or planets around the pulsar \citep{WolszczanFrail1992, BailesBates2011, SpiewakBailes2018, BehrensRansom2020, NituKeith2022}.
Two CW searches targeting globular clusters specifically have been carried out in the past \cite{Wette2023}.
\citet{AbbottAbbott2017b} carried out a search with 10 days of Initial Laser Interferometer Gravitational-Wave Observatory (LIGO) Science Run 6 data targeting unknown neutron stars in the globular cluster NGC 6544.
\citet{DergachevPapa2019} carried out a search with Advanced LIGO Observing Run 1 data targeting a small sky region containing both the Galactic Center and the globular cluster Terzan 5.
Searches directed at the Galactic Center are broadly similar to globular cluster searches, being wide parameter space searches targeting an interesting sky region, and a number of these searches have been performed previously \citep{AasiAbadie2013, DergachevPapa2019, PiccinniAstone2020, AbbottAbe2022c}.
Besides the two specific searches, globular cluster MSPs have been targeted in searches for known pulsars \cite{AbbottAbbott2010,AasiAbadie2014,PitkinGill2015,AbbottAbbott2017c,AbbottAbbott2019b,AbbottAbe2022c}.

This paper presents a search for CW emission from unknown neutron stars in five globular clusters in the Milky Way.
We allow for the possibility that the phase model of the signal does not follow a simple Taylor expansion but includes some additional stochasticity (called ``spin wandering'').
The search uses a hidden Markov model (HMM) to track both the frequency and phase of the wandering signal \cite{SuvorovaSun2016, SuvorovaMelatos2018, MelatosClearwater2021}.
HMMs have been used in a number of CW searches in the past \cite{AbbottAbbott2017a, AbbottAbbott2019a, MillhouseStrang2020, MiddletonClearwater2020, SunBrito2020, AbbottAbbott2021a,JonesSun2021, BeniwalClearwater2021, BeniwalClearwater2022,AbbottAbe2022b, AbbottAbbott2022c, AbbottAbe2022a, VargasMelatos2023} but previously tracked only the frequency, typically using the \fstat{} for isolated targets \cite{JaranowskiKrolak1998, CutlerSchutz2005} or the $\mathcal{J}$-statistic for targets in binaries \cite{SuvorovaClearwater2017}.
In contrast, the HMM implementation used in this paper, developed by \citet{MelatosClearwater2021}, employs a modified version of the \bstat{} \cite{PrixKrishnan2009, MelatosClearwater2021}, which is a function of both frequency and phase.
Phase tracking improves the sensitivity (by $\lesssim 10\%$ as it turns out) by downweighting noise features, that are not approximately phase-coherent over the search.
We opt to use the phase-tracking HMM in this search both to take advantage of its enhanced sensitivity and as a technology demonstration in its own right.
This is the first time that a phase-tracking HMM has been used in an astrophysical search, and its successful application here shows that the technique can be feasibly applied to searches over a wide parameter domain.

The structure of the paper is as follows.
In Section \ref{sec:method} we describe the HMM framework in general, its implementation in this specific search, and the modified \bstat{} which serves as the coherent detection statistic.
In Section \ref{sec:setup} we discuss the setup of the search, including target selection and the parameter domain.
In Section \ref{sec:results} we discuss the candidates returned by the search, the veto procedures used to reject spurious candidates, and the results of applying the vetoes.
In Section \ref{sec:sens} we estimate the sensitivity of the search to CW emission from each cluster.
In Section \ref{sec:conclusion} we summarise the results and look to future work.

\section{Algorithms}
\label{sec:method}
In this section we briefly recap the HMM formalism and its algorithmic components. 
We define the signal model (Section \ref{subsec:alg_signal}), derive the form of the phase-sensitive \bstat{} which serves as the coherent detection statistic throughout this search (Section \ref{subsec:alg_bstat}), and explain how the \bstat{} is combined with phase-frequency transition probabilities as part of a HMM (Section \ref{subsec:alg_hmm}).

\subsection{Signal model}
\label{subsec:alg_signal}
Following \citet{JaranowskiKrolak1998}, the mass quadrupole gravitational wave signal from an isolated rotating neutron star emitting at twice the spin frequency can be written as a linear combination of four components \begin{equation} h(t) = A^\mu h_{\mu}(t), \label{eqn:signal_amuhmu}\end{equation}
with $\mu \in \{1,2,3,4\}$, where we adopt the Einstein summation convention.
The $A^\mu$ are the ``amplitude parameters'' given by Eqs. (32)--(35) in \citet{JaranowskiKrolak1998}.
They do not evolve and depend on the characteristic strain amplitude $h_0$, the cosine of the inclination angle $\cos\iota$, the polarisation angle $\psi$ and the initial signal phase $\Phi_0$.\footnote{Strictly speaking the $A^\mu$ also depend on the angle between the total angular momentum of the star and the axis of symmetry, $\theta$, as well as the angle between the interferometer arms, $\zeta$ \cite{JaranowskiKrolak1998}. The $A^{\mu}$ are multiplied by a factor of $\sin\zeta\sin^2\theta$, but $\zeta = \pi/2$ for the LIGO detectors, and we assume $\theta = \pi/2$ also, so this factor is unity.}
The $h_\mu(t)$ are combinations of the antenna pattern functions [Eqs. (12) and (13) of \citet{JaranowskiKrolak1998}] and the phase of the signal $\Phi(t)$, whose evolution is expanded as a Taylor series, viz. \begin{equation} \Phi(t) = 2\pi\sum_{k=0}^{s} f^{(k)} \frac{t^{k+1}}{(k+1)!} + \Phi_\text{w}(t) + \Phi_\text{SSB}(t). \label{eqn:signal_model} \end{equation} 
In Eq. (\ref{eqn:signal_model}), $f^{(k)}$ is the $k$-th time derivative of the frequency evaluated at $t = 0$, $\Phi_{\text{w}}(t)$ is an additional stochastic wandering term, the nature of which is discussed in detail in Section \ref{subsubsec:alg_application}, and $\Phi_{\text{SSB}}(t)$ is the phase contribution from the transformation between the solar system barycentre (SSB) and the detector frame, cf. Eq. (14) of \citet{JaranowskiKrolak1998}.
For the form of $h_\mu(t)$ see Eq. (36) in Ref. \cite{JaranowskiKrolak1998}.

\subsection{Phase-dependent $\mathcal{B}$-statistic}
\label{subsec:alg_bstat}
The $\mathcal{B}$-statistic was introduced by \citet{PrixKrishnan2009} as a Bayesian alternative to the well-known $\mathcal{F}$-statistic \citep{JaranowskiKrolak1998}.
The $\mathcal{F}$-statistic is obtained by analytically maximizing the likelihood function over $A^\mu$ and hence $h_0$, $\cos\iota$, $\psi$\ and $\Phi_0$.
The $\mathcal{B}$-statistic is obtained by instead \emph{marginalising} over these parameters with a physically motivated prior which is uniform in $h_0$, $\cos\iota$, $\psi$, and $\Phi_0$.
\citet{MelatosClearwater2021} modified this approach to leave $\Phi_0$ as a free parameter, so that we can compute the $\mathcal{B}$-statistic for many coherent segments and then track $f^{(0)}$ and $\Phi_0$ across these segments using a HMM as discussed in Section \ref{subsec:alg_hmm}.
Here we briefly justify the form of the phase-dependent $\mathcal{B}$-statistic and discuss how to compute it in practice.
The discussion draws on Refs. \cite{JaranowskiKrolak1998,PrixKrishnan2009,MelatosClearwater2021}.

We begin with the log likelihood \begin{equation} \ln\Lambda = \braket{x}{h} - \frac{1}{2}\braket{h}{h}, \end{equation}
where $x(t)$ is the detector data and $h(t)$ is the signal as defined in Eqs. (\ref{eqn:signal_amuhmu}) and (\ref{eqn:signal_model}).
The inner product is defined as \begin{align} 
\braket{x}{y} &= 4\Re\int_0^\infty\mathrm{d}f\,\frac{x(f)y^*(f)}{S_h(f)}\\ 
              &\approx \frac{2}{S_h(f_0)}\int_0^T \mathrm{d}t\, x(t)y(t),
\end{align}
where $S_h(f)$ is the one-sided power spectral density of the noise at a frequency $f$, $T$ is the total length of the data, and the approximate equality holds for finite-length signals localized spectrally to a narrow band around $f_0$ in which $S_h(f) \approx S_h(f_0)$ can be taken as constant (which is a good approximation for the quasi-monochromatic signals of interest here \citep{JaranowskiKrolak1998}).
Defining $x_\mu = \braket{x}{h_\mu}$ and $\braket{h_\mu}{h_\nu} = \mathcal{M}_{\mu\nu}$ we write \begin{equation} \ln\Lambda = A^\mu x_\mu - \frac{1}{2}A^\mu\mathcal{M}_{\mu\nu}A^\nu \label{eqn:bstat_loglike}.\end{equation}
The phase-dependent $\mathcal{B}$-statistic is obtained by marginalising over $\cos\iota$, $\psi$, and $h_0$: \begin{equation} \mathcal{B} = \int_{-1}^{1}\mathrm{d}(\cos\iota)\int_{-\pi/4}^{\pi/4}\mathrm{d}\psi\int_0^{h_0^{\mathrm{max}}} \mathrm{d}h_0\, p(\cos\iota,\psi,h_0)\Lambda. \label{eqn:bstat_full_integral}\end{equation}
In Eq. (\ref{eqn:bstat_full_integral}), $p(\cos\iota,\psi,h_0) = C/h_0^{\mathrm{max}}$ is the prior, where $h_0^{\mathrm{max}}$ is a large but arbitrary maximum $h_0$ and $C$ is an irrelevant normalising constant.
Since $p(\cos\iota,\psi,h_0)$ is a constant and can be taken outside the integral it is dropped going forward.

We now substitute Eq. (\ref{eqn:bstat_loglike}) into Eq. (\ref{eqn:bstat_full_integral}) and perform the triple integral.
Writing $\mathcal{A}^\mu = h_0\mathcal{A}'^{\mu}$, with $\mathcal{A}'^{\mu}$ independent of $h_0$, and collapsing the triple integral $\int[\ldots]$ for notational convenience, we obtain \begin{equation}\mathcal{B} = \int[\ldots] \exp\left(h_0\mathcal{A}'^\mu x_\mu - \frac{h_0^2}{2}\mathcal{A}'^\mu \mathcal{M}_{\mu\nu} \mathcal{A}'^\nu\right). \end{equation}
We use the following identity to perform the integral over $h_0$:
\begin{widetext}\begin{equation} \int_0^X \mathrm{d}x\,\exp(Ax-Bx^2) = \left(\frac{\pi}{4B}\right)^{1/2}\exp\left(\frac{A^2}{4B}\right)\left[\erf\left(\frac{A}{2\sqrt{B}}\right) + \erf\left(\frac{2BX-A}{2\sqrt{B}}\right)\right]. \end{equation}
We also take $h_0^{\mathrm{max}}$, which is arbitrary, to be sufficiently large that the second $\erf(\ldots)$ term approaches unity.
The result is \begin{equation} \mathcal{B} = \int_{-1}^{-1}\,\mathrm{d}(\cos\iota)\int_{-\pi/4}^{\pi/4} \,\mathrm{d}\psi \left(\frac{\pi}{2\mathcal{A}'^\mu \mathcal{M}_{\mu\nu}\mathcal{A}'^\nu}\right)^{1/2}\exp\left[\frac{(\mathcal{A}'^\mu x_\mu)^2}{2\mathcal{A}'^\mu \mathcal{M}_{\mu\nu}\mathcal{A}'^\nu}\right]\left[\erf\left(\frac{\mathcal{A}'^\mu x_\mu}{\sqrt{2\mathcal{A}'^\mu \mathcal{M}_{\mu\nu}\mathcal{A}'^\nu}}\right) + 1\right]. \label{eqn:bstat_final_integral}\end{equation}\end{widetext}
The ingredients of the integrand --- $\mathcal{A}'^\mu$, $\mathcal{M}_{\mu\nu}$ and $x_\mu$ --- are all straightforward to compute using existing functionality in \textsc{lalsuite}, the standard software suite for analysis of LIGO, Virgo, and KAGRA data \citep{lalsuite}.
We use the existing graphics processing unit (GPU) framework \cite{DunnClearwater2022} to compute the $\mathcal{F}$-statistic and hence $x_\mu$, the most expensive step.
We also integrate Eq. (\ref{eqn:bstat_final_integral}) using GPUs, running a simple Simpson integrator on a $11 \times 11$ grid of $\cos\iota$ and $\psi$ values.
We verify using synthetic data that the relative error in the value of the \bstat{} when computed using an $11 \times 11$ grid compared to a finer $501 \times 501$ grid is typically small, with a mean fractional error of 1.9\% and a fractional error exceeding 10\% occurring in 1.2\% of the $1.7 \times 10^7$ \bstat{} values (corresponding to five $0.5\,\mathrm{Hz}$ bands with a coherence time of five days and eight phase bins) computed for the purposes of this test.

When computing $\mathcal{B}$ we ignore $\Phi_{\text{w}}(t)$ in Eq. (\ref{eqn:signal_model}).
By assumption this term is small compared to the other terms in Eq. (\ref{eqn:signal_model}) over each coherent segment during which $\mathcal{B}$ is computed, amounting to $\lesssim 1\,\mathrm{rad}$.
The stochastic wandering is tracked instead by the HMM, as discussed in Section \ref{subsubsec:alg_application}.

\subsection{HMMs and the Viterbi algorithm}
\label{subsec:alg_hmm}
The number of signal templates needed to cover an astrophysically relevant parameter domain grows quickly for a phase-coherent search, as the observational timespan $T_{\text{obs}}$ increases.
Hence a number of ``semicoherent'' techniques for CW searches have been developed, which perform phase-coherent computations on subdivided segments of length $T_{\text{coh}}$ and combine the segments without enforcing phase coherence between them.
Most semicoherent methods completely discard the phase information between segments (see e.g. Refs. \citep{SuvorovaSun2016, KrishnanSintes2004, BayleyMessenger2019, AstoneColla2014}).
Some, termed ``loosely coherent'', retain some phase information but do not enforce perfect phase continuity \citep{Dergachev2010, Dergachev2012, DergachevPapa2023}.
While there is some loss of sensitivity inherent in this approach --- the sensitivity of semicoherent searches scales as $(T_{\mathrm{coh}}T_{\mathrm{obs}})^{-1/4}$ \citep{KrishnanSintes2004,Wette2012}, compared to $T_{\mathrm{obs}}^{-1/2}$ for coherent searches --- there is a large computational saving.
Some semicoherent methods also offer flexibility in the face of the unexpected.
For example, although fully coherent methods can accommodate deviations away from an assumed signal model through explicit searches over the parameters determining the phase evolution \citep{AbbottAbbott2008, AshokBeheshtipour2021, AbbottAbbott2022a}, if there is a stochastic component to the phase evolution which operates on timescales shorter than the observing timespan then a semicoherent approach may be needed.
A HMM solved by the Viterbi algorithm to find the most likely sequence of signal frequencies $f(t)$ is one such approach, which has been used in a number of CW searches in the past \cite{AbbottAbbott2017a, AbbottAbbott2019a, MillhouseStrang2020, SunBrito2020, MiddletonClearwater2020, AbbottAbbott2021a,JonesSun2021, BeniwalClearwater2021, BeniwalClearwater2022,AbbottAbe2022b, AbbottAbbott2022c, AbbottAbe2022a, VargasMelatos2023}.
This approach was extended to include tracking of $\Phi(t)$ by \citet{MelatosClearwater2021}, and is the subject of the present paper. 
In this section we review the basic framework of HMMs and the way it is put into practice for the present, loosely coherent search.

\subsubsection{HMMs}
HMMs model the evolution and observation of systems fulfilling two important criteria: a) the evolution of the internal state of the system is Markovian, and b) the internal state is hidden from the observer but may be probed probabilistically through measurement.
This section briefly describes the formalism involved; for a more detailed review of HMMs see \citet{Rabiner1989}.

The hidden state of the system is assumed here to be discretised and bounded, so that each state is in the set $\mathcal{Q} = \{q_1, q_2, \ldots, q_{N_Q}\}$, where $N_Q$ is the total number of hidden states.
Time is likewise discretised, with the HMM occupying the hidden state $q(t_i) \in \mathcal{Q}$ at each timestep $t_i \in \{t_0, t_1, t_2, \ldots, t_{N_T}\}$ ,where $t_0$ is an ``initial timestep'' before any observations have been made.
A sequence of possible hidden states occupied during a particular realisation of the HMM is denoted by $Q = [q(t_0), q(t_1), q(t_2), \ldots, q(t_{N_T})]$.
The dynamics of the system are captured by the transition matrix $A(q_j, q_i)$, whose entries are defined as \begin{equation} A(q_j, q_i) = \Pr[q(t_{n+1}) = q_j \mid q(t_n) = q_i]. \end{equation}

At each timestep we assume that the system is observed.
For a given realisation we write the sequence of observations as $O = [o(t_1), o(t_2), \ldots, o(t_{N_T})]$.
The set of possible observations need not be finite, discretised, or bounded.
All we require is that there exists an emission probability function $L(o, q_i)$ defined as \begin{equation} L(o, q_i) = \Pr[o(t_n) = o \mid q(t_n) = q_i], \end{equation} i.e. the probability of observing $o$ at the timestep $t_n$, if the hidden state $q_i$ is occupied at $t_n$.

Finally we specify a belief about the state of the system at the beginning of the analysis, i.e. \begin{equation} \Pi(q_i) = \Pr[q(t_0) = q_i]. \end{equation}
In this paper we have no reason to impose anything other than a uniform prior, viz. $\Pi(q_i) = N_Q^{-1}$.

Given the above ingredients, it is straightforward to compute the probability of a hidden state sequence $Q$, given a sequence of observations $O$: \begin{equation} \Pr(Q\mid O) = \Pi[q(t_0)] \prod_{n=1}^{N_T} L[o(t_n), q(t_n)] A[q(t_n), q(t_{n-1})]. \label{eqn:hmm_path_prob}\end{equation}
The goal is to compute $\argmax_Q \Pr(Q\mid O)$.
\subsubsection{Phase tracking}
\label{subsubsec:alg_application}
In this paper, the full observation is divided into segments of length $T_{\text{coh}} = T_{\mathrm{obs}}/N_T$, which are analysed coherently.
Each coherent segment is a timestep in the HMM.
The hidden states are pairs of frequencies and phases at the start of a segment, $q(t_n) = [f(t_n), \Phi(t_n)]$.
The phase-dependent $\mathcal{B}$-statistic, defined in Section \ref{subsec:alg_bstat}, plays the role of $\ln L(o, q_i)$.
It then remains to specify the transition matrix.
We follow \citet{MelatosClearwater2021} and adopt the following model for the evolution of the signal frequency $f$ and phase $\Phi$: \begin{align} \dv{f}{t} &= \sigma \xi(t), \label{eqn:trans_f_evo}\\ \dv{\Phi}{t} &= f, \label{eqn:trans_phi_evo} \end{align} where $\xi(t)$ is a zero-mean white noise term, i.e. \begin{align} \langle \xi(t)\rangle &= 0 \\ \langle \xi(t)\xi(t')\rangle &= \delta(t-t') \label{eqn:trans_wn_corr} \end{align} and $\sigma$ parametrises the noise strength.
The forward transition matrix $A[(f_j, \Phi_j), (f_i, \Phi_i)]$ is readily obtained by solving the Fokker-Planck equation corresponding to Eqs. (\ref{eqn:trans_f_evo})--(\ref{eqn:trans_wn_corr}).
For practical reasons it is useful to also know the \emph{backward} transition matrix \begin{equation} A^{\mathrm{b}}(q_i, q_j) = \Pr[q(t_n) = q_i \mid q(t_{n+1}) = q_j], \end{equation} which is similarly obtained by solving the backward Fokker-Planck equation for this system.

\begin{figure}
    \centering
    \includegraphics[width=\columnwidth]{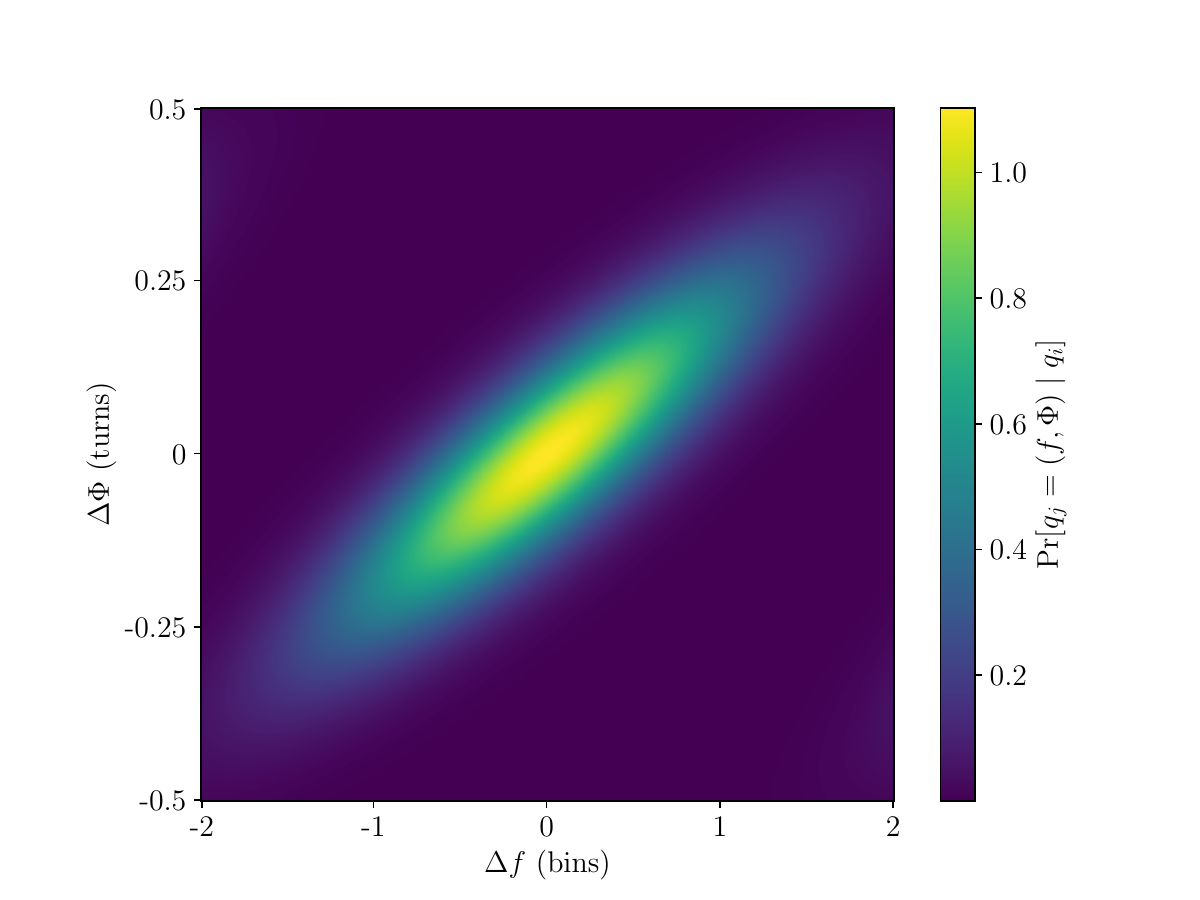}\\
    \includegraphics[width=\columnwidth]{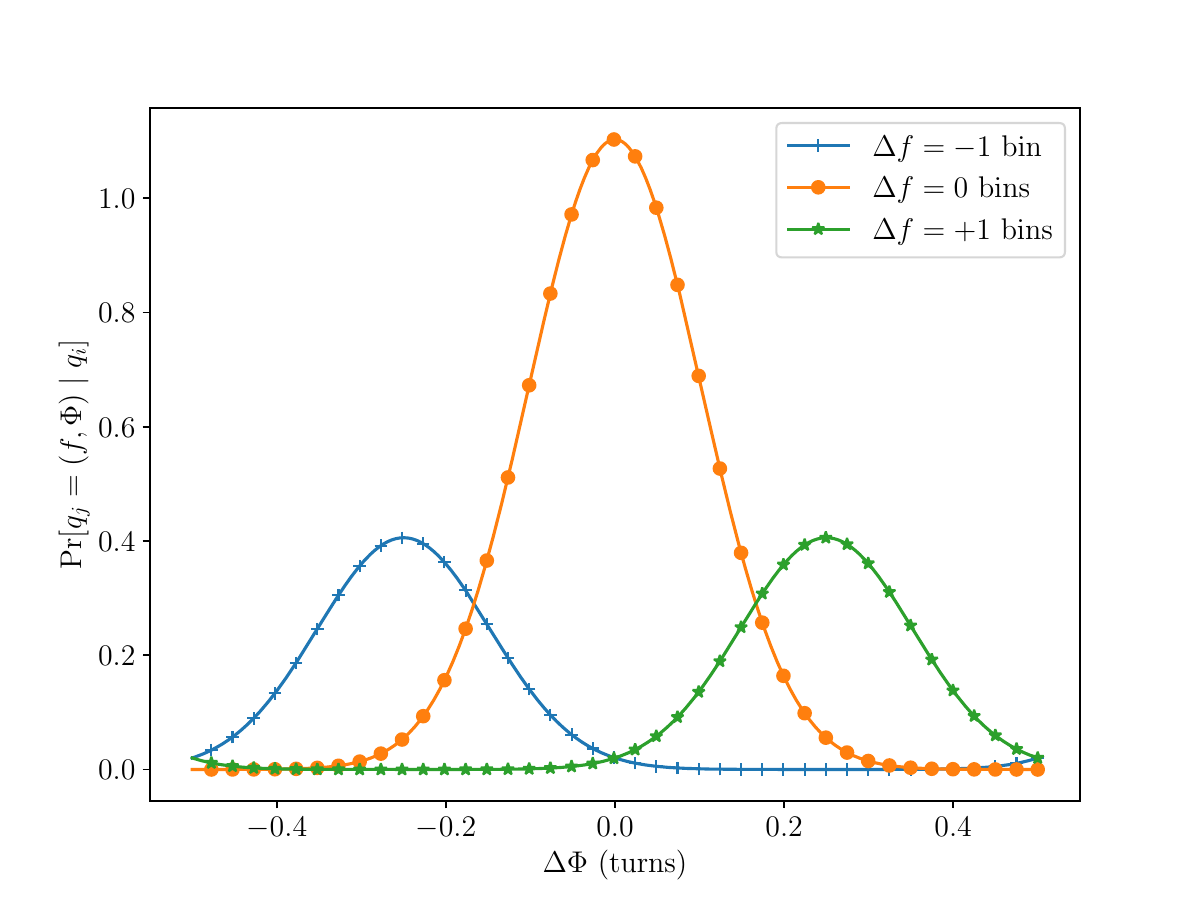}
    \caption{Illustrative plots of the transition matrix $A(q_j, q_i) = \Pr[q(t_{n+1}) = q_j \mid q(t_n) = q_i]$ according to the model described in Section \ref{subsubsec:alg_application}. The top panel shows a heatmap of $A(q_j, q_i)$ where $q_i = [\Phi(t_n), f(t_n)]$ is an arbitrary initial state and $q_j = [\Phi(t_{n+1}), f(t_{n+1})]$ is the final state.
    The heatmap axes are $\Delta\Phi = \Phi(t_{n+1}) - \Phi(t_n)$ and $\Delta f = f(t_{n+1}) - f(t_n)$ in units of turns (cycles of $2\pi$ radians) and bins [of width $(2T_{\mathrm{coh}})^{-1}$] respectively. The bottom panel shows slices of $A(q_j, q_i)$ as a function of $\Delta\Phi$ for $\Delta f = -1,\,0,\,+1$ bins.
    Parameters: $T_{\text{coh}} = 5\,\mathrm{d}$, $\sigma = 1.76 \times 10^{-9}\,\mathrm{s}^{-3/2}$.}
    \label{fig:trans_pdf}
\end{figure}

Details of the transition matrices are given in Appendix \ref{apdx:trans_fpe_calc}, and an illustrative figure is shown in Fig. \ref{fig:trans_pdf}.
The top panel of Fig. \ref{fig:trans_pdf} shows a heatmap of $A(q_j, q_i)$ as a function of $\Delta\Phi = \Phi(t_{n+1}) - \Phi(t_n)$ and $\Delta f = f(t_{n+1}) - f(t_n)$.
As shown in \citet{MelatosClearwater2021} and Appendix \ref{apdx:trans_fpe_calc}, it is approximately a truncated Gaussian centred on $(\Delta \Phi,\Delta f) = (0,0)$, with positive covariance between $\Delta\Phi$ and $\Delta f$.
The bottom panel shows slices through the distribution at $\Delta f = -1\,,0\,,+1\,\mathrm{bins}$.
These slices are the distributions used to compute the transition probabilities in the HMM.
The $\Delta f = \pm 1$ bin distributions are identical up to the location of the peak and each contain approximately 20\% of the total probability, while the $\Delta f = 0$ bin contains the other 60\%.
The transitions probabilities are broad in phase.
All three distributions have similar widths.
From this figure we see that the probability is concentrated in the same bin as the previous timestep, but some leakage into neighboring bins does occur.

\citet{MelatosClearwater2021} included a damping term $-\gamma f$ in Eq. (\ref{eqn:trans_f_evo}).
This approximates the tendency of the spin frequency of the neutron star to approach zero in the long term, i.e. it represents the long-term spin down process.
In this search we explicitly search over the spin-down rate $f^{(1)}$ as a parameter in the deterministic part of the signal model (see Sections \ref{subsubsec:param_ranges_f_fdot} and \ref{subsubsec:setup_param_spacing_f_fdot}), and assume that once the deterministic spin down is taken into account, the mean drift of the signal frequency is negligible.
As such, we impose the condition $\gamma \ll (2fT_{\text{coh}}^2)^{-1}$ which implies $|f(t_{n+1}) - f(t_n)| \ll (2T_\mathrm{coh})^{-1}$ \cite{MelatosClearwater2021}.
This renders the effect of $-\gamma f$ in Eq. (\ref{eqn:trans_f_evo}) negligible, and in this work we do not include it.
The approximation is justified in more detail in Appendix \ref{apdx:trans_fpe_calc}.

\subsubsection{Optimal Viterbi path}
\label{subsubsec:alg_viterbi_path}
With the HMM fully specified, we turn to the question of to use it in a CW search.
To find signal candidates we use the Viterbi algorithm \citep{Rabiner1989, SuvorovaSun2016} to find the most likely sequence of hidden states, \begin{equation} Q^* = \argmax_Q \Pr(Q\mid O). \end{equation}
More precisely, for each possible terminating frequency bin $f_i$ we find the most likely sequence of hidden states which terminates in that bin, \begin{equation} Q^*(f_i) = \argmax_Q \Pr[Q \mid O, f(t_{N_T}) = f_i]. \end{equation}
These paths are sorted according to their probability $\Pr[Q^*(f_i) \mid O]$.
Paths exceeding a predetermined threshold are marked for follow-up analysis.
The determination of the threshold as a function of the desired false alarm probability is discussed in Section \ref{subsec:res_thresh}.
To maximize numerical accuracy, we work with log probabilities so that the products in Eq. (\ref{eqn:hmm_path_prob}) become sums.
Hence we threshold candidates according to their log likelihood $\mathcal{L} = \ln \Pr[Q^*(f_i) \mid O]$.

To maximize computational performance the HMM tracking step is performed on GPUs, as with the computation of the $\mathcal{B}$-statistic described in Section \ref{subsec:alg_bstat}.

\section{Setting up a globular cluster search}
\label{sec:setup}
In this section we describe the astronomical targets and data inputs (Sections \ref{subsec:setup_targets} and \ref{subsec:setup_data} respectively) for the globular cluster search performed in this paper, the choices of certain control parameters which dictate the behaviour of the HMM (Sections \ref{subsec:setup_tcoh} and \ref{subsec:setup_hmm} respectively), and the astrophysical and computational motivations for the parameter domain and its associated grid spacing (Sections \ref{subsec:param_ranges} and \ref{subsec:param_spacing} respectively).

\subsection{Target selection}
\label{subsec:setup_targets}
Globular clusters are dense stellar environments, with correspondingly high rates of dynamical interactions.
They are prodigious manufacturers of low-mass X-ray binaries (LMXBs) \citep{Katz1975, Clark1975, PooleyLewin2003}, which are the progenitors of MSPs, neutron stars which have been ``recycled'' to  millisecond spin periods via accretion \citep{AlparCheng1982}.
Although accretion from a binary companion offers several mechanisms to increase the mass quadrupole moment of a neutron star \citep{Wagoner1984, Bildsten1998, VigeliusMelatos2009}, the search in this paper is not tuned to sources in binaries, as the \bstat{} in Section \ref{subsec:alg_bstat} does not include the effects of binary motion on the frequency modulation of the signal (doing so would require a blind search over the binary parameters, adding greatly to the cost of the search).
Hence the ideal target in this paper is a neutron star that has spent some time in a binary system historically, so as to be recycled to spin frequencies in the LIGO band above $100\,\mathrm{Hz}$ (see Section \ref{subsubsec:param_ranges_f_fdot}), but was disrupted since by a secondary encounter and became an isolated source.

The catalog of Milky Way globular clusters compiled by \citet{Harris1996} (2010 edition) contains parameters for 157 clusters.
With limited resources, we are unable to search every known globular cluster for CW emission.
Therefore we develop a simple ranking statistic to guide target selection.
The rest of this section describes that statistic and the resulting target list.

\citet{VerbuntFreire2014} introduced a parameter\footnote{\citet{VerbuntFreire2014} labelled this parameter $\gamma$, here we relabel it as $\eta$ to avoid confusion with the $\gamma$ introduced in Section \ref{subsubsec:alg_application}.} $\eta$ which they refer to as the ``single-binary encounter rate''.
Roughly speaking it equals the reciprocal of the lifetime of a binary, before it is disrupted by a secondary encounter.
It scales with globular cluster parameters as \begin{equation} \eta \propto \frac{\sqrt{\rho_\text{c}}}{r_\text{c}}, \end{equation} where $\rho_\text{c}$ is the core luminosity density and $r_\text{c}$ is the core radius.
Larger values of $\eta$ are correlated with properties which are encouraging for the present search \citep{VerbuntFreire2014}.
First and foremost, isolated MSPs are more prevalent in high-$\eta$ clusters, perhaps due to recycling followed by ejection from the binary by secondary encounters.
Second, high-$\eta$ clusters contain pulsars far from the core.
A pulsar kicked out of the core by a dynamical interaction should sink back to the core due to dynamical friction, as long as it remains bound \citep{Phinney1992}.
Hence detections of MSPs far from the core suggests that the secondary encounter which ejected the pulsar from the binary happened relatively recently so the overall rate of such encounters is indeed enhanced in high-$\eta$ clusters.

We follow \citet{AbbottAbbott2017b} and prioritize targets according to the cluster figure of merit $\eta^{1/2}D^{-1}$, where $D$ is the distance to the cluster.
The figure of merit follows from the age-based limit on strain amplitude previously used in searches for CWs from supernova remnants \citep{WetteOwen2008, AbbottAbbott2021a, AbbottAbbott2022b}, viz. \begin{equation} h_0^{\mathrm{max}} = 1.26 \times 10^{-24} \left(\frac{3.3\,\mathrm{kpc}}{D}\right)\sqrt{\frac{300\,\mathrm{yr}}{a}}, \end{equation} where $a$ is the age of the object.
Recalling that $\eta$ is the reciprocal of the binary disruption timescale, and that shorter timescales are preferred for our application, we replace $a$ by $\eta^{-1}$, following Ref. \cite{AbbottAbbott2017b}.
This approach yields the figure of merit $\eta^{1/2}D^{-1} = \rho_{\mathrm{c}}^{1/4}r_{\mathrm{c}}^{-1/2}D^{-1}$ (up to a multiplicative constant which we ignore, as we are only interested in the ranking of the clusters).

Ranking the clusters listed in the Harris catalog \cite{Harris1996} according to this figure of merit and taking the top five, we arrive at the list in Table \ref{tbl:gc_topcands}.
\begin{table*}[]
    \centering
    \begin{tabular}{lllllll}\toprule
    Name & Right ascension & Declination & $r_{\mathrm{c}}$ [pc] & $\log_{10}(\rho_{\mathrm{c}}/L_{\odot}\,\mathrm{pc}^{-3})$ & $D$ [kpc] & $\eta^{1/2}D^{-1}$\\\hline
NGC 6325 & $17^{\mathrm{h}}17^{\mathrm{m}}59.21^{\mathrm{s}}$ & $-23^\circ45'57.6''$ & $6.8 \times 10^{-2}$ & 5.52 & 7.8 & 11.8\\
Terzan 6 & $17^{\mathrm{h}}50^{\mathrm{m}}46.38^{\mathrm{s}}$ & $-31^\circ16'31.4''$ & $9.9 \times 10^{-2}$ & 5.83 & 6.8 & 13.4\\
NGC 6540 & $18^{\mathrm{h}}06^{\mathrm{m}}08.60^{\mathrm{s}}$ & $-27^\circ45'55.0''$ & $4.6 \times 10^{-2}$ & 5.85 & 5.3 & 25.4\\
NGC 6544 & $18^{\mathrm{h}}07^{\mathrm{m}}20.58^{\mathrm{s}}$ & $-24^\circ59'50.4''$ & $4.4 \times 10^{-2}$ & 6.06 & 3.0 & 52.2\\
NGC 6397 & $17^{\mathrm{h}}40^{\mathrm{m}}42.09^{\mathrm{s}}$ & $-53^\circ40'27.6''$ & $3.3 \times 10^{-2}$ & 5.76 & 2.3 & 65.5\\\hline
    \end{tabular}
    \caption{Cluster parameters for the five clusters targeted in this paper, from \citet{Harris1996}. The sky position (columns two and three) refers to the centroid of the light distribution. The core radius is denoted by $r_{\mathrm{c}}$, the central luminosity density is denoted by $\rho_{\mathrm{c}}$, and the distance is denoted by $D$. The final column is the figure of merit used to rank the clusters, as discussed in Section \ref{subsec:setup_targets}.}
    \label{tbl:gc_topcands}
\end{table*}
The cluster with the second highest figure of merit is NGC 6544, which was previously targeted in the 10-day coherent search using data from the sixth LIGO science run performed by \citet{AbbottAbbott2017b}.
We caution that the figure of merit is \emph{not} directly related to an expected strain amplitude, despite its origin in the age-based limit for supernova remnants.
It should instead be regarded as a convenient encapsulation of the two primary criteria for this search: isolated sources (higher $\eta$) are better, as are closer sources (lower $D$).
The target selection strategy employed here is relatively simple and by no means optimal.
A comprehensive study of how CW sources might form in globular clusters, and how observable cluster parameters correlate with these formation processes, is certainly worthwhile but beyond the scope of this work.

\subsection{Data}
\label{subsec:setup_data}
This paper analyzes data from the third observing run (O3) of the Advanced LIGO detectors \citep{AbbottAbe2023}, two dual-recycled Fabry-P\'{e}rot Michelson interferometers with $4\,\mathrm{km}$ arms located in Hanford, Washington (H1) and Livingston, Louisiana (L1) in the United States of America \citep{BuikemaCahillane2020}.
The data cover the time period 1st April 2019 15:00 UTC (GPS 1238166018) to 27th March 2020 17:00 UTC (GPS 1269363618), with a commissioning break between 1st October 2019 and 1st November 2019 \citep{DCCsegments}.
We analyze a calibrated data stream \citep{SunGoetz2020, SunGoetz2021}, and so are working with measurements of the dimensionless strain $h$.

The O3 LIGO data contain noise transients.
A self-gating procedure mitigates their impact on continuous wave searches \citep{DavisAreeda2021, DCCgating}.
The detector data are packaged into ``Short Fourier Transforms'' (SFTs) generated from $1800\,\mathrm{s}$ of data.
Multiple SFTs are combined to construct the \bstat{} in each coherent segment of length $T_{\mathrm{coh}}$ (see Section \ref{subsec:alg_bstat}).
In the event that there are not enough data available to compute the $\mathcal{B}$-statistic in a given coherent segment, due to detector downtime, the value of $\ln L(o, q_i)$ in that segment is set to zero for all $q_i$.
In this paper the only data gaps significant enough to prevent the computation of the \bstat{} for a coherent segment occur during the commissioning break mentioned above.

\subsection{Coherence time}
\label{subsec:setup_tcoh}
The duration of each coherent step $T_{\mathrm{coh}}$ is an important parameter.
It controls the sensitivity of the search, although the dependence is fairly weak, scaling roughly as $(T_{\mathrm{obs}}T_{\mathrm{coh}})^{-1/4}$ \citep{KrishnanSintes2004,Wette2012}.
It also plays a role in determining the HMM's tolerance to deviations away from a deterministic signal model, either via $\Phi_{\mathrm{w}}(t)$ in Eq. (\ref{eqn:signal_model}), or via errors in the parameters which determine the first and third terms of Eq. (\ref{eqn:signal_model}).
In this paper we cover a wide parameter space, so it makes sense to set $T_{\mathrm{coh}}$ relatively low to reduce the number of templates and hence the computational cost
In this search we adopt $T_{\mathrm{coh}} = 5\,\mathrm{d}$ , c.f. $T_{\mathrm{coh}} = 10\,\mathrm{d}$ in some HMM searches published previously \cite{AbbottAbbott2019a,MiddletonClearwater2020,AbbottAbbott2022c,AbbottAbe2022b}.

Tracking intrinsic spin wandering is a secondary priority in this paper.
Electromagnetic measurements of timing noise in non-accreting pulsars, which are the objects primarily targeted in this search, suggest that $T_{\mathrm{coh}} \sim 5\,\mathrm{d}$ with the transition matrix described in Section \ref{subsec:alg_hmm} allows conservatively for more spin wandering than is observed, assuming that the spin wandering in the CW signal tracks the electromagnetic observations.
The timescale of decoherence due to spin wandering is typically $\gtrsim 1\,\mathrm{yr}$ \citep{Jones2004}.
Hence any plausible choice of $T_{\mathrm{coh}}$ with regard to computational cost ($T_{\mathrm{coh}} \lesssim 30\,\mathrm{d}$) allows for more spin wandering than what is expected astrophysically.
As a precaution, we inject spin wandering above the astrophysically expected levels when assessing the performance of the HMM method in Sections \ref{subsec:setup_hmm} and \ref{sec:sens}.
This allows us to verify that we remain sensitive to a wide range of signal models, including some that are unlikely \emph{a priori}.

\subsection{HMM control parameters}
\label{subsec:setup_hmm}
\subsubsection{Timing noise strength}
\label{subsubsec:setup_hmm_sigma}
The phase-tracking implementation of the HMM assumes that the evolution of the rotational phase is driven by a random walk in the frequency, with a strength parametrised by a control parameter $\sigma$ [see Equations (\ref{eqn:trans_f_evo})--(\ref{eqn:trans_wn_corr})].
Our approach here is to fix $T_{\mathrm{coh}}$ based on the particular details of the search, whether that involves prior belief about the nature of the spin wandering, or a desire to keep the number of sky position/spindown templates under control (see Section (\ref{subsec:setup_tcoh}).
With $T_{\mathrm{coh}}$ fixed, we then set $\sigma$ to be consistent with the signal model implied by our choice of $T_{\mathrm{coh}}$.
Setting $\sigma$ appropriately is a subtle task.
If $\sigma$ is too large, the transition probabilities become uniform in both frequency and phase, the dynamical constraints imposed by the noise model are lost, and the method reduces approximately to the phase-maximized \fstat{} \cite{JaranowskiKrolak1998}. 
If $\sigma$ is too small then the HMM struggles to track stochastic spin wandering or other unexpected phase evolution.
\citet{MelatosClearwater2021} recommended \begin{equation} \sigma = \sigma^* = (4T_{\text{coh}}^3)^{-1/2} \label{eqn:sigma_thumb}\end{equation}
but did not validate Eq. (\ref{eqn:sigma_thumb}) empirically.

There is no unique prescription for choosing $\sigma$.
The uncertainty about the physical origin of the time-varying mass-quadrupole moment makes it difficult to argue in general that one choice of $\sigma$ (reflecting the assumed nature of the underlying noise process) is optimal for a given search.
However, $\sigma$ endows the signal model with flexibility and allows the HMM to absorb small errors in parameters determining the secular spin evolution, e.g. sky position and spin-down rate.
This in turn reduces the number of templates and hence the computational cost.
As the HMM is restricted by construction to moving by at most one frequency bin per timestep, there is a limit to the parameter error which can be absorbed.\footnote{For sufficiently large mismatches in sky position or spindown a significant fraction of the total power is also lost in each coherent step, whether or not the HMM is allowed to track the frequency evolution of the signal across coherent steps.}
The aim, then, is to check whether Eq. (\ref{eqn:sigma_thumb}) strikes an acceptable balance, allowing for both secular and stochastic deviations away from a Taylor-expanded signal model, as reflected in our choice of $T_{\mathrm{coh}}$, without discarding the advantage gained by phase tracking.

We test the suitability of Eq. (\ref{eqn:sigma_thumb}) via synthetic data injections into both Gaussian noise and the LIGO O3 data.
For a range of $h_0$ values we compute the detection probability $P_{\text{d}}$ for $\sigma = 0.2\sigma^*$, $\sigma^*$, and $5\sigma^*$ as well as for no phase tracking, at a fixed false alarm probability of $10^{-13}$ per frequency bin.
We inject moderate frequency wandering via stepwise changes in $\ddot{f}$; in each of 52 7-day chunks of injected data $\ddot{f}$ is randomly chosen from a uniform distribution spanning $-1.13 \times 10^{-18} \leq \ddot{f}/(1\,\mathrm{Hz}\,\mathrm{s}^{-2}) \leq 1.13 \times 10^{-18}$.
The values of $\phi$, $f$, and $\dot{f}$ are updated self-consistently as the signal evolves.
With the chosen $\ddot{f}$ distribution, $f(t)$ wanders by approximately 28 frequency bins on average, after correcting for the mean $\dot{f} = T_{\mathrm{obs}}^{-1}[f(T_{\mathrm{obs}})-f(0)]$.
Note that the timescale of $\ddot{f}$ evolution is deliberately chosen not to coincide with $T_{\text{coh}} = 5\,\mathrm{d}$ to challenge the algorithm.
The sky position (right ascension $\alpha$, declination $\delta$) and initial value of $\dot{f}$ is chosen at random.
The value of $f$ is chosen uniformly over the full search range when injecting into synthetic Gaussian noise, but fixed at $645.1442\,\mathrm{Hz}$ when injecting into LIGO O3 data to avoid intersecting with non-Gaussian artefacts in the data.
Signal recovery is performed at the injected sky position, with a $10^{-4}\,\mathrm{Hz}$ frequency band centred on $f=f(0)$ and $\dot{f} = \left[f(T_{\text{obs}}) - f(0)\right]/T_{\text{obs}}$.
The injection and recovery setup is specified in Table \ref{tbl:sigma_test_params}.
\begin{table}[h]
    \centering
    \begin{tabular}{ll}
      \toprule
      Parameter  &  Value\\\hline
      $\alpha$   & $\mathcal{U}(0, 2\pi)\,\mathrm{rad}$\\
      $\cos\delta$ & $\mathcal{U}(-1,1)$\\
      $f(0)$ & $\mathcal{U}(100, 700)\,\mathrm{Hz}$ \\
      $\dot{f}(0)$ & $\mathcal{U}(-5\times 10^{-10}, 0)\,\mathrm{Hz}\,\mathrm{s}^{-1}$ \\
      $\cos\iota$ & $\mathcal{U}(-1,1)$ \\
      $\psi$ & $0.49\,\mathrm{rad}$\\
      $\phi_0$ & $0\,\mathrm{rad}$\\
      Detectors & H1, L1\\
      $S_h(f)^{1/2}$ & $4 \times 10^{-24}\,\mathrm{Hz}^{-1/2}$\\
      Spin wandering timescale & $7\,\mathrm{d}$\\
      $T_{\mathrm{coh}}$ & $5\,\mathrm{d}$\\
      $T_{\mathrm{obs}}$ & $364\,\mathrm{d}$\\\toprule
    \end{tabular}
    \caption{Injection and recovery parameters for the synthetic data tests used to motivate the choices of HMM control parameters as described in Section \ref{subsec:setup_hmm}. $\mathcal{U}(a,b)$ denotes a uniform distribution between $a$ and $b$.}
    \label{tbl:sigma_test_params}
\end{table}

\begin{figure}
    \centering
    \includegraphics[width=\columnwidth]{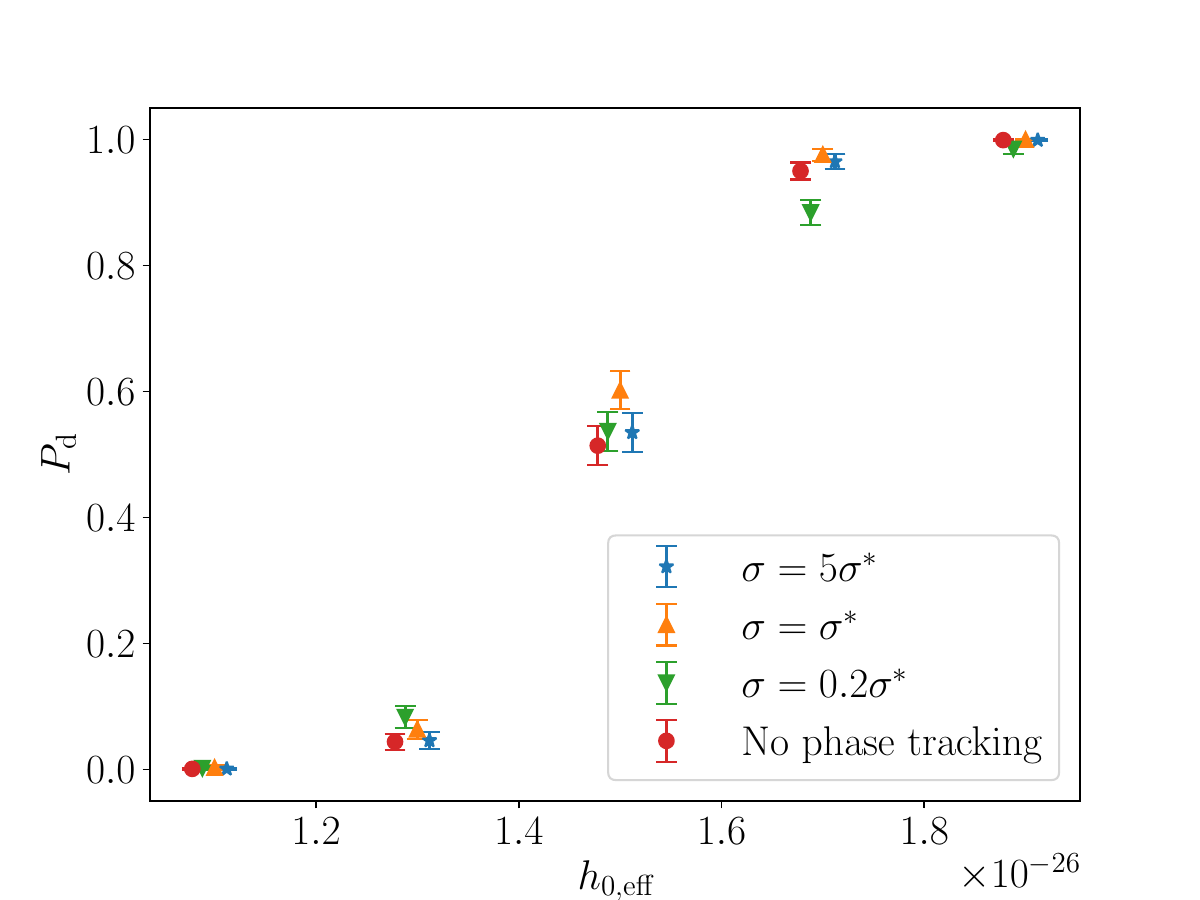}\\
    \includegraphics[width=\columnwidth]{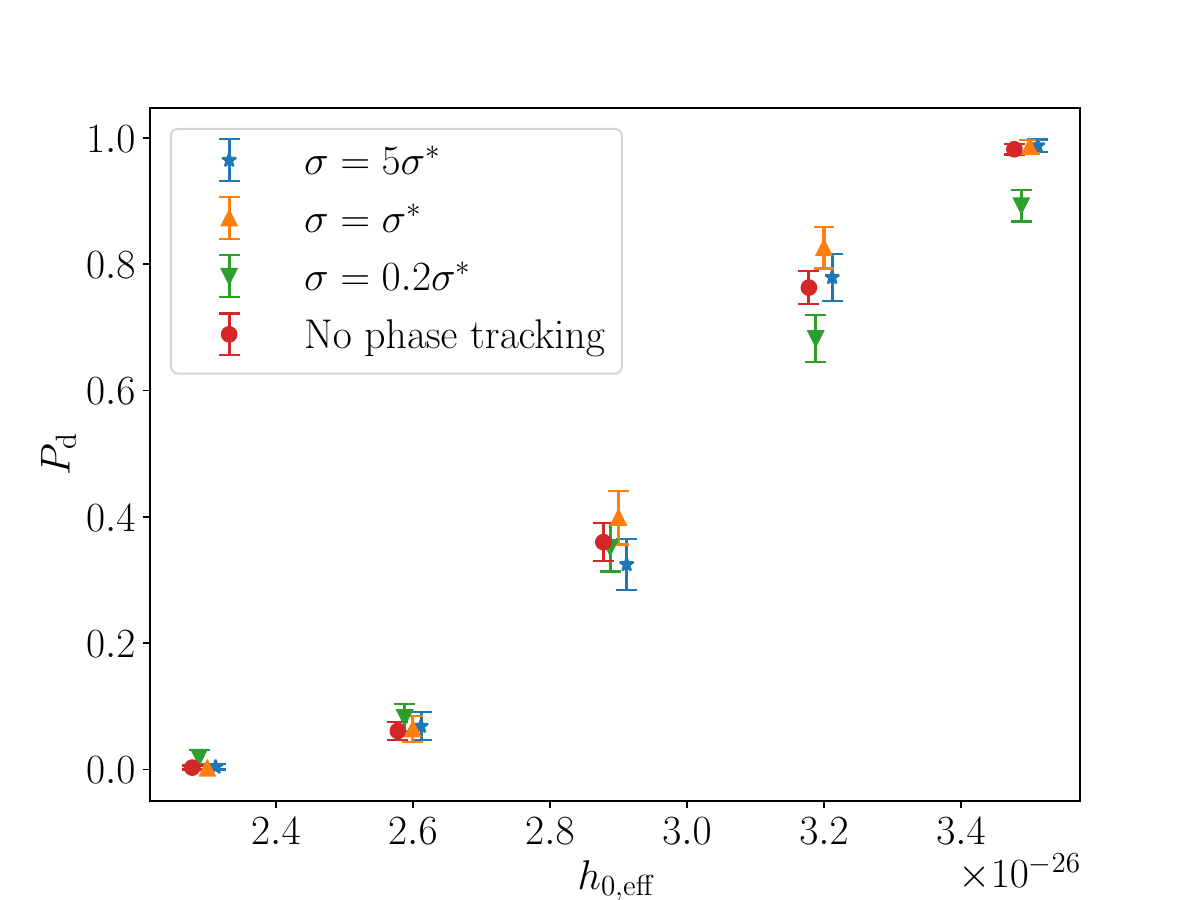}
    \caption{Detection probability $P_{\text{d}}$ versus effective wave strain $h_{0,\text{eff}}$ for three choices of the HMM control parameter $\sigma$ as well as a control experiment with zero phase tracking (see legend), as described in Section \ref{subsubsec:setup_hmm_sigma}. The top panels shows the results for injections into synthetic Gaussian noise, while the bottom panel shows the results for injections into the real LIGO O3 data. The $h_{0,\mathrm{eff}}$ scales differ, because the noise floor is not the same for the two cases.}
    \label{fig:sigma_h0}
\end{figure}

In order to find the detection probability $P_{\text{d}}$ at a given probability of false alarm $P_{\text{fa}}$ we require an estimate of the noise-only distribution of the log likelihoods returned by the HMM.
To obtain this we fit an exponential tail to the distribution of log likelihoods returned from a large number of simulations, using the procedure outlined in Appendix \ref{apdx:thresholds} and adopted in previous searches \citep{AbbottAbbott2022c, KneeDu2024}.
The injected amplitude of the signal is characterised by $h_{0,\mathrm{eff}}$, a combination of $h_0$ and $\cos\iota$ which determines approximately the signal-to-noise ratio for many CW search methods and is defined as \citep{MessengerBulten2015} \begin{equation} \left(h_{0,\text{eff}}\right)^2 = h_0^2\frac{[(1+\cos^2\iota)/2]^2 + (\cos\iota)^2}{2}. \label{eqn:h0eff}\end{equation}
For $\cos\iota = 1$ we have $h_{0,\text{eff}} = h_0$, for $\cos\iota = 0$ we have $h_{0,\text{eff}} = h_0/\sqrt{8}$, and averaging over a uniform distribution in $\cos\iota$ gives $h_{0,\text{eff}} = (2/5)^{1/2}h_0$.
Figure \ref{fig:sigma_h0} graphs $P_{\text{d}}$ as a function of $h_{0,\mathrm{eff}}$ for the choices of $\sigma$ outlined above, showing both the case where the signals are injected into simulated Gaussian noise and the case where the signals are injected into the LIGO O3 data.
In both cases the choice $\sigma = \sigma^*$ performs as well as or better than the alternatives as well as no phase tracking across the full $h_{0,\mathrm{eff}}$ range.
We adopt $\sigma = \sigma^*$ throughout the rest of this paper.


\subsubsection{Number of phase bins}
\label{subsubsec:setup_hmm_nphi}
Another important tunable parameter in the HMM is the number of bins used to track the signal phase, $N_\Phi$.
\citet{MelatosClearwater2021} fixed $N_\Phi = 32$.
It is worth checking how reducing $N_\Phi$ affects performance.
We seek to reduce $N_\Phi$ without sacrificing sensitivity, because the computational cost is proportional to $N_\Phi$.

\begin{figure}
    \centering
    \includegraphics[width=\columnwidth]{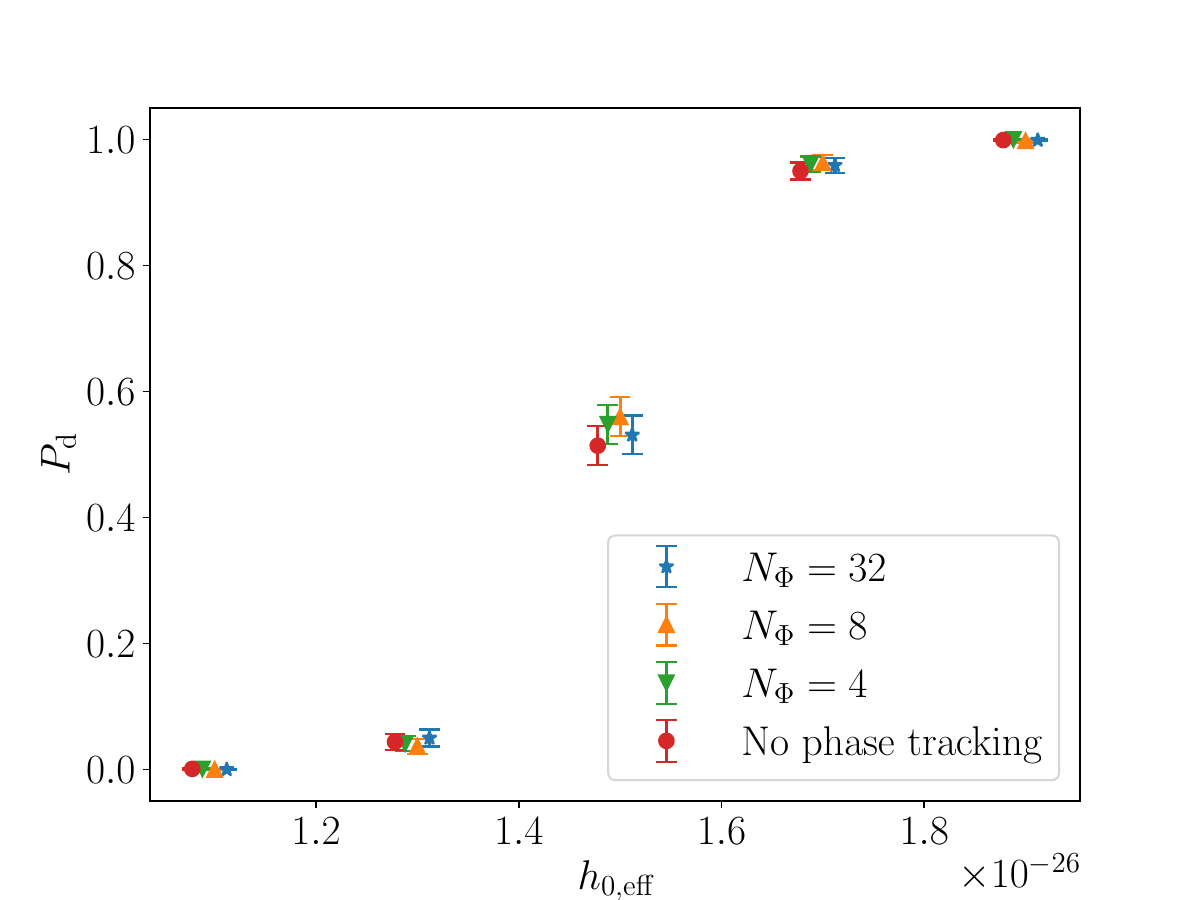}
    \caption{Detection probability $P_{\text{d}}$ versus effective wave strain $h_{0,\text{eff}}$ for three choices of the number of phase bins $N_\Phi$ as well as a control experiment with zero phase tracking, as described in Section \ref{subsubsec:setup_hmm_nphi}. The three choices of $N_\Phi$ perform similarly.}
    \label{fig:phibins_h0}
\end{figure}

To this end we test $N_\Phi = 4$, $8$, and $32$ (c.f. Section \ref{subsubsec:setup_hmm_sigma}).
Sensitivity curves showing $P_{\text{d}}$ as a function of $h_{0,\mathrm{eff}}$ are displayed in Figure \ref{fig:phibins_h0}.
The differences are relatively minor, so we proceed with $N_\Phi = 8$ in what follows.

\subsection{Parameter ranges}
\label{subsec:param_ranges}
Ranges must be set for four parameters: the frequency and frequency derivative of the CW signal, and the sky regions covered for each cluster.
Here we discuss each in turn, and justify the choices made in the implementation of the search.

\subsubsection{Frequency and frequency derivative}
\label{subsubsec:param_ranges_f_fdot}
In principle the possible frequency of CW emission from neutron stars spans a wide range.
Observed pulsar spin frequencies range from less than $1\,\mathrm{Hz}$ \citep{TanBassa2018, Reig2011} up to $716\,\mathrm{Hz}$ \citep{HesselsRansom2006}.
Although the relationship between the rotation frequency $f_{\text{rot}}$ and the CW signal frequency $f$ depends on the precise nature of the emission mechanism, one normally assumes $f_{\text{rot}} \leq f \leq 2f_{\text{rot}}$ \citep{Riles2023}.
We remind the reader that the notional target for this search is a recycled neutron star, whose rotation frequency is likely to fall between $0.1\,\mathrm{kHz}$ and $0.7\,\mathrm{kHz}$, as for observed millisecond pulsars.
Note that ${\sim}80\%$ of pulsars above $0.1\,\mathrm{kHz}$ fall in the range $0.1 \leq f_{\text{rot}}/1\,\mathrm{kHz} \leq 0.4$ \citep{ManchesterHobbs2005}.
There are also some practical considerations when it comes to selecting the frequency range.
The LIGO interferometers are polluted by Newtonian noise below ${\sim} 20\,\mathrm{Hz}$, and narrowband spectral features (instrumental lines) below $\sim 0.1\,\mathrm{kHz}$ \citep{CovasEffler2018, O3LinesSite}.

\begin{figure}
    \centering
    \includegraphics[width=\columnwidth]{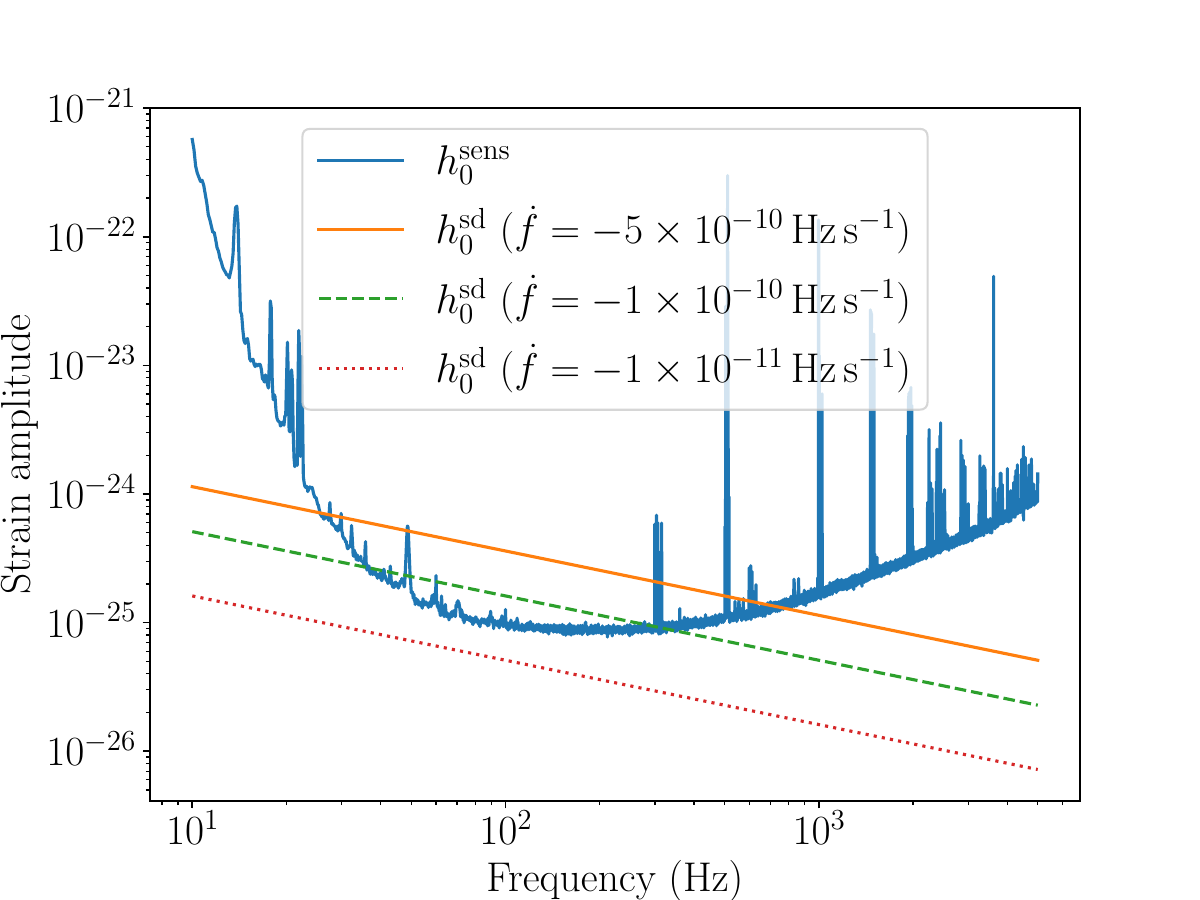}
    \caption{Comparison of $h_0^{\mathrm{sens}}$ [Eq. (\ref{eqn:sens_est})] with $h_0^{\mathrm{sd}}$ [Eq. (\ref{eqn:spindown_limit})], with $h_0^{\mathrm{sd}}$ computed for three values of $\dot{f}$ ($D = 5\,\mathrm{kpc}$ in all cases). For the $\dot{f} = -5 \times10^{-10}\,\mathrm{Hz}\,\mathrm{s}^{-1}$ case, corresponding to the bound adopted in this paper (see Section \ref{subsubsec:param_ranges_f_fdot}), we have $h_0^{\mathrm{sens}} < h_0^{\mathrm{sd}}$ for $f \lesssim 0.8\,\mathrm{kHz}$.}
    \label{fig:h0sens_h0sd}
\end{figure}

The spin-down rate of a CW signal is weakly constrained \emph{a priori}.
Under the optimistic assumption that CW emission is responsible for all of the spindown of the emitting body, the strain amplitude is equal to \begin{align} h_0^{\mathrm{sd}} = &2.6\times 10^{-25} \left(\frac{1\,\mathrm{kpc}}{D}\right)\left(\frac{I_{zz}}{10^{38}\,\mathrm{kg}\,\mathrm{m}^2}\right)^{1/2}\nonumber\\
&\times\left(\frac{100\,\mathrm{Hz}}{f}\right)^{1/2}\left(\frac{|\dot{f}|}{10^{-11}\,\mathrm{Hz}\,\mathrm{s}^{-1}}\right)^{1/2}\label{eqn:spindown_limit}\end{align} where $I_{zz}$ is the component of the moment of inertia tensor about the rotation axis.
The expected sensitivity of this search is roughly \citep{JaranowskiKrolak1998,WetteOwen2008, SunMelatos2018} \begin{equation} h_0^{\mathrm{sens}} \approx 35 S_h^{1/2}(T_{\mathrm{obs}}T_{\mathrm{coh}})^{-1/4}, \label{eqn:sens_est}\end{equation} which implies $\dot{f} \sim -10^{-10}\,\mathrm{Hz}\,\mathrm{s}^{-1}$ for $h_0^{\mathrm{sens}} = h_0^{\mathrm{sd}}$.
To beat the spin-down limit we require $h_0^{\mathrm{sens}} < h_0^{\mathrm{sd}}$. 
We therefore wish to pick a bound on $\dot{f}$ such that the cross-over point $h_0^{\mathrm{sens}} = h_0^{\mathrm{sd}}$ occurs at a high enough frequency that we beat the spin-down limit for a significant fraction of plausible signal frequencies.
Figure \ref{fig:h0sens_h0sd} shows the $h_0^{\mathrm{sens}}$ curve with three $h_0^{\mathrm{sd}}$ curves overplotted with $-\dot{f} = 10^{-11}\,\mathrm{Hz}\,\mathrm{s}^{-1}$, $-\dot{f} = 1 \times 10^{-10}\,\mathrm{Hz}\,\mathrm{s}^{-1}$, and $-\dot{f} = 5 \times 10^{-10}\,\mathrm{Hz}\,\mathrm{s}$.
We set $D = 5\,\mathrm{kpc}$ in all cases.
The $\dot{f} = -5 \times 10^{-10}\,\mathrm{Hz}\,\mathrm{s}^{-1}$ curve crosses over $h_0^{\mathrm{sens}}$ at approximately $f = 0.8\,\mathrm{kHz}$.
Hence for $0.1 \leq f/1\,\mathrm{kHz} \leq 0.8$ and $-5 \leq \dot{f}/10^{-10}\,\mathrm{Hz}\,\mathrm{s}^{-1} \leq 0$ we expect to approach or beat the spin-down limit across the full frequency range for each cluster.
The range includes the bulk of the astrophysical prior on the signal frequency, recalling that the majority of MSPs have $f_{\mathrm{rot}} < 0.4\,\mathrm{kHz}$ and thus $f < 0.8\,\mathrm{kHz}$.
We therefore adopt $0.1 \leq f/1\,\mathrm{kHz} \leq 0.8$ and $-5 \times 10^{-10} \leq \dot{f}/1\,\mathrm{Hz}\,\mathrm{s}^{-1} \leq 0$ in the present paper.
To facilitate data management, the total frequency range is divided into $0.5\,\mathrm{Hz}$ sub-bands which are searched independently.

We note that electromagnetically timed MSPs have spin-down rates orders of magnitude lower than our sensitivity limits according to Fig. \ref{fig:h0sens_h0sd} \citep{ManchesterHobbs2005}.
Their spin down is typically attributed to magnetic dipole braking, with the recycling process suppressing the magnetic dipole moment \citep{KonarBhattacharya1997, MelatosPhinney2001,PayneMelatos2004,MelatosPayne2005,PriymakMelatos2011,Mukherjee2017}.
Hence the signals to which this search is sensitive are most likely to come from a new class of objects with significantly higher spin-down torques, possibly dominated by gravitational radiation reaction \citep{MelatosPayne2005,  Palomba2005}.


\subsubsection{Sky position}
\label{subsubsec:param_ranges_sky}
Four of the clusters targeted in this search (NGC 6325, NGC 6397, NGC 6544, and Terzan 6) are core-collapsed; their density profiles peak sharply at the core.
At the same time, our cluster selection criterion favors clusters which produce disrupted binaries where the neutron star may be ejected from the core.
Indeed clusters similar to those selected are observed to contain millisecond pulsars many core radii away from the centre of the cluster \citep{VerbuntFreire2014}.
Therefore, as a precaution, we extend the search to cover sky positions outside the core.
Specifically, we extend the search out to the tidal radius as recorded in \citet{Harris1996}, ensuring that every point within this radius is covered by at least one sky position template.

\subsection{Parameter spacing}
\label{subsec:param_spacing}
In this paper we search a fairly wide parameter domain (precisely how wide is discussed in Section \ref{subsec:param_ranges}).
In general, multiple templates are needed to cover the domain.
In this section we discuss how closely spaced the templates must be.
Sophisticated work has been done on this problem in the context of CW searches (e.g. \cite{Prix2007a, Prix2007b,Pletsch2010, Wette2015}), but we proceed empirically here, because the Viterbi tracking step makes it difficult to apply previous results.

\subsubsection{Frequency and frequency derivative}
\label{subsubsec:setup_param_spacing_f_fdot}
We follow previous HMM-based searches and take the frequency resolution to be $(2T_{\mathrm{coh}})^{-1}$.
Other choices are possible.
In principle any desired frequency resolution may be achieved by padding or decimating the data, e.g. $\delta f = (10T_{\mathrm{coh}})^{-1}$ \citep{AbbottAbbott2021a, AbbottAbe2022c}, but $\delta f = (2T_{\mathrm{coh}})^{-1}$ has been successfully used many times previously (see e.g. Refs. \citep{SuvorovaSun2016, AbbottAbbott2017a, AbbottAbbott2021a, AbbottAbbott2022c, AbbottAbe2022b}) and is assumed in Eq. (\ref{eqn:sigma_thumb}).

\begin{figure}
    \centering
    \includegraphics[width=\columnwidth]{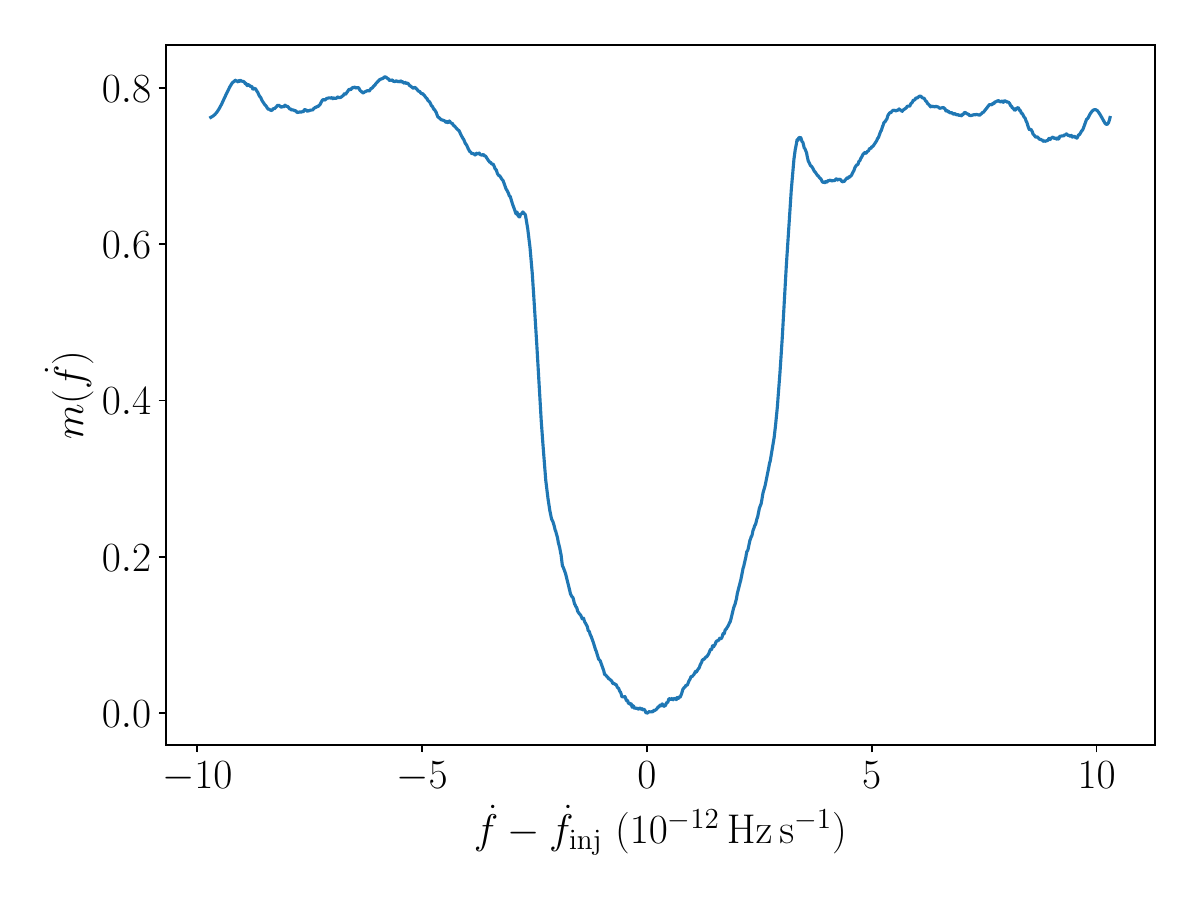}
    \caption{Example mismatch profile $m(\dot{f})$ versus $\dot{f}$, as defined by Equation (\ref{eqn:fdot_mismatch}). 
    The profile is generated from a single random realisation of both noise and signal, with all parameters other than $\dot{f}$ fixed to their injected values. The injected $\dot{f}$ value is denoted by $\dot{f}_{\mathrm{inj}}$.}
    \label{fig:fdot_mismatch_profile}
\end{figure}

The HMM accommodates some displacement between the $\dot{f}$ value used to compute the \bstat{} and the true $\dot{f}$ by letting the signal wander into adjacent frequency bins at the boundaries between coherent segments.
To determine the $\dot{f}$ resolution we empirically estimate the mismatch, i.e. the fractional loss of signal power which is incurred as the value of $\dot{f}$ used in the search is displaced away from the true value, $\dot{f}_{\mathrm{inj}}$, with \begin{equation} m(\dot{f}) = \frac{\mathcal{L}'(\dot{f}) - \mathcal{L}'(\dot{f}_{\mathrm{inj}})}{\mathcal{L}'(\dot{f}_{\mathrm{inj}})}, \label{eqn:fdot_mismatch}\end{equation} where we define $\mathcal{L}' = \mathcal{L} - \mathcal{L}_\text{noise}$ and $\mathcal{L}_{\text{noise}}$ is the mean likelihood returned in noise over a $0.5\,\mathrm{Hz}$ sub-band.
Note that $\mathcal{L}'(\dot{f})$ is the maximum $\mathcal{L}$ value over a $0.5\,\mathrm{Hz}$ sub-band centred on the injected frequency, and so $m(\dot{f})$ does not approach $1$ but plateaus at approximately $0.8$, due to the maximisation over noise-only frequency paths (whereas $\mathcal{L}_{\mathrm{noise}}$ is the mean noise-only $\mathcal{L}$ value in a $0.5\,\mathrm{Hz}$ sub-band, not the maximum).
We compute $m(\dot{f})$ by injecting moderately loud signals into Gaussian noise, with $h_0 = 6 \times 10^{-26}$ and $S_h^{1/2} = 4\times 10^{-24}\,\mathrm{Hz}^{-1/2}$.
An example of $m(\dot{f})$ is displayed in Fig. \ref{fig:fdot_mismatch_profile}, generated from a single realisation of noise and signal.
All mismatch curves are smoothed using a Savitzky-Golay filter.
The shape of $m(\dot{f})$ is consistent across noise realisations.
Hence we take the $\dot{f}$ resolution to be the median difference between the two $\dot{f}$ values satisfying $m(\dot{f}) = 0.2$, drawn from 100 noise realisations.
This gives an $\dot{f}$ resolution of $\delta \dot{f} = 3.7 \times 10^{-12}\,\mathrm{Hz}\,\mathrm{s}^{-1}$.
We verify that the resolution is independent of the injected $f$ and $\dot{f}$ values and adopt it throughout the search.

\subsubsection{Sky location}
\label{subsubsec:setup_param_spacing_sky}
A mismatch in sky location induces a small apparent variation in the signal frequency due to a residual Doppler shift.
Although the locations of globular clusters are well-known, they are extended objects on the sky.
In this search we take a conservative approach and search out to the tidal radius of each cluster, as discussed in Section \ref{subsubsec:param_ranges_sky}.
In general a single sky pointing does not suffice to cover the full sky area of each cluster.
As with a mismatch in $\dot{f}$, we expect the transition matrix in Equation (\ref{eqn:trans_f_evo}) to track some of the residual Doppler variation.
Nevertheless it is important to check how much variation can be accommodated before a significant amount of signal power is lost.

We specify the sky location by right ascension ($\alpha$) and declination ($\delta$), referenced to the J2000 epoch.
We empirically estimate the mismatch profiles $m(\alpha)$ and $m(\delta)$ and thence derive resolutions in $\alpha$ and $\delta$.
As with $\dot{f}$, we smooth the mismatch profiles using a Savitzky-Golay filter.
The treatment is somewhat more involved than in Section \ref{subsubsec:setup_param_spacing_f_fdot}, as $m(\alpha)$ and $m(\delta)$ depend on the injected values $\alpha_{\mathrm{inj}}$ and $\delta_{\mathrm{inj}}$ as well as $f$, i.e. their shape is not a universal function of sky position offset, because the Doppler modulation varies across the sky.
In particular, sky positions at declinations near the ecliptic poles experience relatively little Doppler modulation, and hence proportionally smaller residual Doppler modulations \citep{AbbottAbbott2017b}.
In this paper, we target five relatively small regions of the sky so we do not produce a full skymap of $m(\alpha, \delta)$.
Instead we measure the $\alpha$ and $\delta$ resolutions independently for each cluster.
The clusters themselves are fairly small (diameter $\lesssim 10\,\mathrm{arcmin}$), so we approximate the $\alpha$ and $\delta$ resolutions to be uniform within each cluster.

Each cluster is covered by ellipses arranged in a hexagonal grid, with their semimajor and semiminor axes given by the resolutions in $\delta$ and $\alpha$ corresponding to $m(\delta) \leq 0.2$ and $m(\alpha) \leq 0.2$.
The centre of one ellipse coincides with the cluster core, and the whole cluster out to the tidal radius is covered by at least one ellipse. 
The centres of the ellipses are taken as the sky position templates in the search.
For simplicity we do not consider degeneracy between sky position and $\dot{f}$, which may reduce modestly the number of templates, but adds to the complexity of implementing the search.

\begin{figure}
    \centering
    \includegraphics[width=\columnwidth]{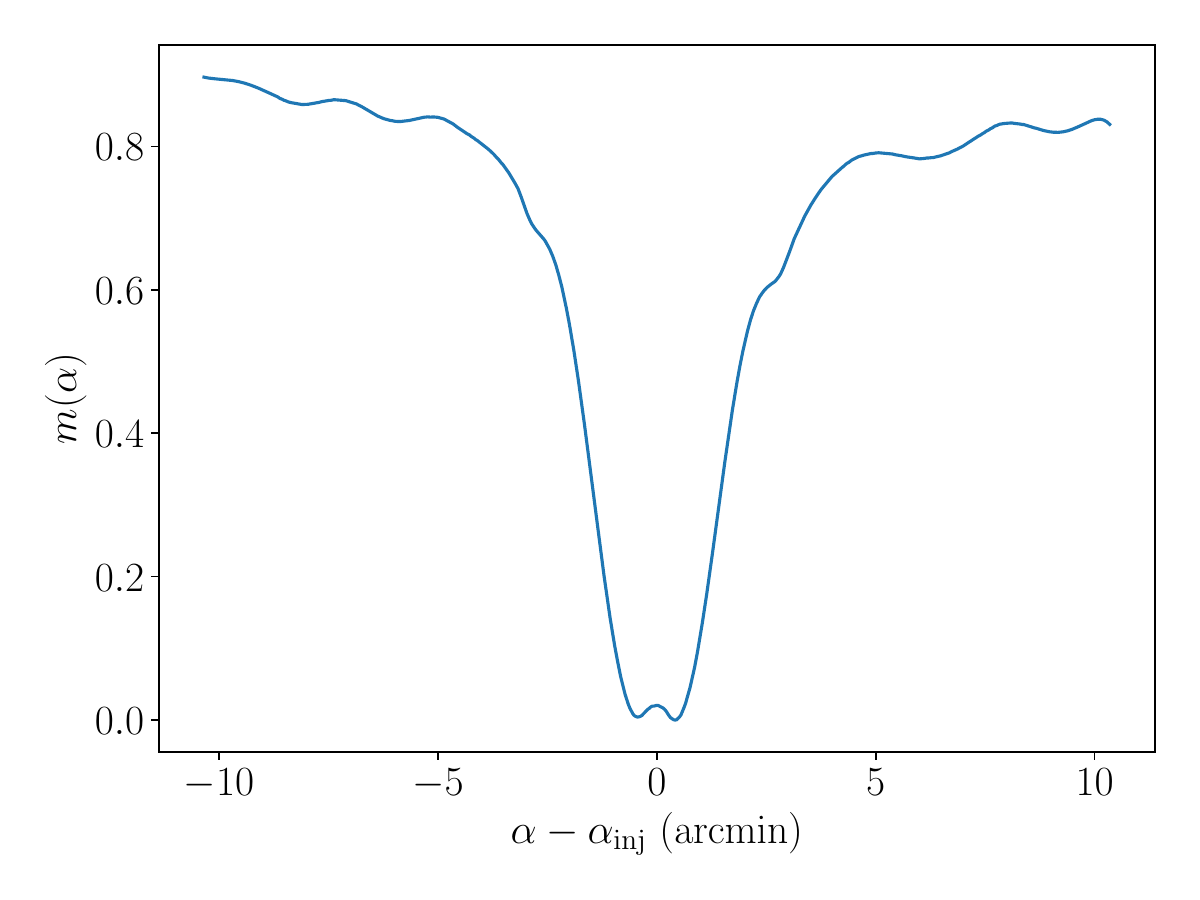}\\
    \includegraphics[width=\columnwidth]{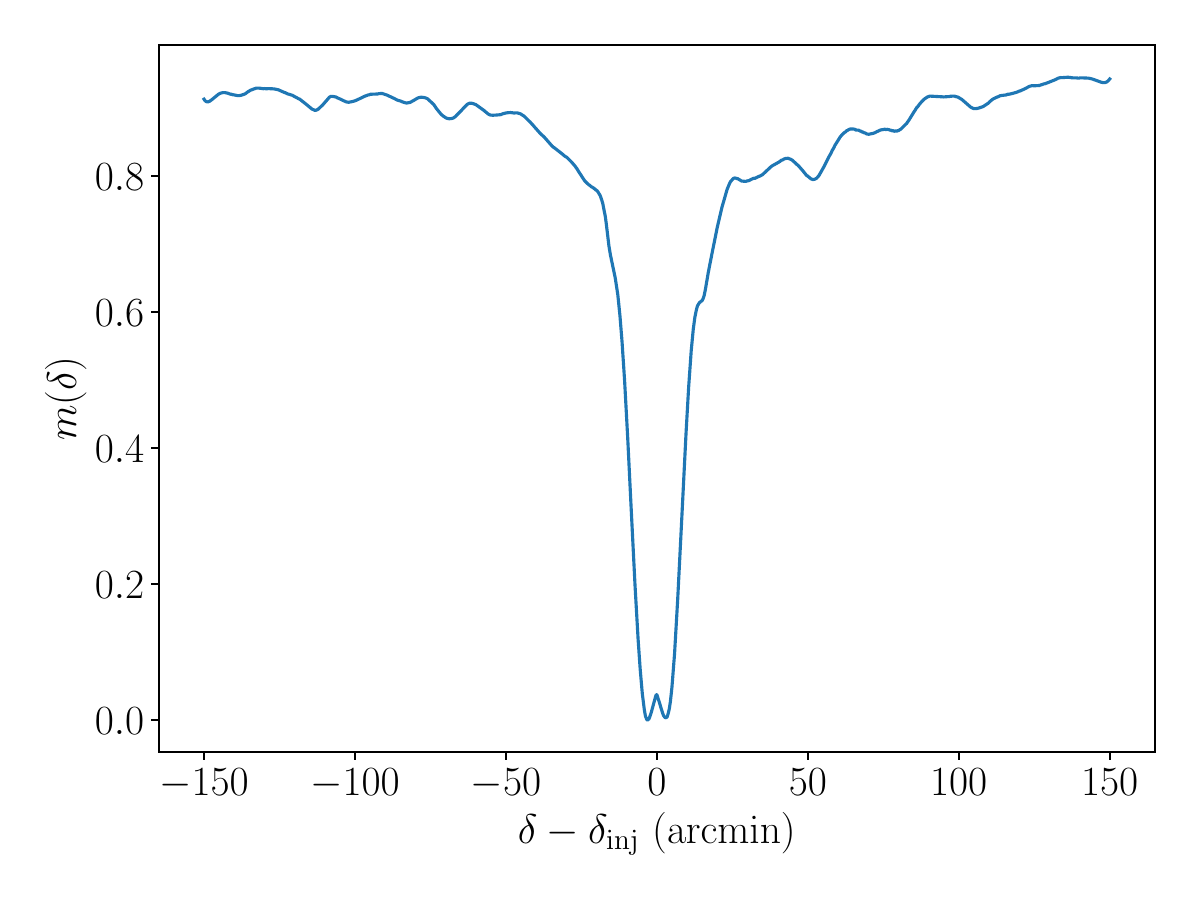}\\
    \includegraphics[width=\columnwidth]{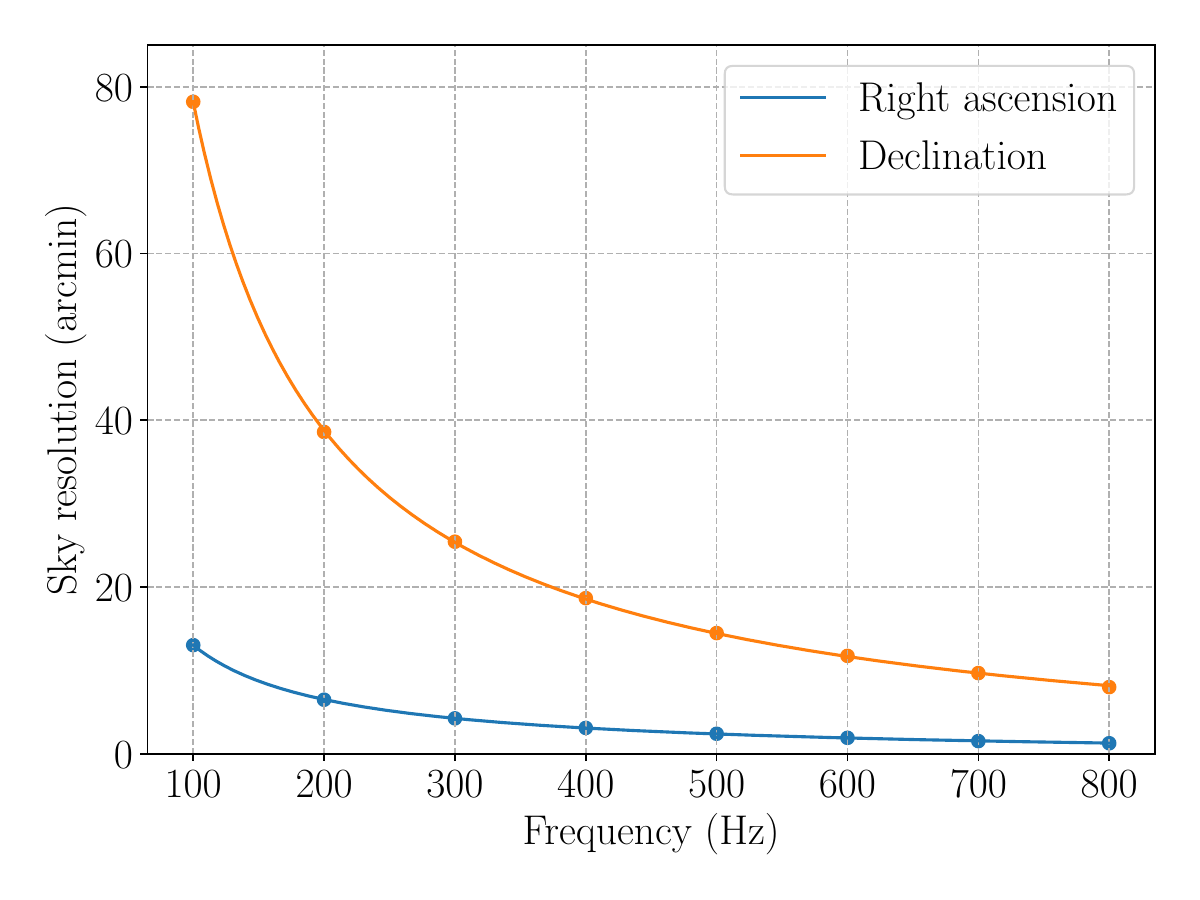}
    \caption{Example sky resolution computations. The top and middle panels show example mismatch profiles $m(\alpha)$ and $m(\delta)$ respectively, calculated for the sky location of Terzan 6 at a frequency of $600\,\mathrm{Hz}$ for a single random realisation of noise and signal. The resolutions are $1.9\,\mathrm{arcmin}$ and $15\,\mathrm{arcmin}$ --- note the difference in scale along the horizontal axes. The bottom panel shows the resolution in right ascension (blue curve) and declination (orange curve) as a function of frequency for the sky location of Terzan 6. Each point is derived by averaging the mismatch measured from 100 random realisations, and the curves are interpolated from the injection results by fitting a power law function, as described in the text.}
    \label{fig:skypos_freq_dependence}
\end{figure}

Examples of $m(\alpha)$ and $m(\delta)$ are shown in the top two panels of Fig. \ref{fig:skypos_freq_dependence}, calculated for Terzan 6 at a frequency of $600\,\mathrm{Hz}$. 
They are qualitatively similar, with well-defined troughs, but the resolutions, again defined as the distances between the two abcissae where $m = 0.2$, differ significantly --- the resolution inferred from the $m(\alpha)$ profile is $1.9\,\mathrm{arcmin}$, compared to $15\,\mathrm{arcmin}$ for $m(\delta)$.
The bottom panel of \ref{fig:skypos_freq_dependence} shows the measured resolution in $\alpha$ and $\delta$ as a function of signal frequency for a series of injections at the sky location of Terzan 6.
Clearly there is a strong frequency dependence, and it would be wasteful to simply apply the finest resolution across a wide band.
Consequently, we produce curves like those in the bottom panel of Fig. \ref{fig:skypos_freq_dependence} for each target.
We interpolate by fitting a power-law function $Af^{k} + B$ for $A$, $k$, and $B$ using the \textsc{lmfit} package \cite{NewvilleStensitzki2016} to find the resolution in each $0.5\,\mathrm{Hz}$ sub-band.

\section{Candidates}
\label{sec:results}
In this section we present the results of the search.
We set the log likelihood threshold in terms of the desired false alarm probaility $P_{\mathrm{fa}}$ in Section \ref{subsec:res_thresh}.
We study the candidates produced by a first-pass analysis in Section \ref{subsec:res_init_cand} and apply vetoes to reject those not of astrophysical origin.
Finally we study the veto survivors in more depth in Section \ref{subsec:res_vetos}.

\subsection{Thresholds}
\label{subsec:res_thresh}
We set a threshold log likelihood $\mathcal{L}_{\text{th}}$ above which candidates are kept for further analysis.
The threshold is determined by the desired $P_{\text{fa}}$, and is computed according to the procedure outlined in Appendix \ref{apdx:thresholds}.
In short, we obtain a sample of $\mathcal{L}$ values calculated in the absence of a signal, and fit an exponential distribution to the tail of this sample.
The exponential distribution is used to calculate the value of $\mathcal{L}_{\mathrm{th}}$ that gives the desired $P_{\mathrm{fa}}$.
Empirical noise-only distributions of $\mathcal{L}$ are obtained from $10^4$ random sky pointings in each of three $0.5\,\mathrm{Hz}$ bands beginning at $263.5\,\mathrm{Hz}$, $322\,\mathrm{Hz}$, and $703\,\mathrm{Hz}$, chosen arbitrarily but checked to ensure that the detector amplitude spectral densities (ASDs) show no obvious spectral artifacts.
We find that $\mathcal{L}_{\mathrm{th}}$ is similar across the three bands (within $\pm 1$ of the mean value per cluster), so we take the mean value over the three bands to be $\mathcal{L}_{\mathrm{th}}$ as a good approximation.
We do not obtain separate noise-only distributions for each cluster.

The choice of $P_{\text{fa}}$ is subjective.
Historically different searches have used a range of values \cite{AbbottAbbott2017b,AbbottAbbott2019b, PiccinniAstone2020,AbbottAbbott2021a,AbbottAbbott2022c,AbbottAbe2022b}.
In this paper, we take a relatively aggressive approach (i.e. $P_{\text{fa}}$ is relatively high), setting $\mathcal{L}_{\mathrm{th}}$ on a per-cluster basis such that across all templates in that cluster, we expect one false alarm on average, and so expect the full search to produce five false alarms.
This total number of expected false alarms is broadly comparable to other published HMM-based searches, which have adopted $P_{\mathrm{fa}}$ values giving false alarm rates ranging from 0.15 per search \citep{AbbottAbbott2021a} to 18 per search \citep{AbbottAbbott2022c}.
The search for CWs from NGC 6544 carried out by \citet{AbbottAbbott2017b} adopted a false alarm rate of 184 per search, while the search targeting Terzan 5 and the galactic center carried out by \citet{DergachevPapa2019} did not set a  threshold explicitly in terms of a false alarm rate.

The total number of sky templates and hence $\mathcal{L}_{\mathrm{th}}$ differs between clusters.
The per-cluster sky template counts and values of $\mathcal{L}_{\text{th}}$ corresponding to a false alarm rate of one per cluster are quoted in Table \ref{tbl:gc_lth}.
\begin{table}
    \centering
    \begin{tabular}{lrr}\toprule
       Cluster  & Sky templates & $\mathcal{L}_{\mathrm{th}}$ \\\hline
        NGC 6544 & 3022 & $278.9$\\
        NGC 6325 & 8416 & $283.9$\\
        NGC 6540 & 8957 & $284.2$ \\
        Terzan 6 & 28462 & $289.8$ \\
        NGC 6397 & 43909 & $291.9$ \\\toprule
    \end{tabular}
    \caption{Number of sky templates and $\mathcal{L}_{\mathrm{th}}$ values for each target cluster as discussed in Section \ref{subsec:res_thresh}, chosen to give a false alarm rate of one per cluster on average. Targets are ordered by the number of sky templates.}
    \label{tbl:gc_lth}
\end{table}
\subsection{First pass candidates and vetoes}
\label{subsec:res_init_cand}
The number of above-threshold candidates in each cluster is listed in the pre-veto column of Table \ref{tbl:veto_res}.
The total across all clusters is $6.4 \times 10^7$, with the candidates distributed approximately evenly among the five clusters\footnote{For comparison, the total number of Viterbi paths evaluated in this search is approximately $1.1 \times 10^{13}$.}.
As discussed in Section \ref{subsubsec:res_disturb_band}, these pre-veto numbers are lower bounds, as most candidates are from heavily disturbed sub-bands and are not saved.
The candidate counts in the subsequent columns are complete.

While $6.4 \times 10^7$ is far greater than the five expected based on the arguments in Section \ref{subsec:res_thresh}, we emphasise that those arguments are based on the assumption that the noise is approximately Gaussian.
All but a handful at most of the candidates in the pre-veto column of Table \ref{tbl:veto_res} are due to non-Gaussian features in the data.
In this section and the next we apply a sequence of veto procedures designed to reject candidates that are consistent with non-Gaussian noise.

\subsubsection{Disturbed sub-band veto}
\label{subsubsec:res_disturb_band}
Some of the $0.5\,\mathrm{Hz}$ sub-bands are so disturbed by non-Gaussian artifacts that they are unusable in the absence of noise cancellation \cite{TiwariDrago2015, DavisMassinger2019, DriggersVitale2019, Kimpson2024}.
In view of this, we impose a blanket veto on sub-bands which produce output files larger than $100\,\mathrm{MB}$ (corresponding to more than $1.67 \times 10^5$ candidates per file).
Between $3.9$\% and $5.4$\% of the full search band is removed thus, with the smallest number of removed sub-bands in NGC 6544 (55 out of 1400), and the highest in NGC 6397 (75 out of 1400).
These percentages are roughly in line with the all-sky O3a search reported by \citet{AbbottAbbott2021b}, for which 13\% of the observing band was removed due to an unmanageable number of candidates (noting that the band in that search extended to $2000\,\mathrm{Hz}$).
After this veto, approximately $4.9 \times 10^5$ candidates remain.

\subsubsection{Known-lines veto}
The known-lines veto is simple and efficient, rejecting many candidates with relatively little computational cost.
The LIGO Scientific Collaboration maintains a list of known instrumental lines \citep{DCClines, O3LinesSite}.
When assessing coincidence we must take into account that the line frequencies and bandwidths in Ref. \citep{DCClines} are quoted in the detector frame, while the HMM frequency path of a candidate is recorded in the Solar System barycentre frame.
The resulting Doppler shift amounts to approximately $\pm 10^{-4}f$ \citep{JonesSun2022}.
For each candidate we check whether any portion of the HMM frequency path overlaps with a known line in either detector, after correcting for the Doppler shift.
In the event of any overlap the candidate is vetoed.
After this veto, $18252$ candidates remain.

\subsection{Second pass survivors and vetoes}
\label{subsec:res_vetos}
The $18252$ survivors from the first pass vetoes are still daunting to follow up manually.
We apply three more vetos to further exclude candidates that are consistent with non-Gaussian noise artifacts.

\subsubsection{Cross-cluster veto}
\label{subsubsec:veto_cross_cluster}
The cross-cluster veto excludes candidates which have frequencies overlapping (up to Doppler modulation) with candidates in another cluster.
Candidates which appear at the same frequency in multiple clusters are consistent with non-Gaussian noise artifacts.

Given the targeted false alarm rate of one per cluster, we expect five candidate which are due to Gaussian noise fluctuations, and a candidate which overlaps with one of these will be falsely dismissed under this veto.
Here we briefly show that the probability of a collision between a Gaussian false alarm and an astrophysical candidate is low.
Our criterion for coincidence between two candidates is $|f_1 - f_2| < 10^{-4} f_1$, where $f_1$ and $f_2$ are the terminal frequencies of the two candidate frequency paths, and $10^{-4}$ is the Doppler broadening factor.
To evaluate the probability of a collision, we simulate a search by picking a candidate frequency uniformly over the search range $100\,\mathrm{Hz} \leq f \leq 800\,\mathrm{Hz}$, which is designated as the frequency of the astrophysical candidate, and then pick five more frequencies, again uniformly, which represent the frequencies of the Gaussian false alarm candidates.
We then check the coincidence criterion for each Gaussian false alarm against the astrophysical candidate.
Simulating $10^5$ searches in this way, we find $67$ collisions, i.e. the probability of falsely dismissing an astrophysical candidate due to coincidence with a Gaussian false alarm is approximately $7 \times 10^{-4}$, which we find to be acceptably low.
We therefore veto any candidate for which there is a candidate in another cluster with respective frequencies $f_1$ and $f_2$ satisfying $|f_1 - f_2| < 10^{-4} f_1$.

$17977$ candidates are vetoed this way.
We find that all of the vetoed candidates have at least 4 cross-cluster counterparts, i.e. distinct combinations of $\alpha$, $\delta$, and $\dot{f}$ with at least one frequency path with $\mathcal{L} > \mathcal{L}_{\mathrm{th}}$ overlapping with the vetoed candidate --- multiple cross-cluster counterparts may be associated with the same cluster.
After this veto, $275$ candidates remain.

\subsubsection{Single-interferometer veto}
\label{subsubsec:veto_single_ifo}
If a candidate is due to a non-Gaussian feature which is present in the data from only one interferometer, we expect the significance of the candidate to increase when data from only that interferometer is used.
We thus veto candidates satisfying $\mathcal{L}_X > \mathcal{L}_{\cup}$, where $\mathcal{L}_X$ is the log likelihood of the candidate template using only data from detector $X$ and $\mathcal{L}_\cup$ is the log likelihood computed using data from both detectors.
Before carrying out this veto we group candidates that have frequencies within $10^{-2}\,\mathrm{Hz}$ of each other, in order to reduce the computation required and enable manual verification of the results of the veto procedure.
There are 10 candidate groups in total, appearing in two clusters, NGC 6397 and NGC 6544.
To compute $\mathcal{L}_X$ over the relevant frequency band we use the loudest sky position and $\dot{f}$ template in each candidate group, and $\mathcal{L}_\cup$ is taken to be the loudest log likelihood value in the candidate group.
Plots showing $\mathcal{L}_X$ in comparison to $\mathcal{L}_\cup$ for each candidate are collected in Appendix \ref{apdx:1ifo_veto}.
In all but one case, one $\mathcal{L}_X$ value peaks above $\mathcal{L}_\cup$, indicating that the candidate group is likely to be due to a disturbance in the corresponding interferometer and is not astrophysical.
These candidate groups are vetoed.

After the single-interferometer veto, two candidates remain.

\subsubsection{Unknown lines veto}
\label{subsubsec:veto_unknown}
The line list used for the known-lines veto \citep{DCClines} includes only lines for which there is good evidence of a non-astrophysical origin.
Other line-like features are visible by eye in plots of the detector ASDs but have not yet been confidently associated with a terrestrial source.

\begin{figure}
    \centering
    \includegraphics[width=\columnwidth]{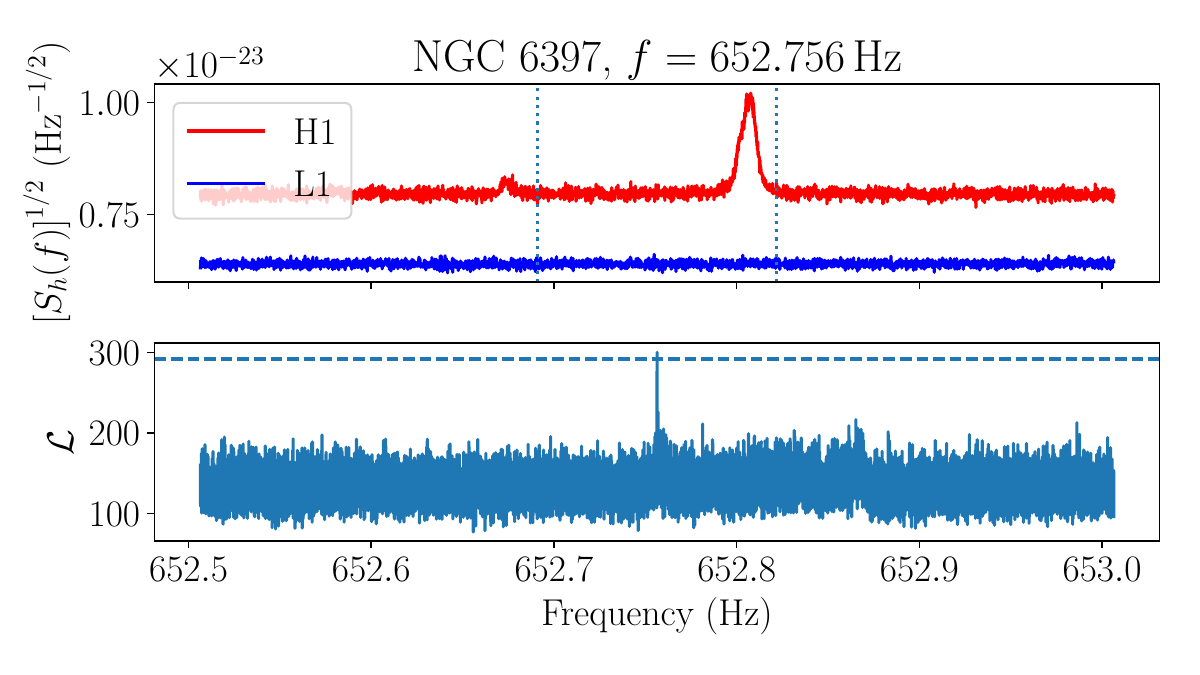}
    \caption{Individual detector ASDs (top panel) and joint-detector log likelihoods (bottom panel) for the candidate group which survived the single-interferometer veto. The vertical dotted lines in the top panel indicate the extent of the Doppler modulation, and the horizontal dashed line in the bottom panel indicates $\mathcal{L}_{\mathrm{th}}$. The cluster name and approximate ending frequency of the loudest candidate within the group are shown at the top of the figure. There is a clear disturbance in the H1 data within the Doppler modulation window, and no corresponding feature in the L1 data. This candidate group is vetoed.}
    \label{fig:ngc6397_last_cand}
\end{figure}

As a further check on the final remaining candidate group, we manually inspect the ASD in each detector to check whether the candidate group overlaps with a loud non-Gaussian feature in the data, taking into account the Doppler modulation.
Candidates which overlap with a clear feature in an ASD are vetoed, on the grounds that we expect any astrophysical signal to be at most marginally visible in the detector ASDs at distances $\gtrsim 2\,\mathrm{kpc}$ \cite{JaumeTenorio2024}.
A plot showing the detector ASDs and corresponding $\mathcal{L}$ values for the loudest candidate in the remaining candidate group, in the cluster NGC $6397$ at $f \approx 652.756\,\mathrm{Hz}$, is shown in Figure \ref{fig:ngc6397_last_cand}.
There is a clear, broad peak in the H1 ASD within the Doppler modulation range of the candidate, with no corresponding feature visible in the L1 ASD. 
We therefore veto this candidate group.

After the unknown lines veto, no candidates remain.

For completeness, in Appendix \ref{apdx:unknown_lines} we present plots analogous to Fig. \ref{fig:ngc6397_last_cand} for all ten candidate groups which survive the cross-cluster veto.

\subsection{Summary of vetoes}
After applying the veto procedures described in Sections \ref{subsec:res_init_cand} and \ref{subsec:res_vetos}, we are left with no surviving candidates.
A summary of the results of the veto procedures is given in Table \ref{tbl:veto_res}.
\begin{table}
    \centering
    \begin{tabular}{lrrrrrr}\toprule
       Cluster  & Pre-veto & DS & KL & CC & 1IFO & UL \\\hline
        NGC 6544 & $9.3 \times 10^6$ & $162509$ & $2918$ & $86$ & $0$ & $0$ \\
        NGC 6325 & $1.1 \times 10^7$ & $68338$ & $3354$ & $0$ & $0$ & $0$ \\
        NGC 6540 & $1.1 \times 10^7$ & $68656$ & $235$ & $0$ & $0$ & $0$ \\
        Terzan 6 & $9.9 \times 10^6$ & $145782$ & $9952$ & $0$ & $0$ & $0$ \\
        NGC 6397 & $1.2 \times 10^7$ & $43566$ & $1793$ & $189$ & $2$ & $0$ \\\hline
        Total & $6.4 \times 10^7$ & $488851$ & $18252$ & $275$ & $2$ & $0$ \\\toprule
    \end{tabular}
    \caption{Results of applying the veto procdures described in Sections \ref{subsec:res_init_cand} and  \ref{subsec:res_vetos}. Each column gives the number of candidates surviving after the corresponding veto. The abbreviated column headings are: disturbed sub-band (DS), known-lines (KL), cross-cluster (CC), single-interferometer (1IFO) and unknown lines (UL). Note that the number of pre-veto candidates is a lower bound, as the candidate list from any sub-band which produces more than $1.67 \times 10^5$ candidates is not saved and is instead marked as vetoed by the disturbed sub-band veto (see Section \ref{subsubsec:res_disturb_band}).}
    \label{tbl:veto_res}
\end{table}
The disturbed sub-band and known-lines vetos reject most of the candidates but leave $18252$, still a significant number.
The cross-cluster veto further rejects all but $275$ candidates, which are grouped into ten narrow frequency bands.
We find that for nine of the ten groups, one of the single-detector log likelihoods peaks at a higher value than the joint-detector peak, indicating that these candidates are due to a non-Gaussian feature which is peculiar to one detector.
Hence these groups are vetoed.
The remaining group is in the vicinity of a broad, loud disturbance in the H1 detector and is therefore also vetoed.


\section{Sensitivity}
\label{sec:sens}
In this section we estimate the sensitivity of the search to CW emission from each targeted globular cluster.
Specifically, for each cluster we estimate $h_{0,\text{eff}}^{95\%}$, the effective strain amplitude [defined in Eq. (\ref{eqn:h0eff})] at which we would expect to detect a signal 95\% of the time.
We draw the distinction between the search sensitivity, defined as above, and an upper limit.
An upper limit is a limit on the loudest astrophysical signal actually present in the data, and is necessarily estimated with reference to the results of the search (i.e. estimation of the upper limit in a given sub-band depends on the loudest candidate in that sub-band).
By contrast the search sensitivity is not a function of the results of the search, only the search configuration.
We prefer to estimate the search sensitivity here as it is straightforward to interpret and requires less computation.
Computing upper limits requires estimating a relationship between the noise ASD in a given sub-band, $S_h(f)^{1/2}$, the log likelihood of the loudest candidate in that sub-band, $\mathcal{L}_{\mathrm{max}}$, and the effective strain amplitude which produces a candidate at least that loud 95\% of the time, $h_{0,\text{eff}}^{95\%}$.
This can be done, of course, but it is more involved and requires costly signal injection studies.
Here we estimate the search sensitivity in a small number of sub-bands and subsequently interpolate across the full search band, as described below.

In order to estimate the search sensitivity we inject synthetic signals into the real O3 data.
We inject at sky positions slightly offset in right ascension from the true cluster position, so that the sky templates do not overlap with the search, so as to avoid the possibility of contamination by an astrophysical signal from the cluster.
In a given $0.5\,\mathrm{Hz}$ sub-band we inject signals at ten evenly spaced $h_{0,\text{eff}}$ levels between $1.7 \times 10^{-26}$ and $3.5 \times 10^{-26}$ and determine the detection rate as a function of $h_{0,\text{eff}}$.
We then fit these rates as a function of $h_{0,\text{eff}}$ to a sigmoid curve, following \citet{AbbottAbbott2021a}.
We use the best-fit parameters to determine the $h_{0,\text{eff}}$ value yielding $P_{\text{d}} = 0.95$, in that sub-band.

The off-target injections and recoveries are treated much the same as in the actual search.
The key difference is that the centre of the injection-containing mock cluster is offset from the true cluster centre by five tidal radii in right ascension.
The placement of $\alpha$, $\delta$, and $\dot{f}$ templates mimics that of the real search, but to save computational resources we only search $\dot{f}$ templates within $\pm3\delta\dot{f}$ of the injected $\dot{f}$.
We also search a reduced frequency range $\pm 5 \times 10^{-4}\,\mathrm{Hz}$ around the injection frequency.
The value of $\mathcal{L}_{\text{th}}$ is the same as in the real search\footnote{In the searches around these mock injections the number of templates is smaller than the number used in real searches, implying a smaller $\mathcal{L}_{\mathrm{th}}$ for the targeted false alarm rate. Nonetheless the aim is to ascertain the sensitivity of the real searches, and so we use the $\mathcal{L}_{\mathrm{th}}$ values quoted in Table \ref{tbl:gc_lth}.}, and we claim a successful recovery if the maximum recovered value of $\mathcal{L}$ exceeds $\mathcal{L}_{\text{th}}$ regardless of which template returns the maximum.
As we set $\mathcal{L}_{\text{th}}$ on a per-cluster basis (see Section \ref{subsec:res_thresh}), we expect different sensitivity curves for each cluster, with clusters requiring fewer sky position templates (and hence smaller $\mathcal{L}_{\text{th}}$ values) reaching deeper sensitivities.

We determine the value of $h_{0,\text{eff}}^{95\%}$ directly following the above recipe in a small number of $0.5\,\mathrm{Hz}$ bands (beginning at $168.2\,\mathrm{Hz}$, $242.6\,\mathrm{Hz}$, $424.3\,\mathrm{Hz}$, $562.8\,\mathrm{Hz}$, $622.2\,\mathrm{Hz}$, and $702.0\,\mathrm{Hz}$) and then interpolate to cover the full observing band.
The value of $h_{0,\mathrm{eff}}^{95\%}$ is a strong function of frequency, but to a good approximation it is directly proportional to the detector-averaged ASD $S_h(f)^{1/2}$.
Therefore in order to interpolate the estimated $h_{0,\text{eff}}^{95\%}$ values in the six sub-bands to a $h_{0,\mathrm{eff}}^{95\%}$ curve which covers the entire observing band we estimate the sensitivity depth $\mathcal{D}^{95\%}$, which is the constant of proportionality between $h_{0,\mathrm{eff}}^{95\%}$ and $S_h(f)^{1/2}$ and is defined as \begin{equation} \mathcal{D}^{95\%} = \frac{S_h(f)^{1/2}}{h_{0,\text{eff}}^{95\%}}.\label{eqn:sens_depth}\end{equation}
We compute $S_h(f)$ using \textsc{lalpulsar\_ComputePSD}, averaging over all data used in the search from both detectors.
There is some scatter in the computed $\mathcal{D}^{95\%}$ values in the six $0.5\mathrm{Hz}$ sub-bands, with $0.96 < \mathcal{D}^{95\%}/\overline{\mathcal{D}}^{95\%} < 1.04$ for each cluster, where $\overline{\mathcal{D}}^{95\%}$ is the mean value of $\mathcal{D}^{95\%}$ across the six sub-bands.
There is no clear trend in $\mathcal{D}^{95\%}$ as a function of frequency.
As $\mathcal{D}^{95\%}$ is defined to be the constant of proportionality between $S_h(f)^{1/2}$ and $h_{0,\mathrm{eff}}^{95\%}$, to obtain a curve of $h_{0,\mathrm{eff}}^{95\%}$ for each cluster over the full band we divide $S_h(f)^{1/2}$ by the $\overline{\mathcal{D}}^{95\%}$ value for that cluster.

\begin{figure}
    \centering
    \includegraphics[width=\columnwidth]{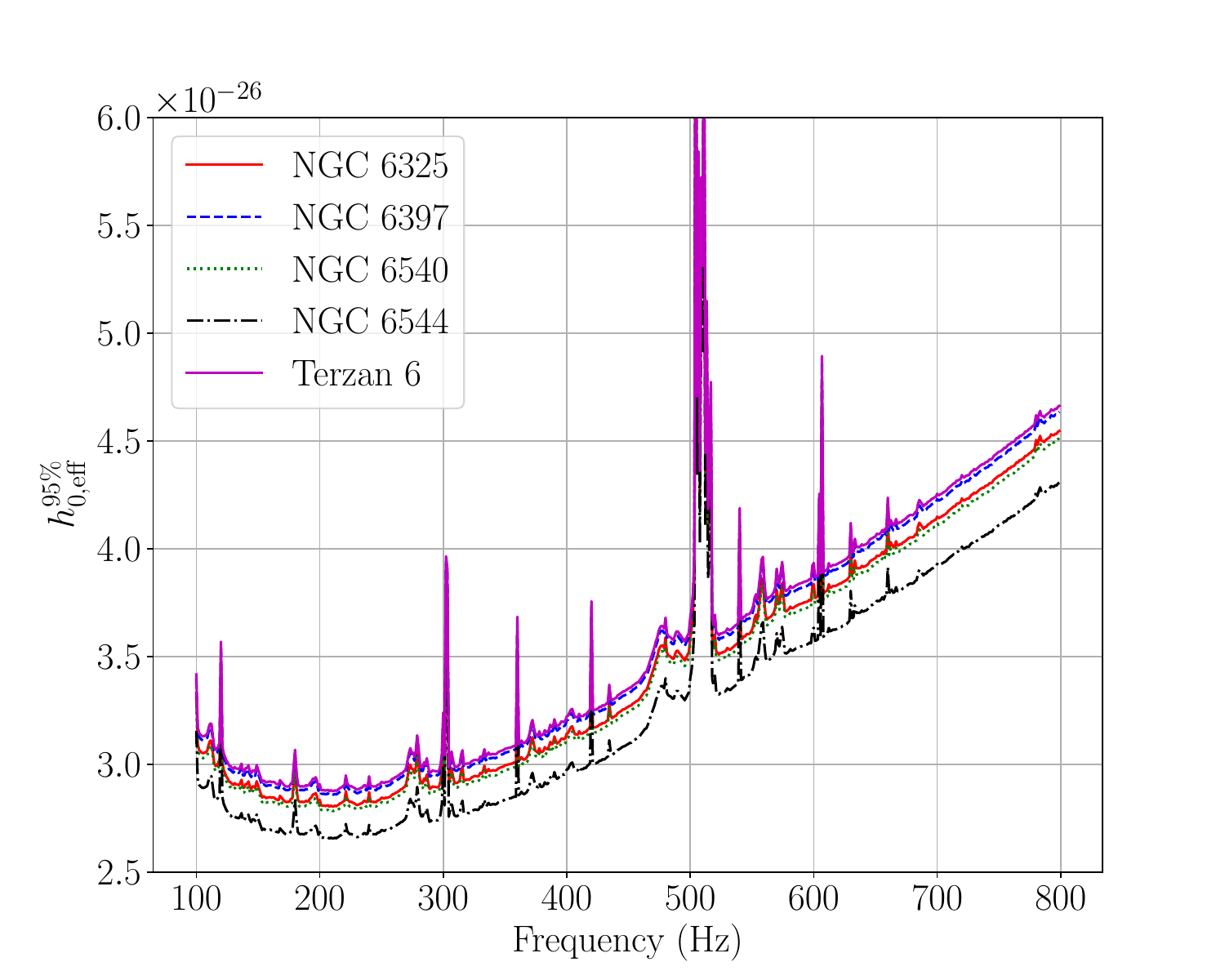}\\
    \includegraphics[width=\columnwidth]{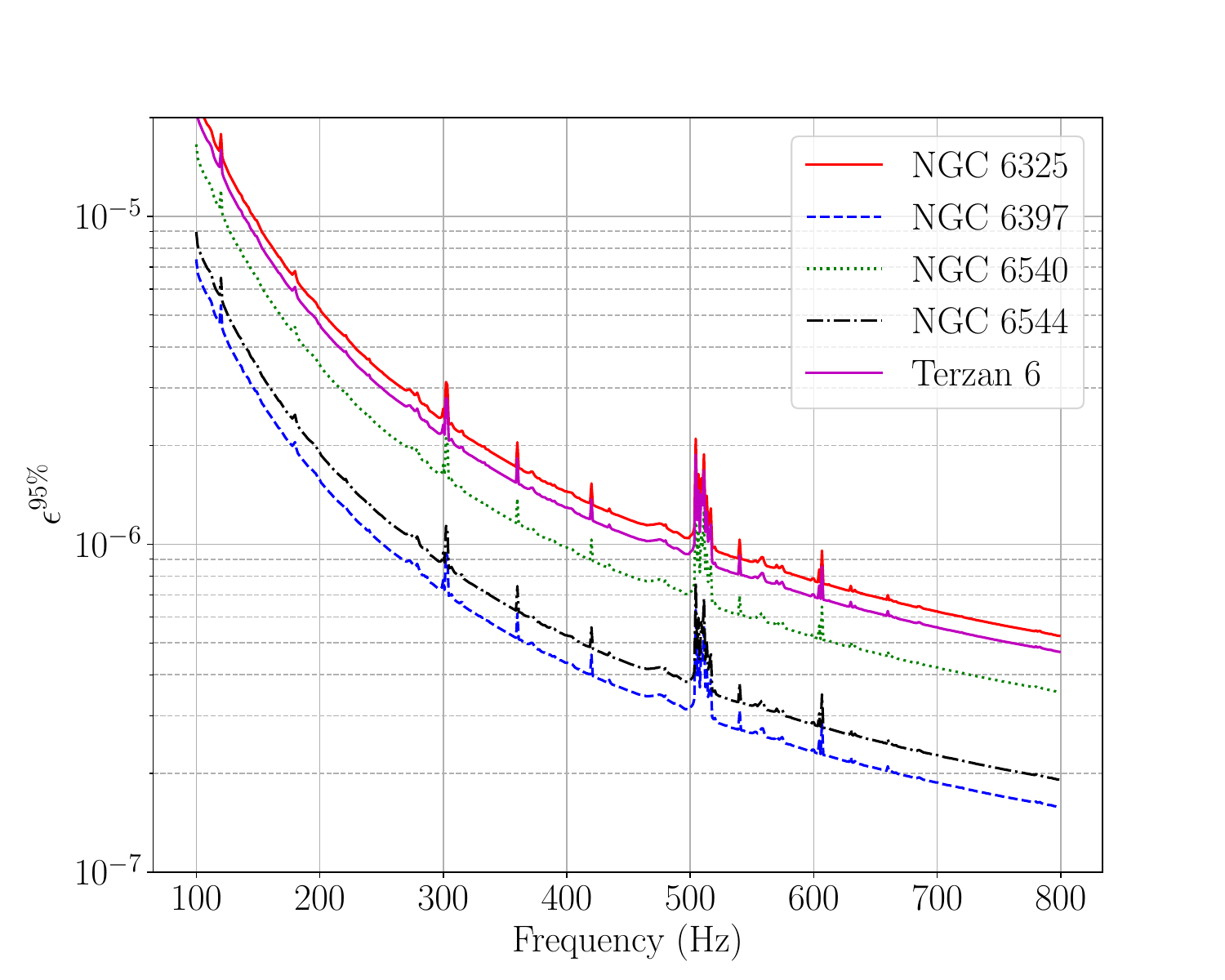}
    \caption{Sensitivity estimates for the five targeted clusters.
    \emph{Top panel}: characteristic wave strain at 95\% detection probability, $h_{0,\mathrm{eff}}^{95\%}$, as a function of signal frequency, $f$ (in Hz), interpolated from the average estimated sensitivity depth defined in Eq. (\ref{eqn:sens_depth}) over six $0.5\,\mathrm{Hz}$ bands. The curves for NGC 6397 and Terzan 6 are too close to be distinguished, as are the curves for NGC 6540 and NGC 6325.
    \emph{Bottom panel}: minimum ellipticity $\epsilon^{95\%}$ as a function of $f$ for the values of $D$ listed in Table \ref{tbl:gc_topcands} and the fiducial value of $I_{zz}$ in Eq. (\ref{eqn:ellipticity}).
    The clusters are color-coded according to the legend.}
    \label{fig:sensitivities}
\end{figure}

The resulting $h_{0,\text{eff}}^{95\%}$ sensitivity curves are shown in the top panel of Figure \ref{fig:sensitivities}.
The $\mathcal{D}^{95\%}$ values used to produce the $h_{0,\mathrm{eff}}^{95\%}$ curves are listed in Table \ref{tbl:depths}.
\begin{table}[]
    \centering
    \begin{tabular}{ll}\toprule
       Cluster  &  $\overline{\mathcal{D}}^{95\%}$ \\\hline
       Terzan 6 & 116.3 \\
        NGC 6397 & 117.0 \\
        NGC 6325 & 119.3 \\
        NGC 6540 & 120.2 \\
        NGC 6544 & 125.9 \\
        \toprule
    \end{tabular}
    \caption{Estimated sensitivity depths $\overline{\mathcal{D}}^{95\%}$ for the five targeted clusters. The value of $\overline{\mathcal{D}}^{95\%}$ is calculated by taking the mean of the $\mathcal{D}^{95\%}$ values [see Eq. (\ref{eqn:sens_depth})] computed in six sub-bands.}
    \label{tbl:depths}
\end{table}
The differences in sensitivity between the five clusters are due to the varying $\mathcal{L}_{\mathrm{th}}$ values for each cluster (see Table \ref{tbl:gc_lth}), which follow from the number of sky position templates (see Sections \ref{subsubsec:param_ranges_sky} and \ref{subsubsec:setup_param_spacing_sky}).
The variation between clusters is fairly small: the least sensitive cluster (Terzan 6) is approximately 10\% worse than the most sensitive (NGC 6544), achieving $h_{0,\mathrm{eff}}^{95\%} = 2.9 \times 10^{-26}$ and $2.7\times 10^{-26}$ respectively at the most sensitive frequency of $f = 211\,\mathrm{Hz}$.

Out of astrophysical interest, we also convert $h_{0,\mathrm{eff}}^{95\%}$ in Figure \ref{fig:sensitivities} into estimates of the minimum ellipticity of a source which would have been detected by this search assuming emission at twice the rotational frequency, according to the relation \cite{BeniwalClearwater2022} \begin{align} \epsilon^{95\%} =& 9.5 \times 10^{-6} \left(\frac{h_{0,\text{eff}}^{95\%}}{10^{-25}}\right)\left(\frac{I_{zz}}{10^{38}\,\mathrm{kg}\,\mathrm{m}^2}\right)^{-1}\nonumber\\
&\times\left(\frac{f}{100\,\mathrm{Hz}}\right)^{-2}\left(\frac{D}{1\,\mathrm{kpc}}\right). \label{eqn:ellipticity}\end{align}
Curves of $\epsilon^{95\%}$ versus signal frequency for the five clusters are shown in the bottom panel of Figure \ref{fig:sensitivities}.
The smallest minimum ellipticity estimates are achieved at the top of the observing band, $800\,\mathrm{Hz}$, due to the $f^{-2}$ factor in Eq. (\ref{eqn:ellipticity}).
The factor of $D$ introduces larger variation between the clusters compared to the variation in $h_0^{95\%}$ values, and modifies the relative ordering --- the smallest minimum ellipticity estimate is $1.6 \times 10^{-7}$, for the nearest cluster NGC 6397 ($D = 2.3\,\mathrm{kpc}$), while the largest is $5.2 \times 10^{-7}$ for the farthest cluster NGC 6325 ($D = 7.8\,\mathrm{kpc}$).

\section{Conclusion}
\label{sec:conclusion}
We carry out a search for continuous gravitational radiation from unknown neutron stars in five globular clusters.
This is the first time that all but one of the chosen clusters have been targeted.
For NGC 6544 which was previously searched using LIGO S6 data, the sensitivity is improved approximately one order of magnitude \citep{AbbottAbbott2017b} (although note that the signal models and parameter domain differ, so a direct, like-for-like comparison is impossible).
This is also the first search to use a phase-tracking HMM in conjunction with a phase-dependent version of the $\mathcal{B}$-statistic \citep{PrixKrishnan2009, MelatosClearwater2021}; previous HMM searches tracked frequency only, not phase.
Phase tracking brings with it an increase in sensitivity, but also increases the computational cost.
For searches aiming to optimize sensitivity at fixed computing cost, a careful analysis of the tradeoff between sensitivity and cost is crucial.
We do not undertake this exercise in this paper, as a key aim is to demonstrate the use of the phase-tracking HMM in an astrophysical search for the first time.

We find no significant candidates.
We estimate the $95\%$ effective strain amplitude sensitivity for each cluster, along with a corresponding estimate of the $95\%$ ellipticity sensitivity.
The minimum per-cluster values of $h_{0,\text{eff}}^{95\%}$ lie in the range $2.5 \leq h_{0,\mathrm{eff}}^{95\%} / 10^{-26} \leq 3.0$, achieved at approximately $211\,\mathrm{Hz}$, while the minimum values of $\epsilon^{95\%}$ are achieved at the top end of the search band ($800\,\mathrm{Hz}$), and lie in the range $1 \leq \epsilon^{95\%} / 10^{-7} \leq 6$.
The lowest $h_{0,\text{eff}}^{95\%}$ values are obtained for the cluster NGC 6544, while the lowest $\epsilon^{95\%}$ values are obtained for NGC 6397.

We surpass the spin-down limit for $|\dot{f}| \gtrsim 10^{-11}\,\mathrm{Hz}\,\mathrm{s}^{-1}$, with variation between clusters due primarily to the difference in distance.
As discussed in Section \ref{subsec:param_ranges}, electromagnetically visible MSPs have spin-down rates well below this limit, so this search is primarily sensitive to ``gravitars'' --- neutron stars whose spin-down behaviour is dominated by gravitational rather than electromagnetic torques \citep{Palomba2005}.
The inferred limits on the ellipticity of the star probe up to an order of magnitude below theoretical upper limits, viz. $\epsilon \lesssim 10^{-6}$ \citep{Johnson-McDanielOwen2013, MoralesHorowitz2022}.

\section*{Acknowledgments}
Parts of this research are supported by the Australian Research Council (ARC) Centre of Excellence for Gravitational Wave Discovery (OzGrav) (project numbers CE170100004 and CE230100016) and ARC Discovery Project DP170103625.
This material is based upon work supported by NSF's LIGO Laboratory which is a major facility fully funded by the National Science Foundation.
LD is supported by an Australian Government Research Training Program Scholarship.
LS is supported by the ARC Discovery Early Career Researcher Award, Project Number DE240100206.
This work was performed in part on the OzSTAR national facility at Swinburne University of Technology. The OzSTAR program receives funding in part from the Astronomy National Collaborative Research Infrastructure Strategy allocation provided by the Australian Government.

\bibliography{refs, refs_extra}

\begin{thebibliography}{116}%
\makeatletter
\providecommand \@ifxundefined [1]{%
 \@ifx{#1\undefined}
}%
\providecommand \@ifnum [1]{%
 \ifnum #1\expandafter \@firstoftwo
 \else \expandafter \@secondoftwo
 \fi
}%
\providecommand \@ifx [1]{%
 \ifx #1\expandafter \@firstoftwo
 \else \expandafter \@secondoftwo
 \fi
}%
\providecommand \natexlab [1]{#1}%
\providecommand \enquote  [1]{``#1''}%
\providecommand \bibnamefont  [1]{#1}%
\providecommand \bibfnamefont [1]{#1}%
\providecommand \citenamefont [1]{#1}%
\providecommand \href@noop [0]{\@secondoftwo}%
\providecommand \href [0]{\begingroup \@sanitize@url \@href}%
\providecommand \@href[1]{\@@startlink{#1}\@@href}%
\providecommand \@@href[1]{\endgroup#1\@@endlink}%
\providecommand \@sanitize@url [0]{\catcode `\\12\catcode `\$12\catcode
  `\&12\catcode `\#12\catcode `\^12\catcode `\_12\catcode `\%12\relax}%
\providecommand \@@startlink[1]{}%
\providecommand \@@endlink[0]{}%
\providecommand \url  [0]{\begingroup\@sanitize@url \@url }%
\providecommand \@url [1]{\endgroup\@href {#1}{\urlprefix }}%
\providecommand \urlprefix  [0]{URL }%
\providecommand \Eprint [0]{\href }%
\providecommand \doibase [0]{https://doi.org/}%
\providecommand \selectlanguage [0]{\@gobble}%
\providecommand \bibinfo  [0]{\@secondoftwo}%
\providecommand \bibfield  [0]{\@secondoftwo}%
\providecommand \translation [1]{[#1]}%
\providecommand \BibitemOpen [0]{}%
\providecommand \bibitemStop [0]{}%
\providecommand \bibitemNoStop [0]{.\EOS\space}%
\providecommand \EOS [0]{\spacefactor3000\relax}%
\providecommand \BibitemShut  [1]{\csname bibitem#1\endcsname}%
\let\auto@bib@innerbib\@empty
\bibitem [{\citenamefont {{Riles}}(2023)}]{Riles2023}%
  \BibitemOpen
  \bibfield  {author} {\bibinfo {author} {\bibfnamefont {K.}~\bibnamefont
  {{Riles}}},\ }\bibfield  {title} {\bibinfo {title} {{Searches for
  continuous-wave gravitational radiation}},\ }\href
  {https://doi.org/10.1007/s41114-023-00044-3} {\bibfield  {journal} {\bibinfo
  {journal} {Living Reviews in Relativity}\ }\textbf {\bibinfo {volume} {26}},\
  \bibinfo {eid} {3} (\bibinfo {year} {2023})},\ \Eprint
  {https://arxiv.org/abs/2206.06447} {arXiv:2206.06447 [astro-ph.HE]}
  \BibitemShut {NoStop}%
\bibitem [{\citenamefont {{Wette}}(2023)}]{Wette2023}%
  \BibitemOpen
  \bibfield  {author} {\bibinfo {author} {\bibfnamefont {K.}~\bibnamefont
  {{Wette}}},\ }\bibfield  {title} {\bibinfo {title} {{Searches for continuous
  gravitational waves from neutron stars: A twenty-year retrospective}},\
  }\href {https://doi.org/10.1016/j.astropartphys.2023.102880} {\bibfield
  {journal} {\bibinfo  {journal} {Astroparticle Physics}\ }\textbf {\bibinfo
  {volume} {153}},\ \bibinfo {eid} {102880} (\bibinfo {year} {2023})},\ \Eprint
  {https://arxiv.org/abs/2305.07106} {arXiv:2305.07106 [gr-qc]} \BibitemShut
  {NoStop}%
\bibitem [{\citenamefont {{Bildsten}}(1998)}]{Bildsten1998}%
  \BibitemOpen
  \bibfield  {author} {\bibinfo {author} {\bibfnamefont {L.}~\bibnamefont
  {{Bildsten}}},\ }\bibfield  {title} {\bibinfo {title} {{Gravitational
  Radiation and Rotation of Accreting Neutron Stars}},\ }\href
  {https://doi.org/10.1086/311440} {\bibfield  {journal} {\bibinfo  {journal}
  {The Astrophysical Journal}\ }\textbf {\bibinfo {volume} {501}},\ \bibinfo
  {pages} {L89} (\bibinfo {year} {1998})},\ \Eprint
  {https://arxiv.org/abs/astro-ph/9804325} {arXiv:astro-ph/9804325 [astro-ph]}
  \BibitemShut {NoStop}%
\bibitem [{\citenamefont {{Ushomirsky}}\ \emph {et~al.}(2000)\citenamefont
  {{Ushomirsky}}, \citenamefont {{Cutler}},\ and\ \citenamefont
  {{Bildsten}}}]{UshomirskyCutler2000}%
  \BibitemOpen
  \bibfield  {author} {\bibinfo {author} {\bibfnamefont {G.}~\bibnamefont
  {{Ushomirsky}}}, \bibinfo {author} {\bibfnamefont {C.}~\bibnamefont
  {{Cutler}}},\ and\ \bibinfo {author} {\bibfnamefont {L.}~\bibnamefont
  {{Bildsten}}},\ }\bibfield  {title} {\bibinfo {title} {{Deformations of
  accreting neutron star crusts and gravitational wave emission}},\ }\href
  {https://doi.org/10.1046/j.1365-8711.2000.03938.x} {\bibfield  {journal}
  {\bibinfo  {journal} {Monthly Notices of the Royal Astronomical Society}\
  }\textbf {\bibinfo {volume} {319}},\ \bibinfo {pages} {902} (\bibinfo {year}
  {2000})},\ \Eprint {https://arxiv.org/abs/astro-ph/0001136}
  {arXiv:astro-ph/0001136 [astro-ph]} \BibitemShut {NoStop}%
\bibitem [{\citenamefont {{Osborne}}\ and\ \citenamefont
  {{Jones}}(2020)}]{OsborneJones2020}%
  \BibitemOpen
  \bibfield  {author} {\bibinfo {author} {\bibfnamefont {E.~L.}\ \bibnamefont
  {{Osborne}}}\ and\ \bibinfo {author} {\bibfnamefont {D.~I.}\ \bibnamefont
  {{Jones}}},\ }\bibfield  {title} {\bibinfo {title} {{Gravitational waves from
  magnetically induced thermal neutron star mountains}},\ }\href
  {https://doi.org/10.1093/mnras/staa858} {\bibfield  {journal} {\bibinfo
  {journal} {Monthly Notices of the Royal Astronomical Society}\ }\textbf
  {\bibinfo {volume} {494}},\ \bibinfo {pages} {2839} (\bibinfo {year}
  {2020})},\ \Eprint {https://arxiv.org/abs/1910.04453} {arXiv:1910.04453
  [astro-ph.HE]} \BibitemShut {NoStop}%
\bibitem [{\citenamefont {{Hutchins}}\ and\ \citenamefont
  {{Jones}}(2023)}]{HutchinsJones2023}%
  \BibitemOpen
  \bibfield  {author} {\bibinfo {author} {\bibfnamefont {T.~J.}\ \bibnamefont
  {{Hutchins}}}\ and\ \bibinfo {author} {\bibfnamefont {D.~I.}\ \bibnamefont
  {{Jones}}},\ }\bibfield  {title} {\bibinfo {title} {{Gravitational radiation
  from thermal mountains on accreting neutron stars: sources of temperature
  non-axisymmetry}},\ }\href {https://doi.org/10.1093/mnras/stad967} {\bibfield
   {journal} {\bibinfo  {journal} {Monthly Notices of the Royal Astronomical
  Society}\ }\textbf {\bibinfo {volume} {522}},\ \bibinfo {pages} {226}
  (\bibinfo {year} {2023})},\ \Eprint {https://arxiv.org/abs/2212.07452}
  {arXiv:2212.07452 [astro-ph.HE]} \BibitemShut {NoStop}%
\bibitem [{\citenamefont {{Bonazzola}}\ and\ \citenamefont
  {{Gourgoulhon}}(1996)}]{BonazzolaGourgoulhon1996}%
  \BibitemOpen
  \bibfield  {author} {\bibinfo {author} {\bibfnamefont {S.}~\bibnamefont
  {{Bonazzola}}}\ and\ \bibinfo {author} {\bibfnamefont {E.}~\bibnamefont
  {{Gourgoulhon}}},\ }\bibfield  {title} {\bibinfo {title} {{Gravitational
  waves from pulsars: emission by the magnetic-field-induced distortion.}},\
  }\href {https://doi.org/10.48550/arXiv.astro-ph/9602107} {\bibfield
  {journal} {\bibinfo  {journal} {Astronomy and Astrophysics}\ }\textbf
  {\bibinfo {volume} {312}},\ \bibinfo {pages} {675} (\bibinfo {year}
  {1996})},\ \Eprint {https://arxiv.org/abs/astro-ph/9602107}
  {arXiv:astro-ph/9602107 [astro-ph]} \BibitemShut {NoStop}%
\bibitem [{\citenamefont {{Cutler}}(2002)}]{Cutler2002}%
  \BibitemOpen
  \bibfield  {author} {\bibinfo {author} {\bibfnamefont {C.}~\bibnamefont
  {{Cutler}}},\ }\bibfield  {title} {\bibinfo {title} {{Gravitational waves
  from neutron stars with large toroidal B fields}},\ }\href
  {https://doi.org/10.1103/PhysRevD.66.084025} {\bibfield  {journal} {\bibinfo
  {journal} {Physical Review D}\ }\textbf {\bibinfo {volume} {66}},\ \bibinfo
  {eid} {084025} (\bibinfo {year} {2002})},\ \Eprint
  {https://arxiv.org/abs/gr-qc/0206051} {arXiv:gr-qc/0206051 [gr-qc]}
  \BibitemShut {NoStop}%
\bibitem [{\citenamefont {{Payne}}\ and\ \citenamefont
  {{Melatos}}(2004)}]{PayneMelatos2004}%
  \BibitemOpen
  \bibfield  {author} {\bibinfo {author} {\bibfnamefont {D.~J.~B.}\
  \bibnamefont {{Payne}}}\ and\ \bibinfo {author} {\bibfnamefont
  {A.}~\bibnamefont {{Melatos}}},\ }\bibfield  {title} {\bibinfo {title}
  {{Burial of the polar magnetic field of an accreting neutron star - I.
  Self-consistent analytic and numerical equilibria}},\ }\href
  {https://doi.org/10.1111/j.1365-2966.2004.07798.x} {\bibfield  {journal}
  {\bibinfo  {journal} {Monthly Notices of the Royal Astronomical Society}\
  }\textbf {\bibinfo {volume} {351}},\ \bibinfo {pages} {569} (\bibinfo {year}
  {2004})},\ \Eprint {https://arxiv.org/abs/astro-ph/0403173}
  {arXiv:astro-ph/0403173 [astro-ph]} \BibitemShut {NoStop}%
\bibitem [{\citenamefont {{Melatos}}\ and\ \citenamefont
  {{Payne}}(2005)}]{MelatosPayne2005}%
  \BibitemOpen
  \bibfield  {author} {\bibinfo {author} {\bibfnamefont {A.}~\bibnamefont
  {{Melatos}}}\ and\ \bibinfo {author} {\bibfnamefont {D.~J.~B.}\ \bibnamefont
  {{Payne}}},\ }\bibfield  {title} {\bibinfo {title} {{Gravitational Radiation
  from an Accreting Millisecond Pulsar with a Magnetically Confined
  Mountain}},\ }\href {https://doi.org/10.1086/428600} {\bibfield  {journal}
  {\bibinfo  {journal} {The Astrophysical Journal}\ }\textbf {\bibinfo {volume}
  {623}},\ \bibinfo {pages} {1044} (\bibinfo {year} {2005})},\ \Eprint
  {https://arxiv.org/abs/astro-ph/0503287} {arXiv:astro-ph/0503287 [astro-ph]}
  \BibitemShut {NoStop}%
\bibitem [{\citenamefont {{Payne}}\ and\ \citenamefont
  {{Melatos}}(2006)}]{PayneMelatos2006}%
  \BibitemOpen
  \bibfield  {author} {\bibinfo {author} {\bibfnamefont {D.~J.~B.}\
  \bibnamefont {{Payne}}}\ and\ \bibinfo {author} {\bibfnamefont
  {A.}~\bibnamefont {{Melatos}}},\ }\bibfield  {title} {\bibinfo {title}
  {{Frequency Spectrum of Gravitational Radiation from Global Hydromagnetic
  Oscillations of a Magnetically Confined Mountain on an Accreting Neutron
  Star}},\ }\href {https://doi.org/10.1086/498855} {\bibfield  {journal}
  {\bibinfo  {journal} {The Astrophysical Journal}\ }\textbf {\bibinfo {volume}
  {641}},\ \bibinfo {pages} {471} (\bibinfo {year} {2006})},\ \Eprint
  {https://arxiv.org/abs/astro-ph/0510053} {arXiv:astro-ph/0510053 [astro-ph]}
  \BibitemShut {NoStop}%
\bibitem [{\citenamefont {{Rossetto}}\ \emph {et~al.}(2023)\citenamefont
  {{Rossetto}}, \citenamefont {{Frauendiener}}, \citenamefont {{Brunet}},\ and\
  \citenamefont {{Melatos}}}]{RossettoFrauendiener2023}%
  \BibitemOpen
  \bibfield  {author} {\bibinfo {author} {\bibfnamefont {P.~H.~B.}\
  \bibnamefont {{Rossetto}}}, \bibinfo {author} {\bibfnamefont
  {J.}~\bibnamefont {{Frauendiener}}}, \bibinfo {author} {\bibfnamefont
  {R.}~\bibnamefont {{Brunet}}},\ and\ \bibinfo {author} {\bibfnamefont
  {A.}~\bibnamefont {{Melatos}}},\ }\bibfield  {title} {\bibinfo {title}
  {{Magnetically confined mountains on accreting neutron stars in general
  relativity}},\ }\href {https://doi.org/10.1093/mnras/stad2850} {\bibfield
  {journal} {\bibinfo  {journal} {Monthly Notices of the Royal Astronomical
  Society}\ }\textbf {\bibinfo {volume} {526}},\ \bibinfo {pages} {2058}
  (\bibinfo {year} {2023})},\ \Eprint {https://arxiv.org/abs/2309.09519}
  {arXiv:2309.09519 [astro-ph.HE]} \BibitemShut {NoStop}%
\bibitem [{\citenamefont {{Caplan}}\ and\ \citenamefont
  {{Horowitz}}(2017)}]{CaplanHorowitz2017}%
  \BibitemOpen
  \bibfield  {author} {\bibinfo {author} {\bibfnamefont {M.~E.}\ \bibnamefont
  {{Caplan}}}\ and\ \bibinfo {author} {\bibfnamefont {C.~J.}\ \bibnamefont
  {{Horowitz}}},\ }\bibfield  {title} {\bibinfo {title} {{Colloquium:
  Astromaterial science and nuclear pasta}},\ }\href
  {https://doi.org/10.1103/RevModPhys.89.041002} {\bibfield  {journal}
  {\bibinfo  {journal} {Reviews of Modern Physics}\ }\textbf {\bibinfo {volume}
  {89}},\ \bibinfo {eid} {041002} (\bibinfo {year} {2017})},\ \Eprint
  {https://arxiv.org/abs/1606.03646} {arXiv:1606.03646 [astro-ph.HE]}
  \BibitemShut {NoStop}%
\bibitem [{\citenamefont {{Gittins}}\ \emph {et~al.}(2021)\citenamefont
  {{Gittins}}, \citenamefont {{Andersson}},\ and\ \citenamefont
  {{Jones}}}]{GittinsAndersson2021b}%
  \BibitemOpen
  \bibfield  {author} {\bibinfo {author} {\bibfnamefont {F.}~\bibnamefont
  {{Gittins}}}, \bibinfo {author} {\bibfnamefont {N.}~\bibnamefont
  {{Andersson}}},\ and\ \bibinfo {author} {\bibfnamefont {D.~I.}\ \bibnamefont
  {{Jones}}},\ }\bibfield  {title} {\bibinfo {title} {{Modelling neutron star
  mountains}},\ }\href {https://doi.org/10.1093/mnras/staa3635} {\bibfield
  {journal} {\bibinfo  {journal} {Monthly Notices of the Royal Astronomical
  Society}\ }\textbf {\bibinfo {volume} {500}},\ \bibinfo {pages} {5570}
  (\bibinfo {year} {2021})},\ \Eprint {https://arxiv.org/abs/2009.12794}
  {arXiv:2009.12794 [astro-ph.HE]} \BibitemShut {NoStop}%
\bibitem [{\citenamefont {{Kerin}}\ and\ \citenamefont
  {{Melatos}}(2022)}]{KerinMelatos2022}%
  \BibitemOpen
  \bibfield  {author} {\bibinfo {author} {\bibfnamefont {A.~D.}\ \bibnamefont
  {{Kerin}}}\ and\ \bibinfo {author} {\bibfnamefont {A.}~\bibnamefont
  {{Melatos}}},\ }\bibfield  {title} {\bibinfo {title} {{Mountain formation by
  repeated, inhomogeneous crustal failure in a neutron star}},\ }\href
  {https://doi.org/10.1093/mnras/stac1351} {\bibfield  {journal} {\bibinfo
  {journal} {Monthly Notices of the Royal Astronomical Society}\ }\textbf
  {\bibinfo {volume} {514}},\ \bibinfo {pages} {1628} (\bibinfo {year}
  {2022})},\ \Eprint {https://arxiv.org/abs/2205.15026} {arXiv:2205.15026
  [astro-ph.HE]} \BibitemShut {NoStop}%
\bibitem [{\citenamefont {{Andersson}}\ \emph {et~al.}(1999)\citenamefont
  {{Andersson}}, \citenamefont {{Kokkotas}},\ and\ \citenamefont
  {{Stergioulas}}}]{AnderssonKokkotas1999}%
  \BibitemOpen
  \bibfield  {author} {\bibinfo {author} {\bibfnamefont {N.}~\bibnamefont
  {{Andersson}}}, \bibinfo {author} {\bibfnamefont {K.~D.}\ \bibnamefont
  {{Kokkotas}}},\ and\ \bibinfo {author} {\bibfnamefont {N.}~\bibnamefont
  {{Stergioulas}}},\ }\bibfield  {title} {\bibinfo {title} {{On the Relevance
  of the R-Mode Instability for Accreting Neutron Stars and White Dwarfs}},\
  }\href {https://doi.org/10.1086/307082} {\bibfield  {journal} {\bibinfo
  {journal} {The Astrophysical Journal}\ }\textbf {\bibinfo {volume} {516}},\
  \bibinfo {pages} {307} (\bibinfo {year} {1999})},\ \Eprint
  {https://arxiv.org/abs/astro-ph/9806089} {arXiv:astro-ph/9806089 [astro-ph]}
  \BibitemShut {NoStop}%
\bibitem [{\citenamefont {{Arras}}\ \emph {et~al.}(2003)\citenamefont
  {{Arras}}, \citenamefont {{Flanagan}}, \citenamefont {{Morsink}},
  \citenamefont {{Schenk}}, \citenamefont {{Teukolsky}},\ and\ \citenamefont
  {{Wasserman}}}]{ArrasFlanagan2003}%
  \BibitemOpen
  \bibfield  {author} {\bibinfo {author} {\bibfnamefont {P.}~\bibnamefont
  {{Arras}}}, \bibinfo {author} {\bibfnamefont {E.~E.}\ \bibnamefont
  {{Flanagan}}}, \bibinfo {author} {\bibfnamefont {S.~M.}\ \bibnamefont
  {{Morsink}}}, \bibinfo {author} {\bibfnamefont {A.~K.}\ \bibnamefont
  {{Schenk}}}, \bibinfo {author} {\bibfnamefont {S.~A.}\ \bibnamefont
  {{Teukolsky}}},\ and\ \bibinfo {author} {\bibfnamefont {I.}~\bibnamefont
  {{Wasserman}}},\ }\bibfield  {title} {\bibinfo {title} {{Saturation of the
  r-Mode Instability}},\ }\href {https://doi.org/10.1086/374657} {\bibfield
  {journal} {\bibinfo  {journal} {The Astrophysical Journal}\ }\textbf
  {\bibinfo {volume} {591}},\ \bibinfo {pages} {1129} (\bibinfo {year}
  {2003})},\ \Eprint {https://arxiv.org/abs/astro-ph/0202345}
  {arXiv:astro-ph/0202345 [astro-ph]} \BibitemShut {NoStop}%
\bibitem [{\citenamefont {{Bondarescu}}\ \emph {et~al.}(2007)\citenamefont
  {{Bondarescu}}, \citenamefont {{Teukolsky}},\ and\ \citenamefont
  {{Wasserman}}}]{BondarescuTeukolsky2007}%
  \BibitemOpen
  \bibfield  {author} {\bibinfo {author} {\bibfnamefont {R.}~\bibnamefont
  {{Bondarescu}}}, \bibinfo {author} {\bibfnamefont {S.~A.}\ \bibnamefont
  {{Teukolsky}}},\ and\ \bibinfo {author} {\bibfnamefont {I.}~\bibnamefont
  {{Wasserman}}},\ }\bibfield  {title} {\bibinfo {title} {{Spin evolution of
  accreting neutron stars: Nonlinear development of the r-mode instability}},\
  }\href {https://doi.org/10.1103/PhysRevD.76.064019} {\bibfield  {journal}
  {\bibinfo  {journal} {Physical Review D}\ }\textbf {\bibinfo {volume} {76}},\
  \bibinfo {eid} {064019} (\bibinfo {year} {2007})},\ \Eprint
  {https://arxiv.org/abs/0704.0799} {arXiv:0704.0799 [astro-ph]} \BibitemShut
  {NoStop}%
\bibitem [{\citenamefont {{Katz}}(1975)}]{Katz1975}%
  \BibitemOpen
  \bibfield  {author} {\bibinfo {author} {\bibfnamefont {J.~I.}\ \bibnamefont
  {{Katz}}},\ }\bibfield  {title} {\bibinfo {title} {{Two kinds of stellar
  collapse}},\ }\href {https://doi.org/10.1038/253698a0} {\bibfield  {journal}
  {\bibinfo  {journal} {Nature}\ }\textbf {\bibinfo {volume} {253}},\ \bibinfo
  {pages} {698} (\bibinfo {year} {1975})}\BibitemShut {NoStop}%
\bibitem [{\citenamefont {{Clark}}(1975)}]{Clark1975}%
  \BibitemOpen
  \bibfield  {author} {\bibinfo {author} {\bibfnamefont {G.~W.}\ \bibnamefont
  {{Clark}}},\ }\bibfield  {title} {\bibinfo {title} {{X-ray binaries in
  globular clusters.}},\ }\href {https://doi.org/10.1086/181869} {\bibfield
  {journal} {\bibinfo  {journal} {The Astrophysical Journal}\ }\textbf
  {\bibinfo {volume} {199}},\ \bibinfo {pages} {L143} (\bibinfo {year}
  {1975})}\BibitemShut {NoStop}%
\bibitem [{\citenamefont {{Alpar}}\ \emph {et~al.}(1982)\citenamefont
  {{Alpar}}, \citenamefont {{Cheng}}, \citenamefont {{Ruderman}},\ and\
  \citenamefont {{Shaham}}}]{AlparCheng1982}%
  \BibitemOpen
  \bibfield  {author} {\bibinfo {author} {\bibfnamefont {M.~A.}\ \bibnamefont
  {{Alpar}}}, \bibinfo {author} {\bibfnamefont {A.~F.}\ \bibnamefont
  {{Cheng}}}, \bibinfo {author} {\bibfnamefont {M.~A.}\ \bibnamefont
  {{Ruderman}}},\ and\ \bibinfo {author} {\bibfnamefont {J.}~\bibnamefont
  {{Shaham}}},\ }\bibfield  {title} {\bibinfo {title} {{A new class of radio
  pulsars}},\ }\href {https://doi.org/10.1038/300728a0} {\bibfield  {journal}
  {\bibinfo  {journal} {Nature}\ }\textbf {\bibinfo {volume} {300}},\ \bibinfo
  {pages} {728} (\bibinfo {year} {1982})}\BibitemShut {NoStop}%
\bibitem [{\citenamefont {{Pooley}}\ \emph {et~al.}(2003)\citenamefont
  {{Pooley}}, \citenamefont {{Lewin}}, \citenamefont {{Anderson}},
  \citenamefont {{Baumgardt}}, \citenamefont {{Filippenko}}, \citenamefont
  {{Gaensler}}, \citenamefont {{Homer}}, \citenamefont {{Hut}}, \citenamefont
  {{Kaspi}}, \citenamefont {{Makino}}, \citenamefont {{Margon}}, \citenamefont
  {{McMillan}}, \citenamefont {{Portegies Zwart}}, \citenamefont {{van der
  Klis}},\ and\ \citenamefont {{Verbunt}}}]{PooleyLewin2003}%
  \BibitemOpen
  \bibfield  {author} {\bibinfo {author} {\bibfnamefont {D.}~\bibnamefont
  {{Pooley}}}, \bibinfo {author} {\bibfnamefont {W.~H.~G.}\ \bibnamefont
  {{Lewin}}}, \bibinfo {author} {\bibfnamefont {S.~F.}\ \bibnamefont
  {{Anderson}}}, \bibinfo {author} {\bibfnamefont {H.}~\bibnamefont
  {{Baumgardt}}}, \bibinfo {author} {\bibfnamefont {A.~V.}\ \bibnamefont
  {{Filippenko}}}, \bibinfo {author} {\bibfnamefont {B.~M.}\ \bibnamefont
  {{Gaensler}}}, \bibinfo {author} {\bibfnamefont {L.}~\bibnamefont {{Homer}}},
  \bibinfo {author} {\bibfnamefont {P.}~\bibnamefont {{Hut}}}, \bibinfo
  {author} {\bibfnamefont {V.~M.}\ \bibnamefont {{Kaspi}}}, \bibinfo {author}
  {\bibfnamefont {J.}~\bibnamefont {{Makino}}}, \bibinfo {author}
  {\bibfnamefont {B.}~\bibnamefont {{Margon}}}, \bibinfo {author}
  {\bibfnamefont {S.}~\bibnamefont {{McMillan}}}, \bibinfo {author}
  {\bibfnamefont {S.}~\bibnamefont {{Portegies Zwart}}}, \bibinfo {author}
  {\bibfnamefont {M.}~\bibnamefont {{van der Klis}}},\ and\ \bibinfo {author}
  {\bibfnamefont {F.}~\bibnamefont {{Verbunt}}},\ }\bibfield  {title} {\bibinfo
  {title} {{Dynamical Formation of Close Binary Systems in Globular
  Clusters}},\ }\href {https://doi.org/10.1086/377074} {\bibfield  {journal}
  {\bibinfo  {journal} {The Astrophysical Journal}\ }\textbf {\bibinfo {volume}
  {591}},\ \bibinfo {pages} {L131} (\bibinfo {year} {2003})},\ \Eprint
  {https://arxiv.org/abs/astro-ph/0305003} {arXiv:astro-ph/0305003 [astro-ph]}
  \BibitemShut {NoStop}%
\bibitem [{\citenamefont {{Wang}}\ \emph {et~al.}(2006)\citenamefont {{Wang}},
  \citenamefont {{Chakrabarty}},\ and\ \citenamefont
  {{Kaplan}}}]{WangChakrabarty2006}%
  \BibitemOpen
  \bibfield  {author} {\bibinfo {author} {\bibfnamefont {Z.}~\bibnamefont
  {{Wang}}}, \bibinfo {author} {\bibfnamefont {D.}~\bibnamefont
  {{Chakrabarty}}},\ and\ \bibinfo {author} {\bibfnamefont {D.~L.}\
  \bibnamefont {{Kaplan}}},\ }\bibfield  {title} {\bibinfo {title} {{A debris
  disk around an isolated young neutron star}},\ }\href
  {https://doi.org/10.1038/nature04669} {\bibfield  {journal} {\bibinfo
  {journal} {Nature}\ }\textbf {\bibinfo {volume} {440}},\ \bibinfo {pages}
  {772} (\bibinfo {year} {2006})},\ \Eprint
  {https://arxiv.org/abs/astro-ph/0604076} {arXiv:astro-ph/0604076 [astro-ph]}
  \BibitemShut {NoStop}%
\bibitem [{\citenamefont {{Abbott}}\ \emph
  {et~al.}(2017{\natexlab{a}})\citenamefont {{Abbott}}, \citenamefont
  {{Abbott}}, \citenamefont {{Abbott}}, \citenamefont {{Abernathy}},
  \citenamefont {{Acernese}}, \citenamefont {{Ackley}}, \citenamefont
  {{Adams}}, \citenamefont {{Adams}}, \citenamefont {{Addesso}}, \citenamefont
  {{Adhikari}},\ and\ \citenamefont {et~al.}}]{AbbottAbbott2017b}%
  \BibitemOpen
  \bibfield  {author} {\bibinfo {author} {\bibfnamefont {B.~P.}\ \bibnamefont
  {{Abbott}}}, \bibinfo {author} {\bibfnamefont {R.}~\bibnamefont {{Abbott}}},
  \bibinfo {author} {\bibfnamefont {T.~D.}\ \bibnamefont {{Abbott}}}, \bibinfo
  {author} {\bibfnamefont {M.~R.}\ \bibnamefont {{Abernathy}}}, \bibinfo
  {author} {\bibfnamefont {F.}~\bibnamefont {{Acernese}}}, \bibinfo {author}
  {\bibfnamefont {K.}~\bibnamefont {{Ackley}}}, \bibinfo {author}
  {\bibfnamefont {C.}~\bibnamefont {{Adams}}}, \bibinfo {author} {\bibfnamefont
  {T.}~\bibnamefont {{Adams}}}, \bibinfo {author} {\bibfnamefont
  {P.}~\bibnamefont {{Addesso}}}, \bibinfo {author} {\bibfnamefont {R.~X.}\
  \bibnamefont {{Adhikari}}},\ and\ \bibinfo {author} {\bibnamefont {et~al.}},\
  }\bibfield  {title} {\bibinfo {title} {{Search for continuous gravitational
  waves from neutron stars in globular cluster NGC 6544}},\ }\href
  {https://doi.org/10.1103/PhysRevD.95.082005} {\bibfield  {journal} {\bibinfo
  {journal} {Physical Review D}\ }\textbf {\bibinfo {volume} {95}},\ \bibinfo
  {eid} {082005} (\bibinfo {year} {2017}{\natexlab{a}})},\ \Eprint
  {https://arxiv.org/abs/1607.02216} {arXiv:1607.02216 [gr-qc]} \BibitemShut
  {NoStop}%
\bibitem [{\citenamefont {{Jennings}}\ \emph {et~al.}(2020)\citenamefont
  {{Jennings}}, \citenamefont {{Cordes}},\ and\ \citenamefont
  {{Chatterjee}}}]{JenningsCordes2020}%
  \BibitemOpen
  \bibfield  {author} {\bibinfo {author} {\bibfnamefont {R.~J.}\ \bibnamefont
  {{Jennings}}}, \bibinfo {author} {\bibfnamefont {J.~M.}\ \bibnamefont
  {{Cordes}}},\ and\ \bibinfo {author} {\bibfnamefont {S.}~\bibnamefont
  {{Chatterjee}}},\ }\bibfield  {title} {\bibinfo {title} {{Pulsar Timing
  Signatures of Circumbinary Asteroid Belts}},\ }\href
  {https://doi.org/10.3847/1538-4357/abc178} {\bibfield  {journal} {\bibinfo
  {journal} {The Astrophysical Journal}\ }\textbf {\bibinfo {volume} {904}},\
  \bibinfo {eid} {191} (\bibinfo {year} {2020})},\ \Eprint
  {https://arxiv.org/abs/2007.10388} {arXiv:2007.10388 [astro-ph.HE]}
  \BibitemShut {NoStop}%
\bibitem [{\citenamefont {{Wolszczan}}\ and\ \citenamefont
  {{Frail}}(1992)}]{WolszczanFrail1992}%
  \BibitemOpen
  \bibfield  {author} {\bibinfo {author} {\bibfnamefont {A.}~\bibnamefont
  {{Wolszczan}}}\ and\ \bibinfo {author} {\bibfnamefont {D.~A.}\ \bibnamefont
  {{Frail}}},\ }\bibfield  {title} {\bibinfo {title} {{A planetary system
  around the millisecond pulsar PSR1257 + 12}},\ }\href
  {https://doi.org/10.1038/355145a0} {\bibfield  {journal} {\bibinfo  {journal}
  {Nature}\ }\textbf {\bibinfo {volume} {355}},\ \bibinfo {pages} {145}
  (\bibinfo {year} {1992})}\BibitemShut {NoStop}%
\bibitem [{\citenamefont {{Bailes}}\ \emph {et~al.}(2011)\citenamefont
  {{Bailes}}, \citenamefont {{Bates}}, \citenamefont {{Bhalerao}},
  \citenamefont {{Bhat}}, \citenamefont {{Burgay}}, \citenamefont
  {{Burke-Spolaor}}, \citenamefont {{D'Amico}}, \citenamefont {{Johnston}},
  \citenamefont {{Keith}}, \citenamefont {{Kramer}}, \citenamefont
  {{Kulkarni}}, \citenamefont {{Levin}}, \citenamefont {{Lyne}}, \citenamefont
  {{Milia}}, \citenamefont {{Possenti}}, \citenamefont {{Spitler}},
  \citenamefont {{Stappers}},\ and\ \citenamefont {{van
  Straten}}}]{BailesBates2011}%
  \BibitemOpen
  \bibfield  {author} {\bibinfo {author} {\bibfnamefont {M.}~\bibnamefont
  {{Bailes}}}, \bibinfo {author} {\bibfnamefont {S.~D.}\ \bibnamefont
  {{Bates}}}, \bibinfo {author} {\bibfnamefont {V.}~\bibnamefont {{Bhalerao}}},
  \bibinfo {author} {\bibfnamefont {N.~D.~R.}\ \bibnamefont {{Bhat}}}, \bibinfo
  {author} {\bibfnamefont {M.}~\bibnamefont {{Burgay}}}, \bibinfo {author}
  {\bibfnamefont {S.}~\bibnamefont {{Burke-Spolaor}}}, \bibinfo {author}
  {\bibfnamefont {N.}~\bibnamefont {{D'Amico}}}, \bibinfo {author}
  {\bibfnamefont {S.}~\bibnamefont {{Johnston}}}, \bibinfo {author}
  {\bibfnamefont {M.~J.}\ \bibnamefont {{Keith}}}, \bibinfo {author}
  {\bibfnamefont {M.}~\bibnamefont {{Kramer}}}, \bibinfo {author}
  {\bibfnamefont {S.~R.}\ \bibnamefont {{Kulkarni}}}, \bibinfo {author}
  {\bibfnamefont {L.}~\bibnamefont {{Levin}}}, \bibinfo {author} {\bibfnamefont
  {A.~G.}\ \bibnamefont {{Lyne}}}, \bibinfo {author} {\bibfnamefont
  {S.}~\bibnamefont {{Milia}}}, \bibinfo {author} {\bibfnamefont
  {A.}~\bibnamefont {{Possenti}}}, \bibinfo {author} {\bibfnamefont
  {L.}~\bibnamefont {{Spitler}}}, \bibinfo {author} {\bibfnamefont
  {B.}~\bibnamefont {{Stappers}}},\ and\ \bibinfo {author} {\bibfnamefont
  {W.}~\bibnamefont {{van Straten}}},\ }\bibfield  {title} {\bibinfo {title}
  {{Transformation of a Star into a Planet in a Millisecond Pulsar Binary}},\
  }\href {https://doi.org/10.1126/science.1208890} {\bibfield  {journal}
  {\bibinfo  {journal} {Science}\ }\textbf {\bibinfo {volume} {333}},\ \bibinfo
  {pages} {1717} (\bibinfo {year} {2011})},\ \Eprint
  {https://arxiv.org/abs/1108.5201} {arXiv:1108.5201 [astro-ph.SR]}
  \BibitemShut {NoStop}%
\bibitem [{\citenamefont {{Spiewak}}\ \emph {et~al.}(2018)\citenamefont
  {{Spiewak}}, \citenamefont {{Bailes}}, \citenamefont {{Barr}}, \citenamefont
  {{Bhat}}, \citenamefont {{Burgay}}, \citenamefont {{Cameron}}, \citenamefont
  {{Champion}}, \citenamefont {{Flynn}}, \citenamefont {{Jameson}},
  \citenamefont {{Johnston}}, \citenamefont {{Keith}}, \citenamefont
  {{Kramer}}, \citenamefont {{Kulkarni}}, \citenamefont {{Levin}},
  \citenamefont {{Lyne}}, \citenamefont {{Morello}}, \citenamefont {{Ng}},
  \citenamefont {{Possenti}}, \citenamefont {{Ravi}}, \citenamefont
  {{Stappers}}, \citenamefont {{van Straten}},\ and\ \citenamefont
  {{Tiburzi}}}]{SpiewakBailes2018}%
  \BibitemOpen
  \bibfield  {author} {\bibinfo {author} {\bibfnamefont {R.}~\bibnamefont
  {{Spiewak}}}, \bibinfo {author} {\bibfnamefont {M.}~\bibnamefont {{Bailes}}},
  \bibinfo {author} {\bibfnamefont {E.~D.}\ \bibnamefont {{Barr}}}, \bibinfo
  {author} {\bibfnamefont {N.~D.~R.}\ \bibnamefont {{Bhat}}}, \bibinfo {author}
  {\bibfnamefont {M.}~\bibnamefont {{Burgay}}}, \bibinfo {author}
  {\bibfnamefont {A.~D.}\ \bibnamefont {{Cameron}}}, \bibinfo {author}
  {\bibfnamefont {D.~J.}\ \bibnamefont {{Champion}}}, \bibinfo {author}
  {\bibfnamefont {C.~M.~L.}\ \bibnamefont {{Flynn}}}, \bibinfo {author}
  {\bibfnamefont {A.}~\bibnamefont {{Jameson}}}, \bibinfo {author}
  {\bibfnamefont {S.}~\bibnamefont {{Johnston}}}, \bibinfo {author}
  {\bibfnamefont {M.~J.}\ \bibnamefont {{Keith}}}, \bibinfo {author}
  {\bibfnamefont {M.}~\bibnamefont {{Kramer}}}, \bibinfo {author}
  {\bibfnamefont {S.~R.}\ \bibnamefont {{Kulkarni}}}, \bibinfo {author}
  {\bibfnamefont {L.}~\bibnamefont {{Levin}}}, \bibinfo {author} {\bibfnamefont
  {A.~G.}\ \bibnamefont {{Lyne}}}, \bibinfo {author} {\bibfnamefont
  {V.}~\bibnamefont {{Morello}}}, \bibinfo {author} {\bibfnamefont
  {C.}~\bibnamefont {{Ng}}}, \bibinfo {author} {\bibfnamefont {A.}~\bibnamefont
  {{Possenti}}}, \bibinfo {author} {\bibfnamefont {V.}~\bibnamefont {{Ravi}}},
  \bibinfo {author} {\bibfnamefont {B.~W.}\ \bibnamefont {{Stappers}}},
  \bibinfo {author} {\bibfnamefont {W.}~\bibnamefont {{van Straten}}},\ and\
  \bibinfo {author} {\bibfnamefont {C.}~\bibnamefont {{Tiburzi}}},\ }\bibfield
  {title} {\bibinfo {title} {{PSR J2322-2650 - a low-luminosity millisecond
  pulsar with a planetary-mass companion}},\ }\href
  {https://doi.org/10.1093/mnras/stx3157} {\bibfield  {journal} {\bibinfo
  {journal} {Monthly Notices of the Royal Astronomical Society}\ }\textbf
  {\bibinfo {volume} {475}},\ \bibinfo {pages} {469} (\bibinfo {year}
  {2018})},\ \Eprint {https://arxiv.org/abs/1712.04445} {arXiv:1712.04445
  [astro-ph.HE]} \BibitemShut {NoStop}%
\bibitem [{\citenamefont {{Behrens}}\ \emph {et~al.}(2020)\citenamefont
  {{Behrens}}, \citenamefont {{Ransom}}, \citenamefont {{Madison}},
  \citenamefont {{Arzoumanian}}, \citenamefont {{Crowter}}, \citenamefont
  {{DeCesar}}, \citenamefont {{Demorest}}, \citenamefont {{Dolch}},
  \citenamefont {{Ellis}}, \citenamefont {{Ferdman}}, \citenamefont
  {{Ferrara}}, \citenamefont {{Fonseca}}, \citenamefont {{Gentile}},
  \citenamefont {{Jones}}, \citenamefont {{Jones}}, \citenamefont {{Lam}},
  \citenamefont {{Levin}}, \citenamefont {{Lorimer}}, \citenamefont {{Lynch}},
  \citenamefont {{McLaughlin}}, \citenamefont {{Ng}}, \citenamefont {{Nice}},
  \citenamefont {{Pennucci}}, \citenamefont {{Perera}}, \citenamefont {{Ray}},
  \citenamefont {{Spiewak}}, \citenamefont {{Stairs}}, \citenamefont
  {{Stovall}}, \citenamefont {{Swiggum}},\ and\ \citenamefont
  {{Zhu}}}]{BehrensRansom2020}%
  \BibitemOpen
  \bibfield  {author} {\bibinfo {author} {\bibfnamefont {E.~A.}\ \bibnamefont
  {{Behrens}}}, \bibinfo {author} {\bibfnamefont {S.~M.}\ \bibnamefont
  {{Ransom}}}, \bibinfo {author} {\bibfnamefont {D.~R.}\ \bibnamefont
  {{Madison}}}, \bibinfo {author} {\bibfnamefont {Z.}~\bibnamefont
  {{Arzoumanian}}}, \bibinfo {author} {\bibfnamefont {K.}~\bibnamefont
  {{Crowter}}}, \bibinfo {author} {\bibfnamefont {M.~E.}\ \bibnamefont
  {{DeCesar}}}, \bibinfo {author} {\bibfnamefont {P.~B.}\ \bibnamefont
  {{Demorest}}}, \bibinfo {author} {\bibfnamefont {T.}~\bibnamefont {{Dolch}}},
  \bibinfo {author} {\bibfnamefont {J.~A.}\ \bibnamefont {{Ellis}}}, \bibinfo
  {author} {\bibfnamefont {R.~D.}\ \bibnamefont {{Ferdman}}}, \bibinfo {author}
  {\bibfnamefont {E.~C.}\ \bibnamefont {{Ferrara}}}, \bibinfo {author}
  {\bibfnamefont {E.}~\bibnamefont {{Fonseca}}}, \bibinfo {author}
  {\bibfnamefont {P.~A.}\ \bibnamefont {{Gentile}}}, \bibinfo {author}
  {\bibfnamefont {G.}~\bibnamefont {{Jones}}}, \bibinfo {author} {\bibfnamefont
  {M.~L.}\ \bibnamefont {{Jones}}}, \bibinfo {author} {\bibfnamefont {M.~T.}\
  \bibnamefont {{Lam}}}, \bibinfo {author} {\bibfnamefont {L.}~\bibnamefont
  {{Levin}}}, \bibinfo {author} {\bibfnamefont {D.~R.}\ \bibnamefont
  {{Lorimer}}}, \bibinfo {author} {\bibfnamefont {R.~S.}\ \bibnamefont
  {{Lynch}}}, \bibinfo {author} {\bibfnamefont {M.~A.}\ \bibnamefont
  {{McLaughlin}}}, \bibinfo {author} {\bibfnamefont {C.}~\bibnamefont {{Ng}}},
  \bibinfo {author} {\bibfnamefont {D.~J.}\ \bibnamefont {{Nice}}}, \bibinfo
  {author} {\bibfnamefont {T.~T.}\ \bibnamefont {{Pennucci}}}, \bibinfo
  {author} {\bibfnamefont {B.~B.~P.}\ \bibnamefont {{Perera}}}, \bibinfo
  {author} {\bibfnamefont {P.~S.}\ \bibnamefont {{Ray}}}, \bibinfo {author}
  {\bibfnamefont {R.}~\bibnamefont {{Spiewak}}}, \bibinfo {author}
  {\bibfnamefont {I.~H.}\ \bibnamefont {{Stairs}}}, \bibinfo {author}
  {\bibfnamefont {K.}~\bibnamefont {{Stovall}}}, \bibinfo {author}
  {\bibfnamefont {J.~K.}\ \bibnamefont {{Swiggum}}},\ and\ \bibinfo {author}
  {\bibfnamefont {W.~W.}\ \bibnamefont {{Zhu}}},\ }\bibfield  {title} {\bibinfo
  {title} {{The NANOGrav 11 yr Data Set: Constraints on Planetary Masses Around
  45 Millisecond Pulsars}},\ }\href {https://doi.org/10.3847/2041-8213/ab8121}
  {\bibfield  {journal} {\bibinfo  {journal} {The Astrophysical Journal}\
  }\textbf {\bibinfo {volume} {893}},\ \bibinfo {eid} {L8} (\bibinfo {year}
  {2020})},\ \Eprint {https://arxiv.org/abs/1912.00482} {arXiv:1912.00482
  [astro-ph.EP]} \BibitemShut {NoStop}%
\bibitem [{\citenamefont {{Ni{\c{t}}u}}\ \emph {et~al.}(2022)\citenamefont
  {{Ni{\c{t}}u}}, \citenamefont {{Keith}}, \citenamefont {{Stappers}},
  \citenamefont {{Lyne}},\ and\ \citenamefont {{Mickaliger}}}]{NituKeith2022}%
  \BibitemOpen
  \bibfield  {author} {\bibinfo {author} {\bibfnamefont {I.~C.}\ \bibnamefont
  {{Ni{\c{t}}u}}}, \bibinfo {author} {\bibfnamefont {M.~J.}\ \bibnamefont
  {{Keith}}}, \bibinfo {author} {\bibfnamefont {B.~W.}\ \bibnamefont
  {{Stappers}}}, \bibinfo {author} {\bibfnamefont {A.~G.}\ \bibnamefont
  {{Lyne}}},\ and\ \bibinfo {author} {\bibfnamefont {M.~B.}\ \bibnamefont
  {{Mickaliger}}},\ }\bibfield  {title} {\bibinfo {title} {{A search for
  planetary companions around 800 pulsars from the Jodrell Bank pulsar timing
  programme}},\ }\href {https://doi.org/10.1093/mnras/stac593} {\bibfield
  {journal} {\bibinfo  {journal} {Monthly Notices of the Royal Astronomical
  Society}\ }\textbf {\bibinfo {volume} {512}},\ \bibinfo {pages} {2446}
  (\bibinfo {year} {2022})},\ \Eprint {https://arxiv.org/abs/2203.01136}
  {arXiv:2203.01136 [astro-ph.EP]} \BibitemShut {NoStop}%
\bibitem [{\citenamefont {{Dergachev}}\ \emph {et~al.}(2019)\citenamefont
  {{Dergachev}}, \citenamefont {{Papa}}, \citenamefont {{Steltner}},\ and\
  \citenamefont {{Eggenstein}}}]{DergachevPapa2019}%
  \BibitemOpen
  \bibfield  {author} {\bibinfo {author} {\bibfnamefont {V.}~\bibnamefont
  {{Dergachev}}}, \bibinfo {author} {\bibfnamefont {M.~A.}\ \bibnamefont
  {{Papa}}}, \bibinfo {author} {\bibfnamefont {B.}~\bibnamefont {{Steltner}}},\
  and\ \bibinfo {author} {\bibfnamefont {H.-B.}\ \bibnamefont {{Eggenstein}}},\
  }\bibfield  {title} {\bibinfo {title} {{Loosely coherent search in LIGO O1
  data for continuous gravitational waves from Terzan 5 and the Galactic
  Center}},\ }\href {https://doi.org/10.1103/PhysRevD.99.084048} {\bibfield
  {journal} {\bibinfo  {journal} {Physical Review D}\ }\textbf {\bibinfo
  {volume} {99}},\ \bibinfo {eid} {084048} (\bibinfo {year} {2019})},\ \Eprint
  {https://arxiv.org/abs/1903.02389} {arXiv:1903.02389 [gr-qc]} \BibitemShut
  {NoStop}%
\bibitem [{\citenamefont {{Aasi}}\ \emph {et~al.}(2013)\citenamefont {{Aasi}},
  \citenamefont {{Abadie}}, \citenamefont {{Abbott}}, \citenamefont {{Abbott}},
  \citenamefont {{Abbott}}, \citenamefont {{Abernathy}}, \citenamefont
  {{Accadia}}, \citenamefont {{Acernese}}, \citenamefont {{Adams}},
  \citenamefont {{Adams}},\ and\ \citenamefont {et~al.}}]{AasiAbadie2013}%
  \BibitemOpen
  \bibfield  {author} {\bibinfo {author} {\bibfnamefont {J.}~\bibnamefont
  {{Aasi}}}, \bibinfo {author} {\bibfnamefont {J.}~\bibnamefont {{Abadie}}},
  \bibinfo {author} {\bibfnamefont {B.~P.}\ \bibnamefont {{Abbott}}}, \bibinfo
  {author} {\bibfnamefont {R.}~\bibnamefont {{Abbott}}}, \bibinfo {author}
  {\bibfnamefont {T.}~\bibnamefont {{Abbott}}}, \bibinfo {author}
  {\bibfnamefont {M.~R.}\ \bibnamefont {{Abernathy}}}, \bibinfo {author}
  {\bibfnamefont {T.}~\bibnamefont {{Accadia}}}, \bibinfo {author}
  {\bibfnamefont {F.}~\bibnamefont {{Acernese}}}, \bibinfo {author}
  {\bibfnamefont {C.}~\bibnamefont {{Adams}}}, \bibinfo {author} {\bibfnamefont
  {T.}~\bibnamefont {{Adams}}},\ and\ \bibinfo {author} {\bibnamefont
  {et~al.}},\ }\bibfield  {title} {\bibinfo {title} {{Directed search for
  continuous gravitational waves from the Galactic center}},\ }\href
  {https://doi.org/10.1103/PhysRevD.88.102002} {\bibfield  {journal} {\bibinfo
  {journal} {Physical Review D}\ }\textbf {\bibinfo {volume} {88}},\ \bibinfo
  {eid} {102002} (\bibinfo {year} {2013})},\ \Eprint
  {https://arxiv.org/abs/1309.6221} {arXiv:1309.6221 [gr-qc]} \BibitemShut
  {NoStop}%
\bibitem [{\citenamefont {{Piccinni}}\ \emph {et~al.}(2020)\citenamefont
  {{Piccinni}}, \citenamefont {{Astone}}, \citenamefont {{D'Antonio}},
  \citenamefont {{Frasca}}, \citenamefont {{Intini}}, \citenamefont {{La
  Rosa}}, \citenamefont {{Leaci}}, \citenamefont {{Mastrogiovanni}},
  \citenamefont {{Miller}},\ and\ \citenamefont
  {{Palomba}}}]{PiccinniAstone2020}%
  \BibitemOpen
  \bibfield  {author} {\bibinfo {author} {\bibfnamefont {O.~J.}\ \bibnamefont
  {{Piccinni}}}, \bibinfo {author} {\bibfnamefont {P.}~\bibnamefont
  {{Astone}}}, \bibinfo {author} {\bibfnamefont {S.}~\bibnamefont
  {{D'Antonio}}}, \bibinfo {author} {\bibfnamefont {S.}~\bibnamefont
  {{Frasca}}}, \bibinfo {author} {\bibfnamefont {G.}~\bibnamefont {{Intini}}},
  \bibinfo {author} {\bibfnamefont {I.}~\bibnamefont {{La Rosa}}}, \bibinfo
  {author} {\bibfnamefont {P.}~\bibnamefont {{Leaci}}}, \bibinfo {author}
  {\bibfnamefont {S.}~\bibnamefont {{Mastrogiovanni}}}, \bibinfo {author}
  {\bibfnamefont {A.}~\bibnamefont {{Miller}}},\ and\ \bibinfo {author}
  {\bibfnamefont {C.}~\bibnamefont {{Palomba}}},\ }\bibfield  {title} {\bibinfo
  {title} {{Directed search for continuous gravitational-wave signals from the
  Galactic Center in the Advanced LIGO second observing run}},\ }\href
  {https://doi.org/10.1103/PhysRevD.101.082004} {\bibfield  {journal} {\bibinfo
   {journal} {Physical Review D}\ }\textbf {\bibinfo {volume} {101}},\ \bibinfo
  {eid} {082004} (\bibinfo {year} {2020})},\ \Eprint
  {https://arxiv.org/abs/1910.05097} {arXiv:1910.05097 [gr-qc]} \BibitemShut
  {NoStop}%
\bibitem [{\citenamefont {{Abbott}}\ \emph
  {et~al.}(2022{\natexlab{a}})\citenamefont {{Abbott}}, \citenamefont {{Abe}},
  \citenamefont {{Acernese}}, \citenamefont {{Ackley}}, \citenamefont
  {{Adhikari}}, \citenamefont {{Adhikari}}, \citenamefont {{Adkins}},
  \citenamefont {{Adya}}, \citenamefont {{Affeldt}}, \citenamefont
  {{Agarwal}},\ and\ \citenamefont {et~al.}}]{AbbottAbe2022c}%
  \BibitemOpen
  \bibfield  {author} {\bibinfo {author} {\bibfnamefont {R.}~\bibnamefont
  {{Abbott}}}, \bibinfo {author} {\bibfnamefont {H.}~\bibnamefont {{Abe}}},
  \bibinfo {author} {\bibfnamefont {F.}~\bibnamefont {{Acernese}}}, \bibinfo
  {author} {\bibfnamefont {K.}~\bibnamefont {{Ackley}}}, \bibinfo {author}
  {\bibfnamefont {N.}~\bibnamefont {{Adhikari}}}, \bibinfo {author}
  {\bibfnamefont {R.~X.}\ \bibnamefont {{Adhikari}}}, \bibinfo {author}
  {\bibfnamefont {V.~K.}\ \bibnamefont {{Adkins}}}, \bibinfo {author}
  {\bibfnamefont {V.~B.}\ \bibnamefont {{Adya}}}, \bibinfo {author}
  {\bibfnamefont {C.}~\bibnamefont {{Affeldt}}}, \bibinfo {author}
  {\bibfnamefont {D.}~\bibnamefont {{Agarwal}}},\ and\ \bibinfo {author}
  {\bibnamefont {et~al.}},\ }\bibfield  {title} {\bibinfo {title} {{Search for
  continuous gravitational wave emission from the Milky Way center in O3
  LIGO-Virgo data}},\ }\href {https://doi.org/10.1103/PhysRevD.106.042003}
  {\bibfield  {journal} {\bibinfo  {journal} {Physical Review D}\ }\textbf
  {\bibinfo {volume} {106}},\ \bibinfo {eid} {042003} (\bibinfo {year}
  {2022}{\natexlab{a}})},\ \Eprint {https://arxiv.org/abs/2204.04523}
  {arXiv:2204.04523 [astro-ph.HE]} \BibitemShut {NoStop}%
\bibitem [{\citenamefont {{Abbott}}\ \emph {et~al.}(2010)\citenamefont
  {{Abbott}}, \citenamefont {{Abbott}}, \citenamefont {{Acernese}},
  \citenamefont {{Adhikari}}, \citenamefont {{Ajith}}, \citenamefont {{Allen}},
  \citenamefont {{Allen}}, \citenamefont {{Alshourbagy}}, \citenamefont
  {{Amin}}, \citenamefont {{Anderson}},\ and\ \citenamefont
  {et~al.}}]{AbbottAbbott2010}%
  \BibitemOpen
  \bibfield  {author} {\bibinfo {author} {\bibfnamefont {B.~P.}\ \bibnamefont
  {{Abbott}}}, \bibinfo {author} {\bibfnamefont {R.}~\bibnamefont {{Abbott}}},
  \bibinfo {author} {\bibfnamefont {F.}~\bibnamefont {{Acernese}}}, \bibinfo
  {author} {\bibfnamefont {R.}~\bibnamefont {{Adhikari}}}, \bibinfo {author}
  {\bibfnamefont {P.}~\bibnamefont {{Ajith}}}, \bibinfo {author} {\bibfnamefont
  {B.}~\bibnamefont {{Allen}}}, \bibinfo {author} {\bibfnamefont
  {G.}~\bibnamefont {{Allen}}}, \bibinfo {author} {\bibfnamefont
  {M.}~\bibnamefont {{Alshourbagy}}}, \bibinfo {author} {\bibfnamefont {R.~S.}\
  \bibnamefont {{Amin}}}, \bibinfo {author} {\bibfnamefont {S.~B.}\
  \bibnamefont {{Anderson}}},\ and\ \bibinfo {author} {\bibnamefont {et~al.}},\
  }\bibfield  {title} {\bibinfo {title} {{Searches for Gravitational Waves from
  Known Pulsars with Science Run 5 LIGO Data}},\ }\href
  {https://doi.org/10.1088/0004-637X/713/1/671} {\bibfield  {journal} {\bibinfo
   {journal} {The Astrophysical Journal}\ }\textbf {\bibinfo {volume} {713}},\
  \bibinfo {pages} {671} (\bibinfo {year} {2010})},\ \Eprint
  {https://arxiv.org/abs/0909.3583} {arXiv:0909.3583 [astro-ph.HE]}
  \BibitemShut {NoStop}%
\bibitem [{\citenamefont {{Aasi}}\ \emph {et~al.}(2014)\citenamefont {{Aasi}},
  \citenamefont {{Abadie}}, \citenamefont {{Abbott}}, \citenamefont {{Abbott}},
  \citenamefont {{Abbott}}, \citenamefont {{Abernathy}}, \citenamefont
  {{Accadia}}, \citenamefont {{Acernese}}, \citenamefont {{Adams}},
  \citenamefont {{Adams}},\ and\ \citenamefont {et~al.}}]{AasiAbadie2014}%
  \BibitemOpen
  \bibfield  {author} {\bibinfo {author} {\bibfnamefont {J.}~\bibnamefont
  {{Aasi}}}, \bibinfo {author} {\bibfnamefont {J.}~\bibnamefont {{Abadie}}},
  \bibinfo {author} {\bibfnamefont {B.~P.}\ \bibnamefont {{Abbott}}}, \bibinfo
  {author} {\bibfnamefont {R.}~\bibnamefont {{Abbott}}}, \bibinfo {author}
  {\bibfnamefont {T.}~\bibnamefont {{Abbott}}}, \bibinfo {author}
  {\bibfnamefont {M.~R.}\ \bibnamefont {{Abernathy}}}, \bibinfo {author}
  {\bibfnamefont {T.}~\bibnamefont {{Accadia}}}, \bibinfo {author}
  {\bibfnamefont {F.}~\bibnamefont {{Acernese}}}, \bibinfo {author}
  {\bibfnamefont {C.}~\bibnamefont {{Adams}}}, \bibinfo {author} {\bibfnamefont
  {T.}~\bibnamefont {{Adams}}},\ and\ \bibinfo {author} {\bibnamefont
  {et~al.}},\ }\bibfield  {title} {\bibinfo {title} {{Gravitational Waves from
  Known Pulsars: Results from the Initial Detector Era}},\ }\href
  {https://doi.org/10.1088/0004-637X/785/2/119} {\bibfield  {journal} {\bibinfo
   {journal} {The Astrophysical Journal}\ }\textbf {\bibinfo {volume} {785}},\
  \bibinfo {eid} {119} (\bibinfo {year} {2014})},\ \Eprint
  {https://arxiv.org/abs/1309.4027} {arXiv:1309.4027 [astro-ph.HE]}
  \BibitemShut {NoStop}%
\bibitem [{\citenamefont {{Pitkin}}\ \emph {et~al.}(2015)\citenamefont
  {{Pitkin}}, \citenamefont {{Gill}}, \citenamefont {{Jones}}, \citenamefont
  {{Woan}},\ and\ \citenamefont {{Davies}}}]{PitkinGill2015}%
  \BibitemOpen
  \bibfield  {author} {\bibinfo {author} {\bibfnamefont {M.}~\bibnamefont
  {{Pitkin}}}, \bibinfo {author} {\bibfnamefont {C.}~\bibnamefont {{Gill}}},
  \bibinfo {author} {\bibfnamefont {D.~I.}\ \bibnamefont {{Jones}}}, \bibinfo
  {author} {\bibfnamefont {G.}~\bibnamefont {{Woan}}},\ and\ \bibinfo {author}
  {\bibfnamefont {G.~S.}\ \bibnamefont {{Davies}}},\ }\bibfield  {title}
  {\bibinfo {title} {{First results and future prospects for dual-harmonic
  searches for gravitational waves from spinning neutron stars}},\ }\href
  {https://doi.org/10.1093/mnras/stv1931} {\bibfield  {journal} {\bibinfo
  {journal} {Monthly Notices of the Royal Astronomical Society}\ }\textbf
  {\bibinfo {volume} {453}},\ \bibinfo {pages} {4399} (\bibinfo {year}
  {2015})},\ \Eprint {https://arxiv.org/abs/1508.00416} {arXiv:1508.00416
  [astro-ph.HE]} \BibitemShut {NoStop}%
\bibitem [{\citenamefont {{Abbott}}\ \emph
  {et~al.}(2017{\natexlab{b}})\citenamefont {{Abbott}}, \citenamefont
  {{Abbott}}, \citenamefont {{Abbott}}, \citenamefont {{Abernathy}},
  \citenamefont {{Acernese}}, \citenamefont {{Ackley}}, \citenamefont
  {{Adams}}, \citenamefont {{Adams}}, \citenamefont {{Addesso}}, \citenamefont
  {{Adhikari}},\ and\ \citenamefont {et~al.}}]{AbbottAbbott2017c}%
  \BibitemOpen
  \bibfield  {author} {\bibinfo {author} {\bibfnamefont {B.~P.}\ \bibnamefont
  {{Abbott}}}, \bibinfo {author} {\bibfnamefont {R.}~\bibnamefont {{Abbott}}},
  \bibinfo {author} {\bibfnamefont {T.~D.}\ \bibnamefont {{Abbott}}}, \bibinfo
  {author} {\bibfnamefont {M.~R.}\ \bibnamefont {{Abernathy}}}, \bibinfo
  {author} {\bibfnamefont {F.}~\bibnamefont {{Acernese}}}, \bibinfo {author}
  {\bibfnamefont {K.}~\bibnamefont {{Ackley}}}, \bibinfo {author}
  {\bibfnamefont {C.}~\bibnamefont {{Adams}}}, \bibinfo {author} {\bibfnamefont
  {T.}~\bibnamefont {{Adams}}}, \bibinfo {author} {\bibfnamefont
  {P.}~\bibnamefont {{Addesso}}}, \bibinfo {author} {\bibfnamefont {R.~X.}\
  \bibnamefont {{Adhikari}}},\ and\ \bibinfo {author} {\bibnamefont {et~al.}},\
  }\bibfield  {title} {\bibinfo {title} {{First Search for Gravitational Waves
  from Known Pulsars with Advanced LIGO}},\ }\href
  {https://doi.org/10.3847/1538-4357/aa677f} {\bibfield  {journal} {\bibinfo
  {journal} {The Astrophysical Journal}\ }\textbf {\bibinfo {volume} {839}},\
  \bibinfo {eid} {12} (\bibinfo {year} {2017}{\natexlab{b}})},\ \Eprint
  {https://arxiv.org/abs/1701.07709} {arXiv:1701.07709 [astro-ph.HE]}
  \BibitemShut {NoStop}%
\bibitem [{\citenamefont {{Abbott}}\ \emph
  {et~al.}(2019{\natexlab{a}})\citenamefont {{Abbott}}, \citenamefont
  {{Abbott}}, \citenamefont {{Abbott}}, \citenamefont {{Abraham}},
  \citenamefont {{Acernese}}, \citenamefont {{Ackley}}, \citenamefont
  {{Adams}}, \citenamefont {{Adhikari}}, \citenamefont {{Adya}}, \citenamefont
  {{Affeldt}},\ and\ \citenamefont {et~al.}}]{AbbottAbbott2019b}%
  \BibitemOpen
  \bibfield  {author} {\bibinfo {author} {\bibfnamefont {B.~P.}\ \bibnamefont
  {{Abbott}}}, \bibinfo {author} {\bibfnamefont {R.}~\bibnamefont {{Abbott}}},
  \bibinfo {author} {\bibfnamefont {T.~D.}\ \bibnamefont {{Abbott}}}, \bibinfo
  {author} {\bibfnamefont {S.}~\bibnamefont {{Abraham}}}, \bibinfo {author}
  {\bibfnamefont {F.}~\bibnamefont {{Acernese}}}, \bibinfo {author}
  {\bibfnamefont {K.}~\bibnamefont {{Ackley}}}, \bibinfo {author}
  {\bibfnamefont {C.}~\bibnamefont {{Adams}}}, \bibinfo {author} {\bibfnamefont
  {R.~X.}\ \bibnamefont {{Adhikari}}}, \bibinfo {author} {\bibfnamefont
  {V.~B.}\ \bibnamefont {{Adya}}}, \bibinfo {author} {\bibfnamefont
  {C.}~\bibnamefont {{Affeldt}}},\ and\ \bibinfo {author} {\bibnamefont
  {et~al.}},\ }\bibfield  {title} {\bibinfo {title} {{All-sky search for
  continuous gravitational waves from isolated neutron stars using Advanced
  LIGO O2 data}},\ }\href {https://doi.org/10.1103/PhysRevD.100.024004}
  {\bibfield  {journal} {\bibinfo  {journal} {Physical Review D}\ }\textbf
  {\bibinfo {volume} {100}},\ \bibinfo {eid} {024004} (\bibinfo {year}
  {2019}{\natexlab{a}})},\ \Eprint {https://arxiv.org/abs/1903.01901}
  {arXiv:1903.01901 [astro-ph.HE]} \BibitemShut {NoStop}%
\bibitem [{\citenamefont {{Suvorova}}\ \emph {et~al.}(2016)\citenamefont
  {{Suvorova}}, \citenamefont {{Sun}}, \citenamefont {{Melatos}}, \citenamefont
  {{Moran}},\ and\ \citenamefont {{Evans}}}]{SuvorovaSun2016}%
  \BibitemOpen
  \bibfield  {author} {\bibinfo {author} {\bibfnamefont {S.}~\bibnamefont
  {{Suvorova}}}, \bibinfo {author} {\bibfnamefont {L.}~\bibnamefont {{Sun}}},
  \bibinfo {author} {\bibfnamefont {A.}~\bibnamefont {{Melatos}}}, \bibinfo
  {author} {\bibfnamefont {W.}~\bibnamefont {{Moran}}},\ and\ \bibinfo {author}
  {\bibfnamefont {R.~J.}\ \bibnamefont {{Evans}}},\ }\bibfield  {title}
  {\bibinfo {title} {{Hidden Markov model tracking of continuous gravitational
  waves from a neutron star with wandering spin}},\ }\href
  {https://doi.org/10.1103/PhysRevD.93.123009} {\bibfield  {journal} {\bibinfo
  {journal} {Physical Review D}\ }\textbf {\bibinfo {volume} {93}},\ \bibinfo
  {eid} {123009} (\bibinfo {year} {2016})},\ \Eprint
  {https://arxiv.org/abs/1606.02412} {arXiv:1606.02412 [astro-ph.IM]}
  \BibitemShut {NoStop}%
\bibitem [{\citenamefont {Suvorova}\ \emph {et~al.}(2018)\citenamefont
  {Suvorova}, \citenamefont {Melatos}, \citenamefont {Evans}, \citenamefont
  {Moran}, \citenamefont {Clearwater},\ and\ \citenamefont
  {Sun}}]{SuvorovaMelatos2018}%
  \BibitemOpen
  \bibfield  {author} {\bibinfo {author} {\bibfnamefont {S.}~\bibnamefont
  {Suvorova}}, \bibinfo {author} {\bibfnamefont {A.}~\bibnamefont {Melatos}},
  \bibinfo {author} {\bibfnamefont {R.~J.}\ \bibnamefont {Evans}}, \bibinfo
  {author} {\bibfnamefont {W.}~\bibnamefont {Moran}}, \bibinfo {author}
  {\bibfnamefont {P.}~\bibnamefont {Clearwater}},\ and\ \bibinfo {author}
  {\bibfnamefont {L.}~\bibnamefont {Sun}},\ }\bibfield  {title} {\bibinfo
  {title} {{Phase-Continuous Frequency Line Track-Before-Detect of a Tone With
  Slow Frequency Variation}},\ }\href
  {https://doi.org/10.1109/TSP.2018.2877176} {\bibfield  {journal} {\bibinfo
  {journal} {IEEE Transactions on Signal Processing}\ }\textbf {\bibinfo
  {volume} {66}},\ \bibinfo {pages} {6434} (\bibinfo {year}
  {2018})}\BibitemShut {NoStop}%
\bibitem [{\citenamefont {{Melatos}}\ \emph {et~al.}(2021)\citenamefont
  {{Melatos}}, \citenamefont {{Clearwater}}, \citenamefont {{Suvorova}},
  \citenamefont {{Sun}}, \citenamefont {{Moran}},\ and\ \citenamefont
  {{Evans}}}]{MelatosClearwater2021}%
  \BibitemOpen
  \bibfield  {author} {\bibinfo {author} {\bibfnamefont {A.}~\bibnamefont
  {{Melatos}}}, \bibinfo {author} {\bibfnamefont {P.}~\bibnamefont
  {{Clearwater}}}, \bibinfo {author} {\bibfnamefont {S.}~\bibnamefont
  {{Suvorova}}}, \bibinfo {author} {\bibfnamefont {L.}~\bibnamefont {{Sun}}},
  \bibinfo {author} {\bibfnamefont {W.}~\bibnamefont {{Moran}}},\ and\ \bibinfo
  {author} {\bibfnamefont {R.~J.}\ \bibnamefont {{Evans}}},\ }\bibfield
  {title} {\bibinfo {title} {{Hidden Markov model tracking of continuous
  gravitational waves from a binary neutron star with wandering spin. III.
  Rotational phase tracking}},\ }\href
  {https://doi.org/10.1103/PhysRevD.104.042003} {\bibfield  {journal} {\bibinfo
   {journal} {Physical Review D}\ }\textbf {\bibinfo {volume} {104}},\ \bibinfo
  {eid} {042003} (\bibinfo {year} {2021})},\ \Eprint
  {https://arxiv.org/abs/2107.12822} {arXiv:2107.12822 [gr-qc]} \BibitemShut
  {NoStop}%
\bibitem [{\citenamefont {{Abbott}}\ \emph
  {et~al.}(2017{\natexlab{c}})\citenamefont {{Abbott}}, \citenamefont
  {{Abbott}}, \citenamefont {{Abbott}}, \citenamefont {{Acernese}},
  \citenamefont {{Ackley}}, \citenamefont {{Adams}}, \citenamefont {{Adams}},
  \citenamefont {{Addesso}}, \citenamefont {{Adhikari}}, \citenamefont
  {{Adya}},\ and\ \citenamefont {et~al.}}]{AbbottAbbott2017a}%
  \BibitemOpen
  \bibfield  {author} {\bibinfo {author} {\bibfnamefont {B.~P.}\ \bibnamefont
  {{Abbott}}}, \bibinfo {author} {\bibfnamefont {R.}~\bibnamefont {{Abbott}}},
  \bibinfo {author} {\bibfnamefont {T.~D.}\ \bibnamefont {{Abbott}}}, \bibinfo
  {author} {\bibfnamefont {F.}~\bibnamefont {{Acernese}}}, \bibinfo {author}
  {\bibfnamefont {K.}~\bibnamefont {{Ackley}}}, \bibinfo {author}
  {\bibfnamefont {C.}~\bibnamefont {{Adams}}}, \bibinfo {author} {\bibfnamefont
  {T.}~\bibnamefont {{Adams}}}, \bibinfo {author} {\bibfnamefont
  {P.}~\bibnamefont {{Addesso}}}, \bibinfo {author} {\bibfnamefont {R.~X.}\
  \bibnamefont {{Adhikari}}}, \bibinfo {author} {\bibfnamefont {V.~B.}\
  \bibnamefont {{Adya}}},\ and\ \bibinfo {author} {\bibnamefont {et~al.}},\
  }\bibfield  {title} {\bibinfo {title} {{Search for gravitational waves from
  Scorpius X-1 in the first Advanced LIGO observing run with a hidden Markov
  model}},\ }\href {https://doi.org/10.1103/PhysRevD.95.122003} {\bibfield
  {journal} {\bibinfo  {journal} {Physical Review D}\ }\textbf {\bibinfo
  {volume} {95}},\ \bibinfo {eid} {122003} (\bibinfo {year}
  {2017}{\natexlab{c}})},\ \Eprint {https://arxiv.org/abs/1704.03719}
  {arXiv:1704.03719 [gr-qc]} \BibitemShut {NoStop}%
\bibitem [{\citenamefont {{Abbott}}\ \emph
  {et~al.}(2019{\natexlab{b}})\citenamefont {{Abbott}}, \citenamefont
  {{Abbott}}, \citenamefont {{Abbott}}, \citenamefont {{Abraham}},
  \citenamefont {{Acernese}}, \citenamefont {{Ackley}}, \citenamefont
  {{Adams}}, \citenamefont {{Adhikari}}, \citenamefont {{Adya}}, \citenamefont
  {{Affeldt}},\ and\ \citenamefont {et~al.}}]{AbbottAbbott2019a}%
  \BibitemOpen
  \bibfield  {author} {\bibinfo {author} {\bibfnamefont {B.~P.}\ \bibnamefont
  {{Abbott}}}, \bibinfo {author} {\bibfnamefont {R.}~\bibnamefont {{Abbott}}},
  \bibinfo {author} {\bibfnamefont {T.~D.}\ \bibnamefont {{Abbott}}}, \bibinfo
  {author} {\bibfnamefont {S.}~\bibnamefont {{Abraham}}}, \bibinfo {author}
  {\bibfnamefont {F.}~\bibnamefont {{Acernese}}}, \bibinfo {author}
  {\bibfnamefont {K.}~\bibnamefont {{Ackley}}}, \bibinfo {author}
  {\bibfnamefont {C.}~\bibnamefont {{Adams}}}, \bibinfo {author} {\bibfnamefont
  {R.~X.}\ \bibnamefont {{Adhikari}}}, \bibinfo {author} {\bibfnamefont
  {V.~B.}\ \bibnamefont {{Adya}}}, \bibinfo {author} {\bibfnamefont
  {C.}~\bibnamefont {{Affeldt}}},\ and\ \bibinfo {author} {\bibnamefont
  {et~al.}},\ }\bibfield  {title} {\bibinfo {title} {{Search for gravitational
  waves from Scorpius X-1 in the second Advanced LIGO observing run with an
  improved hidden Markov model}},\ }\href
  {https://doi.org/10.1103/PhysRevD.100.122002} {\bibfield  {journal} {\bibinfo
   {journal} {Physical Review D}\ }\textbf {\bibinfo {volume} {100}},\ \bibinfo
  {eid} {122002} (\bibinfo {year} {2019}{\natexlab{b}})},\ \Eprint
  {https://arxiv.org/abs/1906.12040} {arXiv:1906.12040 [gr-qc]} \BibitemShut
  {NoStop}%
\bibitem [{\citenamefont {{Millhouse}}\ \emph {et~al.}(2020)\citenamefont
  {{Millhouse}}, \citenamefont {{Strang}},\ and\ \citenamefont
  {{Melatos}}}]{MillhouseStrang2020}%
  \BibitemOpen
  \bibfield  {author} {\bibinfo {author} {\bibfnamefont {M.}~\bibnamefont
  {{Millhouse}}}, \bibinfo {author} {\bibfnamefont {L.}~\bibnamefont
  {{Strang}}},\ and\ \bibinfo {author} {\bibfnamefont {A.}~\bibnamefont
  {{Melatos}}},\ }\bibfield  {title} {\bibinfo {title} {{Search for
  gravitational waves from 12 young supernova remnants with a hidden Markov
  model in Advanced LIGO's second observing run}},\ }\href
  {https://doi.org/10.1103/PhysRevD.102.083025} {\bibfield  {journal} {\bibinfo
   {journal} {Physical Review D}\ }\textbf {\bibinfo {volume} {102}},\ \bibinfo
  {eid} {083025} (\bibinfo {year} {2020})},\ \Eprint
  {https://arxiv.org/abs/2003.08588} {arXiv:2003.08588 [gr-qc]} \BibitemShut
  {NoStop}%
\bibitem [{\citenamefont {{Middleton}}\ \emph {et~al.}(2020)\citenamefont
  {{Middleton}}, \citenamefont {{Clearwater}}, \citenamefont {{Melatos}},\ and\
  \citenamefont {{Dunn}}}]{MiddletonClearwater2020}%
  \BibitemOpen
  \bibfield  {author} {\bibinfo {author} {\bibfnamefont {H.}~\bibnamefont
  {{Middleton}}}, \bibinfo {author} {\bibfnamefont {P.}~\bibnamefont
  {{Clearwater}}}, \bibinfo {author} {\bibfnamefont {A.}~\bibnamefont
  {{Melatos}}},\ and\ \bibinfo {author} {\bibfnamefont {L.}~\bibnamefont
  {{Dunn}}},\ }\bibfield  {title} {\bibinfo {title} {{Search for gravitational
  waves from five low mass x-ray binaries in the second Advanced LIGO observing
  run with an improved hidden Markov model}},\ }\href
  {https://doi.org/10.1103/PhysRevD.102.023006} {\bibfield  {journal} {\bibinfo
   {journal} {Physical Review D}\ }\textbf {\bibinfo {volume} {102}},\ \bibinfo
  {eid} {023006} (\bibinfo {year} {2020})},\ \Eprint
  {https://arxiv.org/abs/2006.06907} {arXiv:2006.06907 [astro-ph.HE]}
  \BibitemShut {NoStop}%
\bibitem [{\citenamefont {{Sun}}\ \emph
  {et~al.}(2020{\natexlab{a}})\citenamefont {{Sun}}, \citenamefont {{Brito}},\
  and\ \citenamefont {{Isi}}}]{SunBrito2020}%
  \BibitemOpen
  \bibfield  {author} {\bibinfo {author} {\bibfnamefont {L.}~\bibnamefont
  {{Sun}}}, \bibinfo {author} {\bibfnamefont {R.}~\bibnamefont {{Brito}}},\
  and\ \bibinfo {author} {\bibfnamefont {M.}~\bibnamefont {{Isi}}},\ }\bibfield
   {title} {\bibinfo {title} {{Search for ultralight bosons in Cygnus X-1 with
  Advanced LIGO}},\ }\href {https://doi.org/10.1103/PhysRevD.101.063020}
  {\bibfield  {journal} {\bibinfo  {journal} {Physical Review D}\ }\textbf
  {\bibinfo {volume} {101}},\ \bibinfo {eid} {063020} (\bibinfo {year}
  {2020}{\natexlab{a}})},\ \Eprint {https://arxiv.org/abs/1909.11267}
  {arXiv:1909.11267 [gr-qc]} \BibitemShut {NoStop}%
\bibitem [{\citenamefont {{Abbott}}\ \emph
  {et~al.}(2021{\natexlab{a}})\citenamefont {{Abbott}}, \citenamefont
  {{Abbott}}, \citenamefont {{Abraham}}, \citenamefont {{Acernese}},
  \citenamefont {{Ackley}}, \citenamefont {{Adams}}, \citenamefont {{Adams}},
  \citenamefont {{Adhikari}}, \citenamefont {{Adya}}, \citenamefont
  {{Affeldt}},\ and\ \citenamefont {et~al.}}]{AbbottAbbott2021a}%
  \BibitemOpen
  \bibfield  {author} {\bibinfo {author} {\bibfnamefont {R.}~\bibnamefont
  {{Abbott}}}, \bibinfo {author} {\bibfnamefont {T.~D.}\ \bibnamefont
  {{Abbott}}}, \bibinfo {author} {\bibfnamefont {S.}~\bibnamefont {{Abraham}}},
  \bibinfo {author} {\bibfnamefont {F.}~\bibnamefont {{Acernese}}}, \bibinfo
  {author} {\bibfnamefont {K.}~\bibnamefont {{Ackley}}}, \bibinfo {author}
  {\bibfnamefont {A.}~\bibnamefont {{Adams}}}, \bibinfo {author} {\bibfnamefont
  {C.}~\bibnamefont {{Adams}}}, \bibinfo {author} {\bibfnamefont {R.~X.}\
  \bibnamefont {{Adhikari}}}, \bibinfo {author} {\bibfnamefont {V.~B.}\
  \bibnamefont {{Adya}}}, \bibinfo {author} {\bibfnamefont {C.}~\bibnamefont
  {{Affeldt}}},\ and\ \bibinfo {author} {\bibnamefont {et~al.}},\ }\bibfield
  {title} {\bibinfo {title} {{Searches for Continuous Gravitational Waves from
  Young Supernova Remnants in the Early Third Observing Run of Advanced LIGO
  and Virgo}},\ }\href {https://doi.org/10.3847/1538-4357/ac17ea} {\bibfield
  {journal} {\bibinfo  {journal} {The Astrophysical Journal}\ }\textbf
  {\bibinfo {volume} {921}},\ \bibinfo {eid} {80} (\bibinfo {year}
  {2021}{\natexlab{a}})},\ \Eprint {https://arxiv.org/abs/2105.11641}
  {arXiv:2105.11641 [astro-ph.HE]} \BibitemShut {NoStop}%
\bibitem [{\citenamefont {{Jones}}\ and\ \citenamefont
  {{Sun}}(2021)}]{JonesSun2021}%
  \BibitemOpen
  \bibfield  {author} {\bibinfo {author} {\bibfnamefont {D.}~\bibnamefont
  {{Jones}}}\ and\ \bibinfo {author} {\bibfnamefont {L.}~\bibnamefont
  {{Sun}}},\ }\bibfield  {title} {\bibinfo {title} {{Search for continuous
  gravitational waves from Fomalhaut b in the second Advanced LIGO observing
  run with a hidden Markov model}},\ }\href
  {https://doi.org/10.1103/PhysRevD.103.023020} {\bibfield  {journal} {\bibinfo
   {journal} {Physical Review D}\ }\textbf {\bibinfo {volume} {103}},\ \bibinfo
  {eid} {023020} (\bibinfo {year} {2021})},\ \Eprint
  {https://arxiv.org/abs/2007.08732} {arXiv:2007.08732 [gr-qc]} \BibitemShut
  {NoStop}%
\bibitem [{\citenamefont {{Beniwal}}\ \emph {et~al.}(2021)\citenamefont
  {{Beniwal}}, \citenamefont {{Clearwater}}, \citenamefont {{Dunn}},
  \citenamefont {{Melatos}},\ and\ \citenamefont
  {{Ottaway}}}]{BeniwalClearwater2021}%
  \BibitemOpen
  \bibfield  {author} {\bibinfo {author} {\bibfnamefont {D.}~\bibnamefont
  {{Beniwal}}}, \bibinfo {author} {\bibfnamefont {P.}~\bibnamefont
  {{Clearwater}}}, \bibinfo {author} {\bibfnamefont {L.}~\bibnamefont
  {{Dunn}}}, \bibinfo {author} {\bibfnamefont {A.}~\bibnamefont {{Melatos}}},\
  and\ \bibinfo {author} {\bibfnamefont {D.}~\bibnamefont {{Ottaway}}},\
  }\bibfield  {title} {\bibinfo {title} {{Search for continuous gravitational
  waves from ten H.E.S.S. sources using a hidden Markov model}},\ }\href
  {https://doi.org/10.1103/PhysRevD.103.083009} {\bibfield  {journal} {\bibinfo
   {journal} {Physical Review D}\ }\textbf {\bibinfo {volume} {103}},\ \bibinfo
  {eid} {083009} (\bibinfo {year} {2021})},\ \Eprint
  {https://arxiv.org/abs/2102.06334} {arXiv:2102.06334 [astro-ph.HE]}
  \BibitemShut {NoStop}%
\bibitem [{\citenamefont {{Beniwal}}\ \emph {et~al.}(2022)\citenamefont
  {{Beniwal}}, \citenamefont {{Clearwater}}, \citenamefont {{Dunn}},
  \citenamefont {{Strang}}, \citenamefont {{Rowell}}, \citenamefont
  {{Melatos}},\ and\ \citenamefont {{Ottaway}}}]{BeniwalClearwater2022}%
  \BibitemOpen
  \bibfield  {author} {\bibinfo {author} {\bibfnamefont {D.}~\bibnamefont
  {{Beniwal}}}, \bibinfo {author} {\bibfnamefont {P.}~\bibnamefont
  {{Clearwater}}}, \bibinfo {author} {\bibfnamefont {L.}~\bibnamefont
  {{Dunn}}}, \bibinfo {author} {\bibfnamefont {L.}~\bibnamefont {{Strang}}},
  \bibinfo {author} {\bibfnamefont {G.}~\bibnamefont {{Rowell}}}, \bibinfo
  {author} {\bibfnamefont {A.}~\bibnamefont {{Melatos}}},\ and\ \bibinfo
  {author} {\bibfnamefont {D.}~\bibnamefont {{Ottaway}}},\ }\bibfield  {title}
  {\bibinfo {title} {{Search for continuous gravitational waves from HESS
  J1427-608 with a hidden Markov model}},\ }\href
  {https://doi.org/10.1103/PhysRevD.106.103018} {\bibfield  {journal} {\bibinfo
   {journal} {Physical Review D}\ }\textbf {\bibinfo {volume} {106}},\ \bibinfo
  {eid} {103018} (\bibinfo {year} {2022})},\ \Eprint
  {https://arxiv.org/abs/2210.09592} {arXiv:2210.09592 [astro-ph.HE]}
  \BibitemShut {NoStop}%
\bibitem [{\citenamefont {{Abbott}}\ \emph
  {et~al.}(2022{\natexlab{b}})\citenamefont {{Abbott}}, \citenamefont {{Abe}},
  \citenamefont {{Acernese}}, \citenamefont {{Ackley}}, \citenamefont
  {{Adhikari}}, \citenamefont {{Adhikari}}, \citenamefont {{Adkins}},
  \citenamefont {{Adya}}, \citenamefont {{Affeldt}}, \citenamefont
  {{Agarwal}},\ and\ \citenamefont {et~al.}}]{AbbottAbe2022b}%
  \BibitemOpen
  \bibfield  {author} {\bibinfo {author} {\bibfnamefont {R.}~\bibnamefont
  {{Abbott}}}, \bibinfo {author} {\bibfnamefont {H.}~\bibnamefont {{Abe}}},
  \bibinfo {author} {\bibfnamefont {F.}~\bibnamefont {{Acernese}}}, \bibinfo
  {author} {\bibfnamefont {K.}~\bibnamefont {{Ackley}}}, \bibinfo {author}
  {\bibfnamefont {N.}~\bibnamefont {{Adhikari}}}, \bibinfo {author}
  {\bibfnamefont {R.~X.}\ \bibnamefont {{Adhikari}}}, \bibinfo {author}
  {\bibfnamefont {V.~K.}\ \bibnamefont {{Adkins}}}, \bibinfo {author}
  {\bibfnamefont {V.~B.}\ \bibnamefont {{Adya}}}, \bibinfo {author}
  {\bibfnamefont {C.}~\bibnamefont {{Affeldt}}}, \bibinfo {author}
  {\bibfnamefont {D.}~\bibnamefont {{Agarwal}}},\ and\ \bibinfo {author}
  {\bibnamefont {et~al.}},\ }\bibfield  {title} {\bibinfo {title} {{Search for
  gravitational waves from Scorpius X-1 with a hidden Markov model in O3 LIGO
  data}},\ }\href {https://doi.org/10.1103/PhysRevD.106.062002} {\bibfield
  {journal} {\bibinfo  {journal} {Physical Review D}\ }\textbf {\bibinfo
  {volume} {106}},\ \bibinfo {eid} {062002} (\bibinfo {year}
  {2022}{\natexlab{b}})},\ \Eprint {https://arxiv.org/abs/2201.10104}
  {arXiv:2201.10104 [gr-qc]} \BibitemShut {NoStop}%
\bibitem [{\citenamefont {{Abbott}}\ \emph
  {et~al.}(2022{\natexlab{c}})\citenamefont {{Abbott}}, \citenamefont
  {{Abbott}}, \citenamefont {{Acernese}}, \citenamefont {{Ackley}},
  \citenamefont {{Adams}}, \citenamefont {{Adhikari}}, \citenamefont
  {{Adhikari}}, \citenamefont {{Adya}}, \citenamefont {{Affeldt}},
  \citenamefont {{Agarwal}},\ and\ \citenamefont {et~al.}}]{AbbottAbbott2022c}%
  \BibitemOpen
  \bibfield  {author} {\bibinfo {author} {\bibfnamefont {R.}~\bibnamefont
  {{Abbott}}}, \bibinfo {author} {\bibfnamefont {T.~D.}\ \bibnamefont
  {{Abbott}}}, \bibinfo {author} {\bibfnamefont {F.}~\bibnamefont
  {{Acernese}}}, \bibinfo {author} {\bibfnamefont {K.}~\bibnamefont
  {{Ackley}}}, \bibinfo {author} {\bibfnamefont {C.}~\bibnamefont {{Adams}}},
  \bibinfo {author} {\bibfnamefont {N.}~\bibnamefont {{Adhikari}}}, \bibinfo
  {author} {\bibfnamefont {R.~X.}\ \bibnamefont {{Adhikari}}}, \bibinfo
  {author} {\bibfnamefont {V.~B.}\ \bibnamefont {{Adya}}}, \bibinfo {author}
  {\bibfnamefont {C.}~\bibnamefont {{Affeldt}}}, \bibinfo {author}
  {\bibfnamefont {D.}~\bibnamefont {{Agarwal}}},\ and\ \bibinfo {author}
  {\bibnamefont {et~al.}},\ }\bibfield  {title} {\bibinfo {title} {{Search for
  continuous gravitational waves from 20 accreting millisecond x-ray pulsars in
  O3 LIGO data}},\ }\href {https://doi.org/10.1103/PhysRevD.105.022002}
  {\bibfield  {journal} {\bibinfo  {journal} {Physical Review D}\ }\textbf
  {\bibinfo {volume} {105}},\ \bibinfo {eid} {022002} (\bibinfo {year}
  {2022}{\natexlab{c}})},\ \Eprint {https://arxiv.org/abs/2109.09255}
  {arXiv:2109.09255 [astro-ph.HE]} \BibitemShut {NoStop}%
\bibitem [{\citenamefont {{Abbott}}\ \emph
  {et~al.}(2022{\natexlab{d}})\citenamefont {{Abbott}}, \citenamefont {{Abe}},
  \citenamefont {{Acernese}}, \citenamefont {{Ackley}}, \citenamefont
  {{Adhikari}}, \citenamefont {{Adhikari}}, \citenamefont {{Adkins}},
  \citenamefont {{Adya}}, \citenamefont {{Affeldt}}, \citenamefont
  {{Agarwal}},\ and\ \citenamefont {et~al.}}]{AbbottAbe2022a}%
  \BibitemOpen
  \bibfield  {author} {\bibinfo {author} {\bibfnamefont {R.}~\bibnamefont
  {{Abbott}}}, \bibinfo {author} {\bibfnamefont {H.}~\bibnamefont {{Abe}}},
  \bibinfo {author} {\bibfnamefont {F.}~\bibnamefont {{Acernese}}}, \bibinfo
  {author} {\bibfnamefont {K.}~\bibnamefont {{Ackley}}}, \bibinfo {author}
  {\bibfnamefont {N.}~\bibnamefont {{Adhikari}}}, \bibinfo {author}
  {\bibfnamefont {R.~X.}\ \bibnamefont {{Adhikari}}}, \bibinfo {author}
  {\bibfnamefont {V.~K.}\ \bibnamefont {{Adkins}}}, \bibinfo {author}
  {\bibfnamefont {V.~B.}\ \bibnamefont {{Adya}}}, \bibinfo {author}
  {\bibfnamefont {C.}~\bibnamefont {{Affeldt}}}, \bibinfo {author}
  {\bibfnamefont {D.}~\bibnamefont {{Agarwal}}},\ and\ \bibinfo {author}
  {\bibnamefont {et~al.}},\ }\bibfield  {title} {\bibinfo {title} {{All-sky
  search for continuous gravitational waves from isolated neutron stars using
  Advanced LIGO and Advanced Virgo O3 data}},\ }\href
  {https://doi.org/10.1103/PhysRevD.106.102008} {\bibfield  {journal} {\bibinfo
   {journal} {Physical Review D}\ }\textbf {\bibinfo {volume} {106}},\ \bibinfo
  {eid} {102008} (\bibinfo {year} {2022}{\natexlab{d}})},\ \Eprint
  {https://arxiv.org/abs/2201.00697} {arXiv:2201.00697 [gr-qc]} \BibitemShut
  {NoStop}%
\bibitem [{\citenamefont {{Vargas}}\ and\ \citenamefont
  {{Melatos}}(2023)}]{VargasMelatos2023}%
  \BibitemOpen
  \bibfield  {author} {\bibinfo {author} {\bibfnamefont {A.~F.}\ \bibnamefont
  {{Vargas}}}\ and\ \bibinfo {author} {\bibfnamefont {A.}~\bibnamefont
  {{Melatos}}},\ }\bibfield  {title} {\bibinfo {title} {{Search for continuous
  gravitational waves from PSR J 0437 -4715 with a hidden Markov model in O3
  LIGO data}},\ }\href {https://doi.org/10.1103/PhysRevD.107.064062} {\bibfield
   {journal} {\bibinfo  {journal} {Physical Review D}\ }\textbf {\bibinfo
  {volume} {107}},\ \bibinfo {eid} {064062} (\bibinfo {year} {2023})},\ \Eprint
  {https://arxiv.org/abs/2208.03932} {arXiv:2208.03932 [gr-qc]} \BibitemShut
  {NoStop}%
\bibitem [{\citenamefont {{Jaranowski}}\ \emph {et~al.}(1998)\citenamefont
  {{Jaranowski}}, \citenamefont {{Kr{\'o}lak}},\ and\ \citenamefont
  {{Schutz}}}]{JaranowskiKrolak1998}%
  \BibitemOpen
  \bibfield  {author} {\bibinfo {author} {\bibfnamefont {P.}~\bibnamefont
  {{Jaranowski}}}, \bibinfo {author} {\bibfnamefont {A.}~\bibnamefont
  {{Kr{\'o}lak}}},\ and\ \bibinfo {author} {\bibfnamefont {B.~F.}\ \bibnamefont
  {{Schutz}}},\ }\bibfield  {title} {\bibinfo {title} {{Data analysis of
  gravitational-wave signals from spinning neutron stars: The signal and its
  detection}},\ }\href {https://doi.org/10.1103/PhysRevD.58.063001} {\bibfield
  {journal} {\bibinfo  {journal} {Physical Review D}\ }\textbf {\bibinfo
  {volume} {58}},\ \bibinfo {eid} {063001} (\bibinfo {year} {1998})},\ \Eprint
  {https://arxiv.org/abs/gr-qc/9804014} {arXiv:gr-qc/9804014 [gr-qc]}
  \BibitemShut {NoStop}%
\bibitem [{\citenamefont {{Cutler}}\ and\ \citenamefont
  {{Schutz}}(2005)}]{CutlerSchutz2005}%
  \BibitemOpen
  \bibfield  {author} {\bibinfo {author} {\bibfnamefont {C.}~\bibnamefont
  {{Cutler}}}\ and\ \bibinfo {author} {\bibfnamefont {B.~F.}\ \bibnamefont
  {{Schutz}}},\ }\bibfield  {title} {\bibinfo {title} {{Generalized
  F-statistic: Multiple detectors and multiple gravitational wave pulsars}},\
  }\href {https://doi.org/10.1103/PhysRevD.72.063006} {\bibfield  {journal}
  {\bibinfo  {journal} {Physical Review D}\ }\textbf {\bibinfo {volume} {72}},\
  \bibinfo {eid} {063006} (\bibinfo {year} {2005})},\ \Eprint
  {https://arxiv.org/abs/gr-qc/0504011} {arXiv:gr-qc/0504011 [gr-qc]}
  \BibitemShut {NoStop}%
\bibitem [{\citenamefont {{Suvorova}}\ \emph {et~al.}(2017)\citenamefont
  {{Suvorova}}, \citenamefont {{Clearwater}}, \citenamefont {{Melatos}},
  \citenamefont {{Sun}}, \citenamefont {{Moran}},\ and\ \citenamefont
  {{Evans}}}]{SuvorovaClearwater2017}%
  \BibitemOpen
  \bibfield  {author} {\bibinfo {author} {\bibfnamefont {S.}~\bibnamefont
  {{Suvorova}}}, \bibinfo {author} {\bibfnamefont {P.}~\bibnamefont
  {{Clearwater}}}, \bibinfo {author} {\bibfnamefont {A.}~\bibnamefont
  {{Melatos}}}, \bibinfo {author} {\bibfnamefont {L.}~\bibnamefont {{Sun}}},
  \bibinfo {author} {\bibfnamefont {W.}~\bibnamefont {{Moran}}},\ and\ \bibinfo
  {author} {\bibfnamefont {R.~J.}\ \bibnamefont {{Evans}}},\ }\bibfield
  {title} {\bibinfo {title} {{Hidden Markov model tracking of continuous
  gravitational waves from a binary neutron star with wandering spin. II.
  Binary orbital phase tracking}},\ }\href
  {https://doi.org/10.1103/PhysRevD.96.102006} {\bibfield  {journal} {\bibinfo
  {journal} {Physical Review D}\ }\textbf {\bibinfo {volume} {96}},\ \bibinfo
  {eid} {102006} (\bibinfo {year} {2017})},\ \Eprint
  {https://arxiv.org/abs/1710.07092} {arXiv:1710.07092 [astro-ph.IM]}
  \BibitemShut {NoStop}%
\bibitem [{\citenamefont {{Prix}}\ and\ \citenamefont
  {{Krishnan}}(2009)}]{PrixKrishnan2009}%
  \BibitemOpen
  \bibfield  {author} {\bibinfo {author} {\bibfnamefont {R.}~\bibnamefont
  {{Prix}}}\ and\ \bibinfo {author} {\bibfnamefont {B.}~\bibnamefont
  {{Krishnan}}},\ }\bibfield  {title} {\bibinfo {title} {{Targeted search for
  continuous gravitational waves: Bayesian versus maximum-likelihood
  statistics}},\ }\href {https://doi.org/10.1088/0264-9381/26/20/204013}
  {\bibfield  {journal} {\bibinfo  {journal} {Classical and Quantum Gravity}\
  }\textbf {\bibinfo {volume} {26}},\ \bibinfo {eid} {204013} (\bibinfo {year}
  {2009})},\ \Eprint {https://arxiv.org/abs/0907.2569} {arXiv:0907.2569
  [gr-qc]} \BibitemShut {NoStop}%
\bibitem [{\citenamefont {{LIGO Scientific Collaboration}}\ \emph
  {et~al.}(2018)\citenamefont {{LIGO Scientific Collaboration}}, \citenamefont
  {{Virgo Collaboration}},\ and\ \citenamefont {{KAGRA
  Collaboration}}}]{lalsuite}%
  \BibitemOpen
  \bibfield  {author} {\bibinfo {author} {\bibnamefont {{LIGO Scientific
  Collaboration}}}, \bibinfo {author} {\bibnamefont {{Virgo Collaboration}}},\
  and\ \bibinfo {author} {\bibnamefont {{KAGRA Collaboration}}},\ }\href
  {https://doi.org/10.7935/GT1W-FZ16} {\bibinfo {title} {{LVK} {A}lgorithm
  {L}ibrary - {LALS}uite}},\ \bibinfo {howpublished} {Free software (GPL)}
  (\bibinfo {year} {2018})\BibitemShut {NoStop}%
\bibitem [{\citenamefont {{Dunn}}\ \emph {et~al.}(2022)\citenamefont {{Dunn}},
  \citenamefont {{Clearwater}}, \citenamefont {{Melatos}},\ and\ \citenamefont
  {{Wette}}}]{DunnClearwater2022}%
  \BibitemOpen
  \bibfield  {author} {\bibinfo {author} {\bibfnamefont {L.}~\bibnamefont
  {{Dunn}}}, \bibinfo {author} {\bibfnamefont {P.}~\bibnamefont
  {{Clearwater}}}, \bibinfo {author} {\bibfnamefont {A.}~\bibnamefont
  {{Melatos}}},\ and\ \bibinfo {author} {\bibfnamefont {K.}~\bibnamefont
  {{Wette}}},\ }\bibfield  {title} {\bibinfo {title} {{Graphics processing unit
  implementation of the \{F\} -statistic for continuous gravitational wave
  searches}},\ }\href {https://doi.org/10.1088/1361-6382/ac4616} {\bibfield
  {journal} {\bibinfo  {journal} {Classical and Quantum Gravity}\ }\textbf
  {\bibinfo {volume} {39}},\ \bibinfo {eid} {045003} (\bibinfo {year}
  {2022})},\ \Eprint {https://arxiv.org/abs/2201.00451} {arXiv:2201.00451
  [gr-qc]} \BibitemShut {NoStop}%
\bibitem [{\citenamefont {{Krishnan}}\ \emph {et~al.}(2004)\citenamefont
  {{Krishnan}}, \citenamefont {{Sintes}}, \citenamefont {{Papa}}, \citenamefont
  {{Schutz}}, \citenamefont {{Frasca}},\ and\ \citenamefont
  {{Palomba}}}]{KrishnanSintes2004}%
  \BibitemOpen
  \bibfield  {author} {\bibinfo {author} {\bibfnamefont {B.}~\bibnamefont
  {{Krishnan}}}, \bibinfo {author} {\bibfnamefont {A.~M.}\ \bibnamefont
  {{Sintes}}}, \bibinfo {author} {\bibfnamefont {M.~A.}\ \bibnamefont
  {{Papa}}}, \bibinfo {author} {\bibfnamefont {B.~F.}\ \bibnamefont
  {{Schutz}}}, \bibinfo {author} {\bibfnamefont {S.}~\bibnamefont {{Frasca}}},\
  and\ \bibinfo {author} {\bibfnamefont {C.}~\bibnamefont {{Palomba}}},\
  }\bibfield  {title} {\bibinfo {title} {{Hough transform search for continuous
  gravitational waves}},\ }\href {https://doi.org/10.1103/PhysRevD.70.082001}
  {\bibfield  {journal} {\bibinfo  {journal} {Physical Review D}\ }\textbf
  {\bibinfo {volume} {70}},\ \bibinfo {eid} {082001} (\bibinfo {year}
  {2004})},\ \Eprint {https://arxiv.org/abs/gr-qc/0407001} {arXiv:gr-qc/0407001
  [gr-qc]} \BibitemShut {NoStop}%
\bibitem [{\citenamefont {{Bayley}}\ \emph {et~al.}(2019)\citenamefont
  {{Bayley}}, \citenamefont {{Messenger}},\ and\ \citenamefont
  {{Woan}}}]{BayleyMessenger2019}%
  \BibitemOpen
  \bibfield  {author} {\bibinfo {author} {\bibfnamefont {J.}~\bibnamefont
  {{Bayley}}}, \bibinfo {author} {\bibfnamefont {C.}~\bibnamefont
  {{Messenger}}},\ and\ \bibinfo {author} {\bibfnamefont {G.}~\bibnamefont
  {{Woan}}},\ }\bibfield  {title} {\bibinfo {title} {{Generalized application
  of the Viterbi algorithm to searches for continuous gravitational-wave
  signals}},\ }\href {https://doi.org/10.1103/PhysRevD.100.023006} {\bibfield
  {journal} {\bibinfo  {journal} {Physical Review D}\ }\textbf {\bibinfo
  {volume} {100}},\ \bibinfo {eid} {023006} (\bibinfo {year} {2019})},\ \Eprint
  {https://arxiv.org/abs/1903.12614} {arXiv:1903.12614 [astro-ph.IM]}
  \BibitemShut {NoStop}%
\bibitem [{\citenamefont {{Astone}}\ \emph {et~al.}(2014)\citenamefont
  {{Astone}}, \citenamefont {{Colla}}, \citenamefont {{D'Antonio}},
  \citenamefont {{Frasca}},\ and\ \citenamefont {{Palomba}}}]{AstoneColla2014}%
  \BibitemOpen
  \bibfield  {author} {\bibinfo {author} {\bibfnamefont {P.}~\bibnamefont
  {{Astone}}}, \bibinfo {author} {\bibfnamefont {A.}~\bibnamefont {{Colla}}},
  \bibinfo {author} {\bibfnamefont {S.}~\bibnamefont {{D'Antonio}}}, \bibinfo
  {author} {\bibfnamefont {S.}~\bibnamefont {{Frasca}}},\ and\ \bibinfo
  {author} {\bibfnamefont {C.}~\bibnamefont {{Palomba}}},\ }\bibfield  {title}
  {\bibinfo {title} {{Method for all-sky searches of continuous gravitational
  wave signals using the frequency-Hough transform}},\ }\href
  {https://doi.org/10.1103/PhysRevD.90.042002} {\bibfield  {journal} {\bibinfo
  {journal} {Physical Review D}\ }\textbf {\bibinfo {volume} {90}},\ \bibinfo
  {eid} {042002} (\bibinfo {year} {2014})},\ \Eprint
  {https://arxiv.org/abs/1407.8333} {arXiv:1407.8333 [astro-ph.IM]}
  \BibitemShut {NoStop}%
\bibitem [{\citenamefont {{Dergachev}}(2010)}]{Dergachev2010}%
  \BibitemOpen
  \bibfield  {author} {\bibinfo {author} {\bibfnamefont {V.}~\bibnamefont
  {{Dergachev}}},\ }\bibfield  {title} {\bibinfo {title} {{On blind searches
  for noise dominated signals: a loosely coherent approach}},\ }\href
  {https://doi.org/10.1088/0264-9381/27/20/205017} {\bibfield  {journal}
  {\bibinfo  {journal} {Classical and Quantum Gravity}\ }\textbf {\bibinfo
  {volume} {27}},\ \bibinfo {eid} {205017} (\bibinfo {year} {2010})},\ \Eprint
  {https://arxiv.org/abs/1003.2178} {arXiv:1003.2178 [gr-qc]} \BibitemShut
  {NoStop}%
\bibitem [{\citenamefont {{Dergachev}}(2012)}]{Dergachev2012}%
  \BibitemOpen
  \bibfield  {author} {\bibinfo {author} {\bibfnamefont {V.}~\bibnamefont
  {{Dergachev}}},\ }\bibfield  {title} {\bibinfo {title} {{Loosely coherent
  searches for sets of well-modeled signals}},\ }\href
  {https://doi.org/10.1103/PhysRevD.85.062003} {\bibfield  {journal} {\bibinfo
  {journal} {Physical Review D}\ }\textbf {\bibinfo {volume} {85}},\ \bibinfo
  {eid} {062003} (\bibinfo {year} {2012})},\ \Eprint
  {https://arxiv.org/abs/1110.3297} {arXiv:1110.3297 [gr-qc]} \BibitemShut
  {NoStop}%
\bibitem [{\citenamefont {{Dergachev}}\ and\ \citenamefont
  {{Papa}}(2023)}]{DergachevPapa2023}%
  \BibitemOpen
  \bibfield  {author} {\bibinfo {author} {\bibfnamefont {V.}~\bibnamefont
  {{Dergachev}}}\ and\ \bibinfo {author} {\bibfnamefont {M.~A.}\ \bibnamefont
  {{Papa}}},\ }\bibfield  {title} {\bibinfo {title} {{Frequency-Resolved Atlas
  of the Sky in Continuous Gravitational Waves}},\ }\href
  {https://doi.org/10.1103/PhysRevX.13.021020} {\bibfield  {journal} {\bibinfo
  {journal} {Physical Review X}\ }\textbf {\bibinfo {volume} {13}},\ \bibinfo
  {eid} {021020} (\bibinfo {year} {2023})},\ \Eprint
  {https://arxiv.org/abs/2202.10598} {arXiv:2202.10598 [gr-qc]} \BibitemShut
  {NoStop}%
\bibitem [{\citenamefont {{Wette}}(2012)}]{Wette2012}%
  \BibitemOpen
  \bibfield  {author} {\bibinfo {author} {\bibfnamefont {K.}~\bibnamefont
  {{Wette}}},\ }\bibfield  {title} {\bibinfo {title} {{Estimating the
  sensitivity of wide-parameter-space searches for gravitational-wave
  pulsars}},\ }\href {https://doi.org/10.1103/PhysRevD.85.042003} {\bibfield
  {journal} {\bibinfo  {journal} {Physical Review D}\ }\textbf {\bibinfo
  {volume} {85}},\ \bibinfo {eid} {042003} (\bibinfo {year} {2012})},\ \Eprint
  {https://arxiv.org/abs/1111.5650} {arXiv:1111.5650 [gr-qc]} \BibitemShut
  {NoStop}%
\bibitem [{\citenamefont {{Abbott}}\ \emph {et~al.}(2008)\citenamefont
  {{Abbott}}, \citenamefont {{Abbott}}, \citenamefont {{Adhikari}},
  \citenamefont {{Ajith}}, \citenamefont {{Allen}}, \citenamefont {{Allen}},
  \citenamefont {{Amin}}, \citenamefont {{Anderson}}, \citenamefont
  {{Anderson}}, \citenamefont {{Arain}},\ and\ \citenamefont
  {et~al.}}]{AbbottAbbott2008}%
  \BibitemOpen
  \bibfield  {author} {\bibinfo {author} {\bibfnamefont {B.}~\bibnamefont
  {{Abbott}}}, \bibinfo {author} {\bibfnamefont {R.}~\bibnamefont {{Abbott}}},
  \bibinfo {author} {\bibfnamefont {R.}~\bibnamefont {{Adhikari}}}, \bibinfo
  {author} {\bibfnamefont {P.}~\bibnamefont {{Ajith}}}, \bibinfo {author}
  {\bibfnamefont {B.}~\bibnamefont {{Allen}}}, \bibinfo {author} {\bibfnamefont
  {G.}~\bibnamefont {{Allen}}}, \bibinfo {author} {\bibfnamefont
  {R.}~\bibnamefont {{Amin}}}, \bibinfo {author} {\bibfnamefont {S.~B.}\
  \bibnamefont {{Anderson}}}, \bibinfo {author} {\bibfnamefont {W.~G.}\
  \bibnamefont {{Anderson}}}, \bibinfo {author} {\bibfnamefont {M.~A.}\
  \bibnamefont {{Arain}}},\ and\ \bibinfo {author} {\bibnamefont {et~al.}},\
  }\bibfield  {title} {\bibinfo {title} {{Beating the Spin-Down Limit on
  Gravitational Wave Emission from the Crab Pulsar}},\ }\href
  {https://doi.org/10.1086/591526} {\bibfield  {journal} {\bibinfo  {journal}
  {The Astrophysical Journal}\ }\textbf {\bibinfo {volume} {683}},\ \bibinfo
  {pages} {L45} (\bibinfo {year} {2008})},\ \Eprint
  {https://arxiv.org/abs/0805.4758} {arXiv:0805.4758 [astro-ph]} \BibitemShut
  {NoStop}%
\bibitem [{\citenamefont {{Ashok}}\ \emph {et~al.}(2021)\citenamefont
  {{Ashok}}, \citenamefont {{Beheshtipour}}, \citenamefont {{Papa}},
  \citenamefont {{Freire}}, \citenamefont {{Steltner}}, \citenamefont
  {{Machenschalk}}, \citenamefont {{Behnke}}, \citenamefont {{Allen}},\ and\
  \citenamefont {{Prix}}}]{AshokBeheshtipour2021}%
  \BibitemOpen
  \bibfield  {author} {\bibinfo {author} {\bibfnamefont {A.}~\bibnamefont
  {{Ashok}}}, \bibinfo {author} {\bibfnamefont {B.}~\bibnamefont
  {{Beheshtipour}}}, \bibinfo {author} {\bibfnamefont {M.~A.}\ \bibnamefont
  {{Papa}}}, \bibinfo {author} {\bibfnamefont {P.~C.~C.}\ \bibnamefont
  {{Freire}}}, \bibinfo {author} {\bibfnamefont {B.}~\bibnamefont
  {{Steltner}}}, \bibinfo {author} {\bibfnamefont {B.}~\bibnamefont
  {{Machenschalk}}}, \bibinfo {author} {\bibfnamefont {O.}~\bibnamefont
  {{Behnke}}}, \bibinfo {author} {\bibfnamefont {B.}~\bibnamefont {{Allen}}},\
  and\ \bibinfo {author} {\bibfnamefont {R.}~\bibnamefont {{Prix}}},\
  }\bibfield  {title} {\bibinfo {title} {{New Searches for Continuous
  Gravitational Waves from Seven Fast Pulsars}},\ }\href
  {https://doi.org/10.3847/1538-4357/ac2582} {\bibfield  {journal} {\bibinfo
  {journal} {The Astrophysical Journal}\ }\textbf {\bibinfo {volume} {923}},\
  \bibinfo {eid} {85} (\bibinfo {year} {2021})},\ \Eprint
  {https://arxiv.org/abs/2107.09727} {arXiv:2107.09727 [astro-ph.HE]}
  \BibitemShut {NoStop}%
\bibitem [{\citenamefont {{Abbott}}\ \emph
  {et~al.}(2022{\natexlab{e}})\citenamefont {{Abbott}}, \citenamefont
  {{Abbott}}, \citenamefont {{Acernese}}, \citenamefont {{Ackley}},
  \citenamefont {{Adams}}, \citenamefont {{Adhikari}}, \citenamefont
  {{Adhikari}}, \citenamefont {{Adya}}, \citenamefont {{Affeldt}},
  \citenamefont {{Agarwal}},\ and\ \citenamefont {et~al.}}]{AbbottAbbott2022a}%
  \BibitemOpen
  \bibfield  {author} {\bibinfo {author} {\bibfnamefont {R.}~\bibnamefont
  {{Abbott}}}, \bibinfo {author} {\bibfnamefont {T.~D.}\ \bibnamefont
  {{Abbott}}}, \bibinfo {author} {\bibfnamefont {F.}~\bibnamefont
  {{Acernese}}}, \bibinfo {author} {\bibfnamefont {K.}~\bibnamefont
  {{Ackley}}}, \bibinfo {author} {\bibfnamefont {C.}~\bibnamefont {{Adams}}},
  \bibinfo {author} {\bibfnamefont {N.}~\bibnamefont {{Adhikari}}}, \bibinfo
  {author} {\bibfnamefont {R.~X.}\ \bibnamefont {{Adhikari}}}, \bibinfo
  {author} {\bibfnamefont {V.~B.}\ \bibnamefont {{Adya}}}, \bibinfo {author}
  {\bibfnamefont {C.}~\bibnamefont {{Affeldt}}}, \bibinfo {author}
  {\bibfnamefont {D.}~\bibnamefont {{Agarwal}}},\ and\ \bibinfo {author}
  {\bibnamefont {et~al.}},\ }\bibfield  {title} {\bibinfo {title} {{Narrowband
  Searches for Continuous and Long-duration Transient Gravitational Waves from
  Known Pulsars in the LIGO-Virgo Third Observing Run}},\ }\href
  {https://doi.org/10.3847/1538-4357/ac6ad0} {\bibfield  {journal} {\bibinfo
  {journal} {The Astrophysical Journal}\ }\textbf {\bibinfo {volume} {932}},\
  \bibinfo {eid} {133} (\bibinfo {year} {2022}{\natexlab{e}})},\ \Eprint
  {https://arxiv.org/abs/2112.10990} {arXiv:2112.10990 [gr-qc]} \BibitemShut
  {NoStop}%
\bibitem [{\citenamefont {Rabiner}(1989)}]{Rabiner1989}%
  \BibitemOpen
  \bibfield  {author} {\bibinfo {author} {\bibfnamefont {L.}~\bibnamefont
  {Rabiner}},\ }\bibfield  {title} {\bibinfo {title} {{A tutorial on hidden
  Markov models and selected applications in speech recognition}},\ }\href
  {https://doi.org/10.1109/5.18626} {\bibfield  {journal} {\bibinfo  {journal}
  {Proceedings of the IEEE}\ }\textbf {\bibinfo {volume} {77}},\ \bibinfo
  {pages} {257} (\bibinfo {year} {1989})}\BibitemShut {NoStop}%
\bibitem [{\citenamefont {{Wagoner}}(1984)}]{Wagoner1984}%
  \BibitemOpen
  \bibfield  {author} {\bibinfo {author} {\bibfnamefont {R.~V.}\ \bibnamefont
  {{Wagoner}}},\ }\bibfield  {title} {\bibinfo {title} {{Gravitational
  radiation from accreting neutron stars}},\ }\href
  {https://doi.org/10.1086/161798} {\bibfield  {journal} {\bibinfo  {journal}
  {The Astrophysical Journal}\ }\textbf {\bibinfo {volume} {278}},\ \bibinfo
  {pages} {345} (\bibinfo {year} {1984})}\BibitemShut {NoStop}%
\bibitem [{\citenamefont {{Vigelius}}\ and\ \citenamefont
  {{Melatos}}(2009)}]{VigeliusMelatos2009}%
  \BibitemOpen
  \bibfield  {author} {\bibinfo {author} {\bibfnamefont {M.}~\bibnamefont
  {{Vigelius}}}\ and\ \bibinfo {author} {\bibfnamefont {A.}~\bibnamefont
  {{Melatos}}},\ }\bibfield  {title} {\bibinfo {title} {{Resistive relaxation
  of a magnetically confined mountain on an accreting neutron star}},\ }\href
  {https://doi.org/10.1111/j.1365-2966.2009.14698.x} {\bibfield  {journal}
  {\bibinfo  {journal} {Monthly Notices of the Royal Astronomical Society}\
  }\textbf {\bibinfo {volume} {395}},\ \bibinfo {pages} {1985} (\bibinfo {year}
  {2009})},\ \Eprint {https://arxiv.org/abs/0902.4484} {arXiv:0902.4484
  [astro-ph.HE]} \BibitemShut {NoStop}%
\bibitem [{\citenamefont {{Harris}}(1996)}]{Harris1996}%
  \BibitemOpen
  \bibfield  {author} {\bibinfo {author} {\bibfnamefont {W.~E.}\ \bibnamefont
  {{Harris}}},\ }\bibfield  {title} {\bibinfo {title} {{A Catalog of Parameters
  for Globular Clusters in the Milky Way}},\ }\href
  {https://doi.org/10.1086/118116} {\bibfield  {journal} {\bibinfo  {journal}
  {The Astronomical Journal}\ }\textbf {\bibinfo {volume} {112}},\ \bibinfo
  {pages} {1487} (\bibinfo {year} {1996})}\BibitemShut {NoStop}%
\bibitem [{\citenamefont {{Verbunt}}\ and\ \citenamefont
  {{Freire}}(2014)}]{VerbuntFreire2014}%
  \BibitemOpen
  \bibfield  {author} {\bibinfo {author} {\bibfnamefont {F.}~\bibnamefont
  {{Verbunt}}}\ and\ \bibinfo {author} {\bibfnamefont {P.~C.~C.}\ \bibnamefont
  {{Freire}}},\ }\bibfield  {title} {\bibinfo {title} {{On the disruption of
  pulsar and X-ray binar ies in globular clusters}},\ }\href
  {https://doi.org/10.1051/0004-6361/201321177} {\bibfield  {journal} {\bibinfo
   {journal} {Astronomy and Astrophysics}\ }\textbf {\bibinfo {volume} {561}},\
  \bibinfo {eid} {A11} (\bibinfo {year} {2014})},\ \Eprint
  {https://arxiv.org/abs/1310.4669} {arXiv:1310.4669 [astro-ph.SR]}
  \BibitemShut {NoStop}%
\bibitem [{\citenamefont {{Phinney}}(1992)}]{Phinney1992}%
  \BibitemOpen
  \bibfield  {author} {\bibinfo {author} {\bibfnamefont {E.~S.}\ \bibnamefont
  {{Phinney}}},\ }\bibfield  {title} {\bibinfo {title} {{Pulsars as Probes of
  Newtonian Dynamical Systems}},\ }\href
  {https://doi.org/10.1098/rsta.1992.0084} {\bibfield  {journal} {\bibinfo
  {journal} {Philosophical Transactions of the Royal Society of London Series
  A}\ }\textbf {\bibinfo {volume} {341}},\ \bibinfo {pages} {39} (\bibinfo
  {year} {1992})}\BibitemShut {NoStop}%
\bibitem [{\citenamefont {{Wette}}\ \emph {et~al.}(2008)\citenamefont
  {{Wette}}, \citenamefont {{Owen}}, \citenamefont {{Allen}}, \citenamefont
  {{Ashley}}, \citenamefont {{Betzwieser}}, \citenamefont {{Christensen}},
  \citenamefont {{Creighton}}, \citenamefont {{Dergachev}}, \citenamefont
  {{Gholami}}, \citenamefont {{Goetz}}, \citenamefont {{Gustafson}},
  \citenamefont {{Hammer}}, \citenamefont {{Jones}}, \citenamefont
  {{Krishnan}}, \citenamefont {{Landry}}, \citenamefont {{Machenschalk}},
  \citenamefont {{McClelland}}, \citenamefont {{Mendell}}, \citenamefont
  {{Messenger}}, \citenamefont {{Papa}}, \citenamefont {{Patel}}, \citenamefont
  {{Pitkin}}, \citenamefont {{Pletsch}}, \citenamefont {{Prix}}, \citenamefont
  {{Riles}}, \citenamefont {{Sancho de la Jordana}}, \citenamefont {{Scott}},
  \citenamefont {{Sintes}}, \citenamefont {{Trias}}, \citenamefont {{Whelan}},\
  and\ \citenamefont {{Woan}}}]{WetteOwen2008}%
  \BibitemOpen
  \bibfield  {author} {\bibinfo {author} {\bibfnamefont {K.}~\bibnamefont
  {{Wette}}}, \bibinfo {author} {\bibfnamefont {B.~J.}\ \bibnamefont {{Owen}}},
  \bibinfo {author} {\bibfnamefont {B.}~\bibnamefont {{Allen}}}, \bibinfo
  {author} {\bibfnamefont {M.}~\bibnamefont {{Ashley}}}, \bibinfo {author}
  {\bibfnamefont {J.}~\bibnamefont {{Betzwieser}}}, \bibinfo {author}
  {\bibfnamefont {N.}~\bibnamefont {{Christensen}}}, \bibinfo {author}
  {\bibfnamefont {T.~D.}\ \bibnamefont {{Creighton}}}, \bibinfo {author}
  {\bibfnamefont {V.}~\bibnamefont {{Dergachev}}}, \bibinfo {author}
  {\bibfnamefont {I.}~\bibnamefont {{Gholami}}}, \bibinfo {author}
  {\bibfnamefont {E.}~\bibnamefont {{Goetz}}}, \bibinfo {author} {\bibfnamefont
  {R.}~\bibnamefont {{Gustafson}}}, \bibinfo {author} {\bibfnamefont
  {D.}~\bibnamefont {{Hammer}}}, \bibinfo {author} {\bibfnamefont {D.~I.}\
  \bibnamefont {{Jones}}}, \bibinfo {author} {\bibfnamefont {B.}~\bibnamefont
  {{Krishnan}}}, \bibinfo {author} {\bibfnamefont {M.}~\bibnamefont
  {{Landry}}}, \bibinfo {author} {\bibfnamefont {B.}~\bibnamefont
  {{Machenschalk}}}, \bibinfo {author} {\bibfnamefont {D.~E.}\ \bibnamefont
  {{McClelland}}}, \bibinfo {author} {\bibfnamefont {G.}~\bibnamefont
  {{Mendell}}}, \bibinfo {author} {\bibfnamefont {C.~J.}\ \bibnamefont
  {{Messenger}}}, \bibinfo {author} {\bibfnamefont {M.~A.}\ \bibnamefont
  {{Papa}}}, \bibinfo {author} {\bibfnamefont {P.}~\bibnamefont {{Patel}}},
  \bibinfo {author} {\bibfnamefont {M.}~\bibnamefont {{Pitkin}}}, \bibinfo
  {author} {\bibfnamefont {H.~J.}\ \bibnamefont {{Pletsch}}}, \bibinfo {author}
  {\bibfnamefont {R.}~\bibnamefont {{Prix}}}, \bibinfo {author} {\bibfnamefont
  {K.}~\bibnamefont {{Riles}}}, \bibinfo {author} {\bibfnamefont
  {L.}~\bibnamefont {{Sancho de la Jordana}}}, \bibinfo {author} {\bibfnamefont
  {S.~M.}\ \bibnamefont {{Scott}}}, \bibinfo {author} {\bibfnamefont {A.~M.}\
  \bibnamefont {{Sintes}}}, \bibinfo {author} {\bibfnamefont {M.}~\bibnamefont
  {{Trias}}}, \bibinfo {author} {\bibfnamefont {J.~T.}\ \bibnamefont
  {{Whelan}}},\ and\ \bibinfo {author} {\bibfnamefont {G.}~\bibnamefont
  {{Woan}}},\ }\bibfield  {title} {\bibinfo {title} {{Searching for
  gravitational waves from Cassiopeia A with LIGO}},\ }\href
  {https://doi.org/10.1088/0264-9381/25/23/235011} {\bibfield  {journal}
  {\bibinfo  {journal} {Classical and Quantum Gravity}\ }\textbf {\bibinfo
  {volume} {25}},\ \bibinfo {eid} {235011} (\bibinfo {year} {2008})},\ \Eprint
  {https://arxiv.org/abs/0802.3332} {arXiv:0802.3332 [gr-qc]} \BibitemShut
  {NoStop}%
\bibitem [{\citenamefont {{Abbott}}\ \emph
  {et~al.}(2022{\natexlab{f}})\citenamefont {{Abbott}}, \citenamefont
  {{Abbott}}, \citenamefont {{Acernese}}, \citenamefont {{Ackley}},
  \citenamefont {{Adams}}, \citenamefont {{Adhikari}}, \citenamefont
  {{Adhikari}}, \citenamefont {{Adya}}, \citenamefont {{Affeldt}},
  \citenamefont {{Agarwal}},\ and\ \citenamefont {et~al.}}]{AbbottAbbott2022b}%
  \BibitemOpen
  \bibfield  {author} {\bibinfo {author} {\bibfnamefont {R.}~\bibnamefont
  {{Abbott}}}, \bibinfo {author} {\bibfnamefont {T.~D.}\ \bibnamefont
  {{Abbott}}}, \bibinfo {author} {\bibfnamefont {F.}~\bibnamefont
  {{Acernese}}}, \bibinfo {author} {\bibfnamefont {K.}~\bibnamefont
  {{Ackley}}}, \bibinfo {author} {\bibfnamefont {C.}~\bibnamefont {{Adams}}},
  \bibinfo {author} {\bibfnamefont {N.}~\bibnamefont {{Adhikari}}}, \bibinfo
  {author} {\bibfnamefont {R.~X.}\ \bibnamefont {{Adhikari}}}, \bibinfo
  {author} {\bibfnamefont {V.~B.}\ \bibnamefont {{Adya}}}, \bibinfo {author}
  {\bibfnamefont {C.}~\bibnamefont {{Affeldt}}}, \bibinfo {author}
  {\bibfnamefont {D.}~\bibnamefont {{Agarwal}}},\ and\ \bibinfo {author}
  {\bibnamefont {et~al.}},\ }\bibfield  {title} {\bibinfo {title} {{Search of
  the early O3 LIGO data for continuous gravitational waves from the Cassiopeia
  A and Vela Jr. supernova remnants}},\ }\href
  {https://doi.org/10.1103/PhysRevD.105.082005} {\bibfield  {journal} {\bibinfo
   {journal} {Physical Review D}\ }\textbf {\bibinfo {volume} {105}},\ \bibinfo
  {eid} {082005} (\bibinfo {year} {2022}{\natexlab{f}})},\ \Eprint
  {https://arxiv.org/abs/2111.15116} {arXiv:2111.15116 [gr-qc]} \BibitemShut
  {NoStop}%
\bibitem [{\citenamefont {{Abbott}}\ \emph {et~al.}(2023)\citenamefont
  {{Abbott}}, \citenamefont {{Abe}}, \citenamefont {{Acernese}}, \citenamefont
  {{Ackley}}, \citenamefont {{Adhicary}}, \citenamefont {{Adhikari}},
  \citenamefont {{Adhikari}}, \citenamefont {{Adkins}}, \citenamefont {{Adya}},
  \citenamefont {{Affeldt}},\ and\ \citenamefont {et~al.}}]{AbbottAbe2023}%
  \BibitemOpen
  \bibfield  {author} {\bibinfo {author} {\bibfnamefont {R.}~\bibnamefont
  {{Abbott}}}, \bibinfo {author} {\bibfnamefont {H.}~\bibnamefont {{Abe}}},
  \bibinfo {author} {\bibfnamefont {F.}~\bibnamefont {{Acernese}}}, \bibinfo
  {author} {\bibfnamefont {K.}~\bibnamefont {{Ackley}}}, \bibinfo {author}
  {\bibfnamefont {S.}~\bibnamefont {{Adhicary}}}, \bibinfo {author}
  {\bibfnamefont {N.}~\bibnamefont {{Adhikari}}}, \bibinfo {author}
  {\bibfnamefont {R.~X.}\ \bibnamefont {{Adhikari}}}, \bibinfo {author}
  {\bibfnamefont {V.~K.}\ \bibnamefont {{Adkins}}}, \bibinfo {author}
  {\bibfnamefont {V.~B.}\ \bibnamefont {{Adya}}}, \bibinfo {author}
  {\bibfnamefont {C.}~\bibnamefont {{Affeldt}}},\ and\ \bibinfo {author}
  {\bibnamefont {et~al.}},\ }\bibfield  {title} {\bibinfo {title} {{Open Data
  from the Third Observing Run of LIGO, Virgo, KAGRA, and GEO}},\ }\href
  {https://doi.org/10.3847/1538-4365/acdc9f} {\bibfield  {journal} {\bibinfo
  {journal} {The Astrophysical Journal Supplement Series}\ }\textbf {\bibinfo
  {volume} {267}},\ \bibinfo {eid} {29} (\bibinfo {year} {2023})},\ \Eprint
  {https://arxiv.org/abs/2302.03676} {arXiv:2302.03676 [gr-qc]} \BibitemShut
  {NoStop}%
\bibitem [{\citenamefont {{Buikema}}\ \emph {et~al.}(2020)\citenamefont
  {{Buikema}}, \citenamefont {{Cahillane}}, \citenamefont {{Mansell}},
  \citenamefont {{Blair}}, \citenamefont {{Abbott}}, \citenamefont {{Adams}},
  \citenamefont {{Adhikari}}, \citenamefont {{Ananyeva}}, \citenamefont
  {{Appert}}, \citenamefont {{Arai}}, \citenamefont {{Areeda}}, \citenamefont
  {{Asali}}, \citenamefont {{Aston}}, \citenamefont {{Austin}}, \citenamefont
  {{Baer}}, \citenamefont {{Ball}}, \citenamefont {{Ballmer}}, \citenamefont
  {{Banagiri}}, \citenamefont {{Barker}}, \citenamefont {{Barsotti}},
  \citenamefont {{Bartlett}}, \citenamefont {{Berger}}, \citenamefont
  {{Betzwieser}}, \citenamefont {{Bhattacharjee}}, \citenamefont
  {{Billingsley}}, \citenamefont {{Biscans}}, \citenamefont {{Blair}},
  \citenamefont {{Bode}}, \citenamefont {{Booker}}, \citenamefont {{Bork}},
  \citenamefont {{Bramley}}, \citenamefont {{Brooks}}, \citenamefont {{Brown}},
  \citenamefont {{Cannon}}, \citenamefont {{Chen}}, \citenamefont {{Ciobanu}},
  \citenamefont {{Clara}}, \citenamefont {{Cooper}}, \citenamefont {{Corley}},
  \citenamefont {{Countryman}}, \citenamefont {{Covas}}, \citenamefont
  {{Coyne}}, \citenamefont {{Datrier}}, \citenamefont {{Davis}}, \citenamefont
  {{Di Fronzo}}, \citenamefont {{Dooley}}, \citenamefont {{Driggers}},
  \citenamefont {{Dupej}}, \citenamefont {{Dwyer}}, \citenamefont {{Effler}},
  \citenamefont {{Etzel}}, \citenamefont {{Evans}}, \citenamefont {{Evans}},
  \citenamefont {{Feicht}}, \citenamefont {{Fernandez-Galiana}}, \citenamefont
  {{Fritschel}}, \citenamefont {{Frolov}}, \citenamefont {{Fulda}},
  \citenamefont {{Fyffe}}, \citenamefont {{Giaime}}, \citenamefont
  {{Giardina}}, \citenamefont {{Godwin}}, \citenamefont {{Goetz}},
  \citenamefont {{Gras}}, \citenamefont {{Gray}}, \citenamefont {{Gray}},
  \citenamefont {{Green}}, \citenamefont {{Gustafson}}, \citenamefont
  {{Gustafson}}, \citenamefont {{Hanks}}, \citenamefont {{Hanson}},
  \citenamefont {{Hardwick}}, \citenamefont {{Hasskew}}, \citenamefont
  {{Heintze}}, \citenamefont {{Helmling-Cornell}}, \citenamefont {{Holland}},
  \citenamefont {{Jones}}, \citenamefont {{Kandhasamy}}, \citenamefont
  {{Karki}}, \citenamefont {{Kasprzack}}, \citenamefont {{Kawabe}},
  \citenamefont {{Kijbunchoo}}, \citenamefont {{King}}, \citenamefont
  {{Kissel}}, \citenamefont {{Kumar}}, \citenamefont {{Landry}}, \citenamefont
  {{Lane}}, \citenamefont {{Lantz}}, \citenamefont {{Laxen}}, \citenamefont
  {{Lecoeuche}}, \citenamefont {{Leviton}}, \citenamefont {{Liu}},
  \citenamefont {{Lormand}}, \citenamefont {{Lundgren}}, \citenamefont
  {{Macas}}, \citenamefont {{MacInnis}}, \citenamefont {{Macleod}},
  \citenamefont {{M{\'a}rka}}, \citenamefont {{M{\'a}rka}}, \citenamefont
  {{Martynov}}, \citenamefont {{Mason}}, \citenamefont {{Massinger}},
  \citenamefont {{Matichard}}, \citenamefont {{Mavalvala}}, \citenamefont
  {{McCarthy}}, \citenamefont {{McClelland}}, \citenamefont {{McCormick}},
  \citenamefont {{McCuller}}, \citenamefont {{McIver}}, \citenamefont
  {{McRae}}, \citenamefont {{Mendell}}, \citenamefont {{Merfeld}},
  \citenamefont {{Merilh}}, \citenamefont {{Meylahn}}, \citenamefont
  {{Mistry}}, \citenamefont {{Mittleman}}, \citenamefont {{Moreno}},
  \citenamefont {{Mow-Lowry}}, \citenamefont {{Mozzon}}, \citenamefont
  {{Mullavey}}, \citenamefont {{Nelson}}, \citenamefont {{Nguyen}},
  \citenamefont {{Nuttall}}, \citenamefont {{Oberling}}, \citenamefont
  {{Oram}}, \citenamefont {{O'Reilly}}, \citenamefont {{Osthelder}},
  \citenamefont {{Ottaway}}, \citenamefont {{Overmier}}, \citenamefont
  {{Palamos}}, \citenamefont {{Parker}}, \citenamefont {{Payne}}, \citenamefont
  {{Pele}}, \citenamefont {{Penhorwood}}, \citenamefont {{Perez}},
  \citenamefont {{Pirello}}, \citenamefont {{Radkins}}, \citenamefont
  {{Ramirez}}, \citenamefont {{Richardson}}, \citenamefont {{Riles}},
  \citenamefont {{Robertson}}, \citenamefont {{Rollins}}, \citenamefont
  {{Romel}}, \citenamefont {{Romie}}, \citenamefont {{Ross}}, \citenamefont
  {{Ryan}}, \citenamefont {{Sadecki}}, \citenamefont {{Sanchez}}, \citenamefont
  {{Sanchez}}, \citenamefont {{Saravanan}}, \citenamefont {{Savage}},
  \citenamefont {{Schaetzl}}, \citenamefont {{Schnabel}}, \citenamefont
  {{Schofield}}, \citenamefont {{Schwartz}}, \citenamefont {{Sellers}},
  \citenamefont {{Shaffer}}, \citenamefont {{Sigg}}, \citenamefont
  {{Slagmolen}}, \citenamefont {{Smith}}, \citenamefont {{Soni}}, \citenamefont
  {{Sorazu}}, \citenamefont {{Spencer}}, \citenamefont {{Strain}},
  \citenamefont {{Sun}}, \citenamefont {{Szczepa{\'n}czyk}}, \citenamefont
  {{Thomas}}, \citenamefont {{Thomas}}, \citenamefont {{Thorne}}, \citenamefont
  {{Toland}}, \citenamefont {{Torrie}}, \citenamefont {{Traylor}},
  \citenamefont {{Tse}}, \citenamefont {{Urban}}, \citenamefont {{Vajente}},
  \citenamefont {{Valdes}}, \citenamefont {{Vander-Hyde}}, \citenamefont
  {{Veitch}}, \citenamefont {{Venkateswara}}, \citenamefont {{Venugopalan}},
  \citenamefont {{Viets}}, \citenamefont {{Vo}}, \citenamefont {{Vorvick}},
  \citenamefont {{Wade}}, \citenamefont {{Ward}}, \citenamefont {{Warner}},
  \citenamefont {{Weaver}}, \citenamefont {{Weiss}}, \citenamefont {{Whittle}},
  \citenamefont {{Willke}}, \citenamefont {{Wipf}}, \citenamefont {{Xiao}},
  \citenamefont {{Yamamoto}}, \citenamefont {{Yu}}, \citenamefont {{Yu}},
  \citenamefont {{Zhang}}, \citenamefont {{Zucker}},\ and\ \citenamefont
  {{Zweizig}}}]{BuikemaCahillane2020}%
  \BibitemOpen
  \bibfield  {author} {\bibinfo {author} {\bibfnamefont {A.}~\bibnamefont
  {{Buikema}}}, \bibinfo {author} {\bibfnamefont {C.}~\bibnamefont
  {{Cahillane}}}, \bibinfo {author} {\bibfnamefont {G.~L.}\ \bibnamefont
  {{Mansell}}}, \bibinfo {author} {\bibfnamefont {C.~D.}\ \bibnamefont
  {{Blair}}}, \bibinfo {author} {\bibfnamefont {R.}~\bibnamefont {{Abbott}}},
  \bibinfo {author} {\bibfnamefont {C.}~\bibnamefont {{Adams}}}, \bibinfo
  {author} {\bibfnamefont {R.~X.}\ \bibnamefont {{Adhikari}}}, \bibinfo
  {author} {\bibfnamefont {A.}~\bibnamefont {{Ananyeva}}}, \bibinfo {author}
  {\bibfnamefont {S.}~\bibnamefont {{Appert}}}, \bibinfo {author}
  {\bibfnamefont {K.}~\bibnamefont {{Arai}}}, \bibinfo {author} {\bibfnamefont
  {J.~S.}\ \bibnamefont {{Areeda}}}, \bibinfo {author} {\bibfnamefont
  {Y.}~\bibnamefont {{Asali}}}, \bibinfo {author} {\bibfnamefont {S.~M.}\
  \bibnamefont {{Aston}}}, \bibinfo {author} {\bibfnamefont {C.}~\bibnamefont
  {{Austin}}}, \bibinfo {author} {\bibfnamefont {A.~M.}\ \bibnamefont
  {{Baer}}}, \bibinfo {author} {\bibfnamefont {M.}~\bibnamefont {{Ball}}},
  \bibinfo {author} {\bibfnamefont {S.~W.}\ \bibnamefont {{Ballmer}}}, \bibinfo
  {author} {\bibfnamefont {S.}~\bibnamefont {{Banagiri}}}, \bibinfo {author}
  {\bibfnamefont {D.}~\bibnamefont {{Barker}}}, \bibinfo {author}
  {\bibfnamefont {L.}~\bibnamefont {{Barsotti}}}, \bibinfo {author}
  {\bibfnamefont {J.}~\bibnamefont {{Bartlett}}}, \bibinfo {author}
  {\bibfnamefont {B.~K.}\ \bibnamefont {{Berger}}}, \bibinfo {author}
  {\bibfnamefont {J.}~\bibnamefont {{Betzwieser}}}, \bibinfo {author}
  {\bibfnamefont {D.}~\bibnamefont {{Bhattacharjee}}}, \bibinfo {author}
  {\bibfnamefont {G.}~\bibnamefont {{Billingsley}}}, \bibinfo {author}
  {\bibfnamefont {S.}~\bibnamefont {{Biscans}}}, \bibinfo {author}
  {\bibfnamefont {R.~M.}\ \bibnamefont {{Blair}}}, \bibinfo {author}
  {\bibfnamefont {N.}~\bibnamefont {{Bode}}}, \bibinfo {author} {\bibfnamefont
  {P.}~\bibnamefont {{Booker}}}, \bibinfo {author} {\bibfnamefont
  {R.}~\bibnamefont {{Bork}}}, \bibinfo {author} {\bibfnamefont
  {A.}~\bibnamefont {{Bramley}}}, \bibinfo {author} {\bibfnamefont {A.~F.}\
  \bibnamefont {{Brooks}}}, \bibinfo {author} {\bibfnamefont {D.~D.}\
  \bibnamefont {{Brown}}}, \bibinfo {author} {\bibfnamefont {K.~C.}\
  \bibnamefont {{Cannon}}}, \bibinfo {author} {\bibfnamefont {X.}~\bibnamefont
  {{Chen}}}, \bibinfo {author} {\bibfnamefont {A.~A.}\ \bibnamefont
  {{Ciobanu}}}, \bibinfo {author} {\bibfnamefont {F.}~\bibnamefont {{Clara}}},
  \bibinfo {author} {\bibfnamefont {S.~J.}\ \bibnamefont {{Cooper}}}, \bibinfo
  {author} {\bibfnamefont {K.~R.}\ \bibnamefont {{Corley}}}, \bibinfo {author}
  {\bibfnamefont {S.~T.}\ \bibnamefont {{Countryman}}}, \bibinfo {author}
  {\bibfnamefont {P.~B.}\ \bibnamefont {{Covas}}}, \bibinfo {author}
  {\bibfnamefont {D.~C.}\ \bibnamefont {{Coyne}}}, \bibinfo {author}
  {\bibfnamefont {L.~E.~H.}\ \bibnamefont {{Datrier}}}, \bibinfo {author}
  {\bibfnamefont {D.}~\bibnamefont {{Davis}}}, \bibinfo {author} {\bibfnamefont
  {C.}~\bibnamefont {{Di Fronzo}}}, \bibinfo {author} {\bibfnamefont {K.~L.}\
  \bibnamefont {{Dooley}}}, \bibinfo {author} {\bibfnamefont {J.~C.}\
  \bibnamefont {{Driggers}}}, \bibinfo {author} {\bibfnamefont
  {P.}~\bibnamefont {{Dupej}}}, \bibinfo {author} {\bibfnamefont {S.~E.}\
  \bibnamefont {{Dwyer}}}, \bibinfo {author} {\bibfnamefont {A.}~\bibnamefont
  {{Effler}}}, \bibinfo {author} {\bibfnamefont {T.}~\bibnamefont {{Etzel}}},
  \bibinfo {author} {\bibfnamefont {M.}~\bibnamefont {{Evans}}}, \bibinfo
  {author} {\bibfnamefont {T.~M.}\ \bibnamefont {{Evans}}}, \bibinfo {author}
  {\bibfnamefont {J.}~\bibnamefont {{Feicht}}}, \bibinfo {author}
  {\bibfnamefont {A.}~\bibnamefont {{Fernandez-Galiana}}}, \bibinfo {author}
  {\bibfnamefont {P.}~\bibnamefont {{Fritschel}}}, \bibinfo {author}
  {\bibfnamefont {V.~V.}\ \bibnamefont {{Frolov}}}, \bibinfo {author}
  {\bibfnamefont {P.}~\bibnamefont {{Fulda}}}, \bibinfo {author} {\bibfnamefont
  {M.}~\bibnamefont {{Fyffe}}}, \bibinfo {author} {\bibfnamefont {J.~A.}\
  \bibnamefont {{Giaime}}}, \bibinfo {author} {\bibfnamefont {K.~D.}\
  \bibnamefont {{Giardina}}}, \bibinfo {author} {\bibfnamefont
  {P.}~\bibnamefont {{Godwin}}}, \bibinfo {author} {\bibfnamefont
  {E.}~\bibnamefont {{Goetz}}}, \bibinfo {author} {\bibfnamefont
  {S.}~\bibnamefont {{Gras}}}, \bibinfo {author} {\bibfnamefont
  {C.}~\bibnamefont {{Gray}}}, \bibinfo {author} {\bibfnamefont
  {R.}~\bibnamefont {{Gray}}}, \bibinfo {author} {\bibfnamefont {A.~C.}\
  \bibnamefont {{Green}}}, \bibinfo {author} {\bibfnamefont {E.~K.}\
  \bibnamefont {{Gustafson}}}, \bibinfo {author} {\bibfnamefont
  {R.}~\bibnamefont {{Gustafson}}}, \bibinfo {author} {\bibfnamefont
  {J.}~\bibnamefont {{Hanks}}}, \bibinfo {author} {\bibfnamefont
  {J.}~\bibnamefont {{Hanson}}}, \bibinfo {author} {\bibfnamefont
  {T.}~\bibnamefont {{Hardwick}}}, \bibinfo {author} {\bibfnamefont {R.~K.}\
  \bibnamefont {{Hasskew}}}, \bibinfo {author} {\bibfnamefont {M.~C.}\
  \bibnamefont {{Heintze}}}, \bibinfo {author} {\bibfnamefont {A.~F.}\
  \bibnamefont {{Helmling-Cornell}}}, \bibinfo {author} {\bibfnamefont {N.~A.}\
  \bibnamefont {{Holland}}}, \bibinfo {author} {\bibfnamefont {J.~D.}\
  \bibnamefont {{Jones}}}, \bibinfo {author} {\bibfnamefont {S.}~\bibnamefont
  {{Kandhasamy}}}, \bibinfo {author} {\bibfnamefont {S.}~\bibnamefont
  {{Karki}}}, \bibinfo {author} {\bibfnamefont {M.}~\bibnamefont
  {{Kasprzack}}}, \bibinfo {author} {\bibfnamefont {K.}~\bibnamefont
  {{Kawabe}}}, \bibinfo {author} {\bibfnamefont {N.}~\bibnamefont
  {{Kijbunchoo}}}, \bibinfo {author} {\bibfnamefont {P.~J.}\ \bibnamefont
  {{King}}}, \bibinfo {author} {\bibfnamefont {J.~S.}\ \bibnamefont
  {{Kissel}}}, \bibinfo {author} {\bibfnamefont {R.}~\bibnamefont {{Kumar}}},
  \bibinfo {author} {\bibfnamefont {M.}~\bibnamefont {{Landry}}}, \bibinfo
  {author} {\bibfnamefont {B.~B.}\ \bibnamefont {{Lane}}}, \bibinfo {author}
  {\bibfnamefont {B.}~\bibnamefont {{Lantz}}}, \bibinfo {author} {\bibfnamefont
  {M.}~\bibnamefont {{Laxen}}}, \bibinfo {author} {\bibfnamefont {Y.~K.}\
  \bibnamefont {{Lecoeuche}}}, \bibinfo {author} {\bibfnamefont
  {J.}~\bibnamefont {{Leviton}}}, \bibinfo {author} {\bibfnamefont
  {J.}~\bibnamefont {{Liu}}}, \bibinfo {author} {\bibfnamefont
  {M.}~\bibnamefont {{Lormand}}}, \bibinfo {author} {\bibfnamefont {A.~P.}\
  \bibnamefont {{Lundgren}}}, \bibinfo {author} {\bibfnamefont
  {R.}~\bibnamefont {{Macas}}}, \bibinfo {author} {\bibfnamefont
  {M.}~\bibnamefont {{MacInnis}}}, \bibinfo {author} {\bibfnamefont {D.~M.}\
  \bibnamefont {{Macleod}}}, \bibinfo {author} {\bibfnamefont {S.}~\bibnamefont
  {{M{\'a}rka}}}, \bibinfo {author} {\bibfnamefont {Z.}~\bibnamefont
  {{M{\'a}rka}}}, \bibinfo {author} {\bibfnamefont {D.~V.}\ \bibnamefont
  {{Martynov}}}, \bibinfo {author} {\bibfnamefont {K.}~\bibnamefont {{Mason}}},
  \bibinfo {author} {\bibfnamefont {T.~J.}\ \bibnamefont {{Massinger}}},
  \bibinfo {author} {\bibfnamefont {F.}~\bibnamefont {{Matichard}}}, \bibinfo
  {author} {\bibfnamefont {N.}~\bibnamefont {{Mavalvala}}}, \bibinfo {author}
  {\bibfnamefont {R.}~\bibnamefont {{McCarthy}}}, \bibinfo {author}
  {\bibfnamefont {D.~E.}\ \bibnamefont {{McClelland}}}, \bibinfo {author}
  {\bibfnamefont {S.}~\bibnamefont {{McCormick}}}, \bibinfo {author}
  {\bibfnamefont {L.}~\bibnamefont {{McCuller}}}, \bibinfo {author}
  {\bibfnamefont {J.}~\bibnamefont {{McIver}}}, \bibinfo {author}
  {\bibfnamefont {T.}~\bibnamefont {{McRae}}}, \bibinfo {author} {\bibfnamefont
  {G.}~\bibnamefont {{Mendell}}}, \bibinfo {author} {\bibfnamefont
  {K.}~\bibnamefont {{Merfeld}}}, \bibinfo {author} {\bibfnamefont {E.~L.}\
  \bibnamefont {{Merilh}}}, \bibinfo {author} {\bibfnamefont {F.}~\bibnamefont
  {{Meylahn}}}, \bibinfo {author} {\bibfnamefont {T.}~\bibnamefont {{Mistry}}},
  \bibinfo {author} {\bibfnamefont {R.}~\bibnamefont {{Mittleman}}}, \bibinfo
  {author} {\bibfnamefont {G.}~\bibnamefont {{Moreno}}}, \bibinfo {author}
  {\bibfnamefont {C.~M.}\ \bibnamefont {{Mow-Lowry}}}, \bibinfo {author}
  {\bibfnamefont {S.}~\bibnamefont {{Mozzon}}}, \bibinfo {author}
  {\bibfnamefont {A.}~\bibnamefont {{Mullavey}}}, \bibinfo {author}
  {\bibfnamefont {T.~J.~N.}\ \bibnamefont {{Nelson}}}, \bibinfo {author}
  {\bibfnamefont {P.}~\bibnamefont {{Nguyen}}}, \bibinfo {author}
  {\bibfnamefont {L.~K.}\ \bibnamefont {{Nuttall}}}, \bibinfo {author}
  {\bibfnamefont {J.}~\bibnamefont {{Oberling}}}, \bibinfo {author}
  {\bibfnamefont {R.~J.}\ \bibnamefont {{Oram}}}, \bibinfo {author}
  {\bibfnamefont {B.}~\bibnamefont {{O'Reilly}}}, \bibinfo {author}
  {\bibfnamefont {C.}~\bibnamefont {{Osthelder}}}, \bibinfo {author}
  {\bibfnamefont {D.~J.}\ \bibnamefont {{Ottaway}}}, \bibinfo {author}
  {\bibfnamefont {H.}~\bibnamefont {{Overmier}}}, \bibinfo {author}
  {\bibfnamefont {J.~R.}\ \bibnamefont {{Palamos}}}, \bibinfo {author}
  {\bibfnamefont {W.}~\bibnamefont {{Parker}}}, \bibinfo {author}
  {\bibfnamefont {E.}~\bibnamefont {{Payne}}}, \bibinfo {author} {\bibfnamefont
  {A.}~\bibnamefont {{Pele}}}, \bibinfo {author} {\bibfnamefont
  {R.}~\bibnamefont {{Penhorwood}}}, \bibinfo {author} {\bibfnamefont {C.~J.}\
  \bibnamefont {{Perez}}}, \bibinfo {author} {\bibfnamefont {M.}~\bibnamefont
  {{Pirello}}}, \bibinfo {author} {\bibfnamefont {H.}~\bibnamefont
  {{Radkins}}}, \bibinfo {author} {\bibfnamefont {K.~E.}\ \bibnamefont
  {{Ramirez}}}, \bibinfo {author} {\bibfnamefont {J.~W.}\ \bibnamefont
  {{Richardson}}}, \bibinfo {author} {\bibfnamefont {K.}~\bibnamefont
  {{Riles}}}, \bibinfo {author} {\bibfnamefont {N.~A.}\ \bibnamefont
  {{Robertson}}}, \bibinfo {author} {\bibfnamefont {J.~G.}\ \bibnamefont
  {{Rollins}}}, \bibinfo {author} {\bibfnamefont {C.~L.}\ \bibnamefont
  {{Romel}}}, \bibinfo {author} {\bibfnamefont {J.~H.}\ \bibnamefont
  {{Romie}}}, \bibinfo {author} {\bibfnamefont {M.~P.}\ \bibnamefont {{Ross}}},
  \bibinfo {author} {\bibfnamefont {K.}~\bibnamefont {{Ryan}}}, \bibinfo
  {author} {\bibfnamefont {T.}~\bibnamefont {{Sadecki}}}, \bibinfo {author}
  {\bibfnamefont {E.~J.}\ \bibnamefont {{Sanchez}}}, \bibinfo {author}
  {\bibfnamefont {L.~E.}\ \bibnamefont {{Sanchez}}}, \bibinfo {author}
  {\bibfnamefont {T.~R.}\ \bibnamefont {{Saravanan}}}, \bibinfo {author}
  {\bibfnamefont {R.~L.}\ \bibnamefont {{Savage}}}, \bibinfo {author}
  {\bibfnamefont {D.}~\bibnamefont {{Schaetzl}}}, \bibinfo {author}
  {\bibfnamefont {R.}~\bibnamefont {{Schnabel}}}, \bibinfo {author}
  {\bibfnamefont {R.~M.~S.}\ \bibnamefont {{Schofield}}}, \bibinfo {author}
  {\bibfnamefont {E.}~\bibnamefont {{Schwartz}}}, \bibinfo {author}
  {\bibfnamefont {D.}~\bibnamefont {{Sellers}}}, \bibinfo {author}
  {\bibfnamefont {T.}~\bibnamefont {{Shaffer}}}, \bibinfo {author}
  {\bibfnamefont {D.}~\bibnamefont {{Sigg}}}, \bibinfo {author} {\bibfnamefont
  {B.~J.~J.}\ \bibnamefont {{Slagmolen}}}, \bibinfo {author} {\bibfnamefont
  {J.~R.}\ \bibnamefont {{Smith}}}, \bibinfo {author} {\bibfnamefont
  {S.}~\bibnamefont {{Soni}}}, \bibinfo {author} {\bibfnamefont
  {B.}~\bibnamefont {{Sorazu}}}, \bibinfo {author} {\bibfnamefont {A.~P.}\
  \bibnamefont {{Spencer}}}, \bibinfo {author} {\bibfnamefont {K.~A.}\
  \bibnamefont {{Strain}}}, \bibinfo {author} {\bibfnamefont {L.}~\bibnamefont
  {{Sun}}}, \bibinfo {author} {\bibfnamefont {M.~J.}\ \bibnamefont
  {{Szczepa{\'n}czyk}}}, \bibinfo {author} {\bibfnamefont {M.}~\bibnamefont
  {{Thomas}}}, \bibinfo {author} {\bibfnamefont {P.}~\bibnamefont {{Thomas}}},
  \bibinfo {author} {\bibfnamefont {K.~A.}\ \bibnamefont {{Thorne}}}, \bibinfo
  {author} {\bibfnamefont {K.}~\bibnamefont {{Toland}}}, \bibinfo {author}
  {\bibfnamefont {C.~I.}\ \bibnamefont {{Torrie}}}, \bibinfo {author}
  {\bibfnamefont {G.}~\bibnamefont {{Traylor}}}, \bibinfo {author}
  {\bibfnamefont {M.}~\bibnamefont {{Tse}}}, \bibinfo {author} {\bibfnamefont
  {A.~L.}\ \bibnamefont {{Urban}}}, \bibinfo {author} {\bibfnamefont
  {G.}~\bibnamefont {{Vajente}}}, \bibinfo {author} {\bibfnamefont
  {G.}~\bibnamefont {{Valdes}}}, \bibinfo {author} {\bibfnamefont {D.~C.}\
  \bibnamefont {{Vander-Hyde}}}, \bibinfo {author} {\bibfnamefont {P.~J.}\
  \bibnamefont {{Veitch}}}, \bibinfo {author} {\bibfnamefont {K.}~\bibnamefont
  {{Venkateswara}}}, \bibinfo {author} {\bibfnamefont {G.}~\bibnamefont
  {{Venugopalan}}}, \bibinfo {author} {\bibfnamefont {A.~D.}\ \bibnamefont
  {{Viets}}}, \bibinfo {author} {\bibfnamefont {T.}~\bibnamefont {{Vo}}},
  \bibinfo {author} {\bibfnamefont {C.}~\bibnamefont {{Vorvick}}}, \bibinfo
  {author} {\bibfnamefont {M.}~\bibnamefont {{Wade}}}, \bibinfo {author}
  {\bibfnamefont {R.~L.}\ \bibnamefont {{Ward}}}, \bibinfo {author}
  {\bibfnamefont {J.}~\bibnamefont {{Warner}}}, \bibinfo {author}
  {\bibfnamefont {B.}~\bibnamefont {{Weaver}}}, \bibinfo {author}
  {\bibfnamefont {R.}~\bibnamefont {{Weiss}}}, \bibinfo {author} {\bibfnamefont
  {C.}~\bibnamefont {{Whittle}}}, \bibinfo {author} {\bibfnamefont
  {B.}~\bibnamefont {{Willke}}}, \bibinfo {author} {\bibfnamefont {C.~C.}\
  \bibnamefont {{Wipf}}}, \bibinfo {author} {\bibfnamefont {L.}~\bibnamefont
  {{Xiao}}}, \bibinfo {author} {\bibfnamefont {H.}~\bibnamefont {{Yamamoto}}},
  \bibinfo {author} {\bibfnamefont {H.}~\bibnamefont {{Yu}}}, \bibinfo {author}
  {\bibfnamefont {H.}~\bibnamefont {{Yu}}}, \bibinfo {author} {\bibfnamefont
  {L.}~\bibnamefont {{Zhang}}}, \bibinfo {author} {\bibfnamefont {M.~E.}\
  \bibnamefont {{Zucker}}},\ and\ \bibinfo {author} {\bibfnamefont
  {J.}~\bibnamefont {{Zweizig}}},\ }\bibfield  {title} {\bibinfo {title}
  {{Sensitivity and performance of the Advanced LIGO detectors in the third
  observing run}},\ }\href {https://doi.org/10.1103/PhysRevD.102.062003}
  {\bibfield  {journal} {\bibinfo  {journal} {Physical Review D}\ }\textbf
  {\bibinfo {volume} {102}},\ \bibinfo {eid} {062003} (\bibinfo {year}
  {2020})},\ \Eprint {https://arxiv.org/abs/2008.01301} {arXiv:2008.01301
  [astro-ph.IM]} \BibitemShut {NoStop}%
\bibitem [{\citenamefont {Goetz}\ and\ \citenamefont {Riles}()}]{DCCsegments}%
  \BibitemOpen
  \bibfield  {author} {\bibinfo {author} {\bibfnamefont {E.}~\bibnamefont
  {Goetz}}\ and\ \bibinfo {author} {\bibfnamefont {K.}~\bibnamefont {Riles}},\
  }\href {https://dcc.ligo.org/T2300068/public} {\emph {\bibinfo {title}
  {{Segments used for creating standard SFTs in O3 data}}}},\ \bibinfo {type}
  {LIGO Document}\ \bibinfo {number} {T2300068}\BibitemShut {NoStop}%
\bibitem [{\citenamefont {{Sun}}\ \emph
  {et~al.}(2020{\natexlab{b}})\citenamefont {{Sun}}, \citenamefont {{Goetz}},
  \citenamefont {{Kissel}}, \citenamefont {{Betzwieser}}, \citenamefont
  {{Karki}}, \citenamefont {{Viets}}, \citenamefont {{Wade}}, \citenamefont
  {{Bhattacharjee}}, \citenamefont {{Bossilkov}}, \citenamefont {{Covas}},
  \citenamefont {{Datrier}}, \citenamefont {{Gray}}, \citenamefont
  {{Kandhasamy}}, \citenamefont {{Lecoeuche}}, \citenamefont {{Mendell}},
  \citenamefont {{Mistry}}, \citenamefont {{Payne}}, \citenamefont {{Savage}},
  \citenamefont {{Weinstein}}, \citenamefont {{Aston}}, \citenamefont
  {{Buikema}}, \citenamefont {{Cahillane}}, \citenamefont {{Driggers}},
  \citenamefont {{Dwyer}}, \citenamefont {{Kumar}},\ and\ \citenamefont
  {{Urban}}}]{SunGoetz2020}%
  \BibitemOpen
  \bibfield  {author} {\bibinfo {author} {\bibfnamefont {L.}~\bibnamefont
  {{Sun}}}, \bibinfo {author} {\bibfnamefont {E.}~\bibnamefont {{Goetz}}},
  \bibinfo {author} {\bibfnamefont {J.~S.}\ \bibnamefont {{Kissel}}}, \bibinfo
  {author} {\bibfnamefont {J.}~\bibnamefont {{Betzwieser}}}, \bibinfo {author}
  {\bibfnamefont {S.}~\bibnamefont {{Karki}}}, \bibinfo {author} {\bibfnamefont
  {A.}~\bibnamefont {{Viets}}}, \bibinfo {author} {\bibfnamefont
  {M.}~\bibnamefont {{Wade}}}, \bibinfo {author} {\bibfnamefont
  {D.}~\bibnamefont {{Bhattacharjee}}}, \bibinfo {author} {\bibfnamefont
  {V.}~\bibnamefont {{Bossilkov}}}, \bibinfo {author} {\bibfnamefont {P.~B.}\
  \bibnamefont {{Covas}}}, \bibinfo {author} {\bibfnamefont {L.~E.~H.}\
  \bibnamefont {{Datrier}}}, \bibinfo {author} {\bibfnamefont {R.}~\bibnamefont
  {{Gray}}}, \bibinfo {author} {\bibfnamefont {S.}~\bibnamefont
  {{Kandhasamy}}}, \bibinfo {author} {\bibfnamefont {Y.~K.}\ \bibnamefont
  {{Lecoeuche}}}, \bibinfo {author} {\bibfnamefont {G.}~\bibnamefont
  {{Mendell}}}, \bibinfo {author} {\bibfnamefont {T.}~\bibnamefont {{Mistry}}},
  \bibinfo {author} {\bibfnamefont {E.}~\bibnamefont {{Payne}}}, \bibinfo
  {author} {\bibfnamefont {R.~L.}\ \bibnamefont {{Savage}}}, \bibinfo {author}
  {\bibfnamefont {A.~J.}\ \bibnamefont {{Weinstein}}}, \bibinfo {author}
  {\bibfnamefont {S.}~\bibnamefont {{Aston}}}, \bibinfo {author} {\bibfnamefont
  {A.}~\bibnamefont {{Buikema}}}, \bibinfo {author} {\bibfnamefont
  {C.}~\bibnamefont {{Cahillane}}}, \bibinfo {author} {\bibfnamefont {J.~C.}\
  \bibnamefont {{Driggers}}}, \bibinfo {author} {\bibfnamefont {S.~E.}\
  \bibnamefont {{Dwyer}}}, \bibinfo {author} {\bibfnamefont {R.}~\bibnamefont
  {{Kumar}}},\ and\ \bibinfo {author} {\bibfnamefont {A.}~\bibnamefont
  {{Urban}}},\ }\bibfield  {title} {\bibinfo {title} {{Characterization of
  systematic error in Advanced LIGO calibration}},\ }\href
  {https://doi.org/10.1088/1361-6382/abb14e} {\bibfield  {journal} {\bibinfo
  {journal} {Classical and Quantum Gravity}\ }\textbf {\bibinfo {volume}
  {37}},\ \bibinfo {eid} {225008} (\bibinfo {year} {2020}{\natexlab{b}})},\
  \Eprint {https://arxiv.org/abs/2005.02531} {arXiv:2005.02531 [astro-ph.IM]}
  \BibitemShut {NoStop}%
\bibitem [{\citenamefont {{Sun}}\ \emph {et~al.}(2021)\citenamefont {{Sun}},
  \citenamefont {{Goetz}}, \citenamefont {{Kissel}}, \citenamefont
  {{Betzwieser}}, \citenamefont {{Karki}}, \citenamefont {{Bhattacharjee}},
  \citenamefont {{Covas}}, \citenamefont {{Datrier}}, \citenamefont
  {{Kandhasamy}}, \citenamefont {{Lecoeuche}}, \citenamefont {{Mendell}},
  \citenamefont {{Mistry}}, \citenamefont {{Payne}}, \citenamefont {{Savage}},
  \citenamefont {{Viets}}, \citenamefont {{Wade}}, \citenamefont {{Weinstein}},
  \citenamefont {{Aston}}, \citenamefont {{Cahillane}}, \citenamefont
  {{Driggers}}, \citenamefont {{Dwyer}},\ and\ \citenamefont
  {{Urban}}}]{SunGoetz2021}%
  \BibitemOpen
  \bibfield  {author} {\bibinfo {author} {\bibfnamefont {L.}~\bibnamefont
  {{Sun}}}, \bibinfo {author} {\bibfnamefont {E.}~\bibnamefont {{Goetz}}},
  \bibinfo {author} {\bibfnamefont {J.~S.}\ \bibnamefont {{Kissel}}}, \bibinfo
  {author} {\bibfnamefont {J.}~\bibnamefont {{Betzwieser}}}, \bibinfo {author}
  {\bibfnamefont {S.}~\bibnamefont {{Karki}}}, \bibinfo {author} {\bibfnamefont
  {D.}~\bibnamefont {{Bhattacharjee}}}, \bibinfo {author} {\bibfnamefont
  {P.~B.}\ \bibnamefont {{Covas}}}, \bibinfo {author} {\bibfnamefont
  {L.~E.~H.}\ \bibnamefont {{Datrier}}}, \bibinfo {author} {\bibfnamefont
  {S.}~\bibnamefont {{Kandhasamy}}}, \bibinfo {author} {\bibfnamefont {Y.~K.}\
  \bibnamefont {{Lecoeuche}}}, \bibinfo {author} {\bibfnamefont
  {G.}~\bibnamefont {{Mendell}}}, \bibinfo {author} {\bibfnamefont
  {T.}~\bibnamefont {{Mistry}}}, \bibinfo {author} {\bibfnamefont
  {E.}~\bibnamefont {{Payne}}}, \bibinfo {author} {\bibfnamefont {R.~L.}\
  \bibnamefont {{Savage}}}, \bibinfo {author} {\bibfnamefont {A.}~\bibnamefont
  {{Viets}}}, \bibinfo {author} {\bibfnamefont {M.}~\bibnamefont {{Wade}}},
  \bibinfo {author} {\bibfnamefont {A.~J.}\ \bibnamefont {{Weinstein}}},
  \bibinfo {author} {\bibfnamefont {S.}~\bibnamefont {{Aston}}}, \bibinfo
  {author} {\bibfnamefont {C.}~\bibnamefont {{Cahillane}}}, \bibinfo {author}
  {\bibfnamefont {J.~C.}\ \bibnamefont {{Driggers}}}, \bibinfo {author}
  {\bibfnamefont {S.~E.}\ \bibnamefont {{Dwyer}}},\ and\ \bibinfo {author}
  {\bibfnamefont {A.}~\bibnamefont {{Urban}}},\ }\bibfield  {title} {\bibinfo
  {title} {{Characterization of systematic error in Advanced LIGO calibration
  in the second half of O3}},\ }\href
  {https://doi.org/10.48550/arXiv.2107.00129} {\bibfield  {journal} {\bibinfo
  {journal} {arXiv e-prints}\ ,\ \bibinfo {eid} {arXiv:2107.00129}} (\bibinfo
  {year} {2021})},\ \Eprint {https://arxiv.org/abs/2107.00129}
  {arXiv:2107.00129 [astro-ph.IM]} \BibitemShut {NoStop}%
\bibitem [{\citenamefont {{Davis}}\ \emph {et~al.}(2021)\citenamefont
  {{Davis}}, \citenamefont {{Areeda}}, \citenamefont {{Berger}}, \citenamefont
  {{Bruntz}}, \citenamefont {{Effler}}, \citenamefont {{Essick}}, \citenamefont
  {{Fisher}}, \citenamefont {{Godwin}}, \citenamefont {{Goetz}}, \citenamefont
  {{Helmling-Cornell}},\ and\ \citenamefont {et~al.}}]{DavisAreeda2021}%
  \BibitemOpen
  \bibfield  {author} {\bibinfo {author} {\bibfnamefont {D.}~\bibnamefont
  {{Davis}}}, \bibinfo {author} {\bibfnamefont {J.~S.}\ \bibnamefont
  {{Areeda}}}, \bibinfo {author} {\bibfnamefont {B.~K.}\ \bibnamefont
  {{Berger}}}, \bibinfo {author} {\bibfnamefont {R.}~\bibnamefont {{Bruntz}}},
  \bibinfo {author} {\bibfnamefont {A.}~\bibnamefont {{Effler}}}, \bibinfo
  {author} {\bibfnamefont {R.~C.}\ \bibnamefont {{Essick}}}, \bibinfo {author}
  {\bibfnamefont {R.~P.}\ \bibnamefont {{Fisher}}}, \bibinfo {author}
  {\bibfnamefont {P.}~\bibnamefont {{Godwin}}}, \bibinfo {author}
  {\bibfnamefont {E.}~\bibnamefont {{Goetz}}}, \bibinfo {author} {\bibfnamefont
  {A.~F.}\ \bibnamefont {{Helmling-Cornell}}},\ and\ \bibinfo {author}
  {\bibnamefont {et~al.}},\ }\bibfield  {title} {\bibinfo {title} {{LIGO
  detector characterization in the second and third observing runs}},\ }\href
  {https://doi.org/10.1088/1361-6382/abfd85} {\bibfield  {journal} {\bibinfo
  {journal} {Classical and Quantum Gravity}\ }\textbf {\bibinfo {volume}
  {38}},\ \bibinfo {eid} {135014} (\bibinfo {year} {2021})},\ \Eprint
  {https://arxiv.org/abs/2101.11673} {arXiv:2101.11673 [astro-ph.IM]}
  \BibitemShut {NoStop}%
\bibitem [{\citenamefont {Zweizig}\ and\ \citenamefont {Riles}()}]{DCCgating}%
  \BibitemOpen
  \bibfield  {author} {\bibinfo {author} {\bibfnamefont {J.}~\bibnamefont
  {Zweizig}}\ and\ \bibinfo {author} {\bibfnamefont {K.}~\bibnamefont
  {Riles}},\ }\href {https://dcc.ligo.org/T2000384/public} {\emph {\bibinfo
  {title} {{Information on self-gating of $h(t)$ used in O3 continuous-wave and
  stochastic searches}}}},\ \bibinfo {type} {LIGO Document}\ \bibinfo {number}
  {T2000384}\BibitemShut {NoStop}%
\bibitem [{\citenamefont {{Jones}}(2004)}]{Jones2004}%
  \BibitemOpen
  \bibfield  {author} {\bibinfo {author} {\bibfnamefont {D.~I.}\ \bibnamefont
  {{Jones}}},\ }\bibfield  {title} {\bibinfo {title} {{Is timing noise
  important in the gravitational wave detection of neutron stars?}},\ }\href
  {https://doi.org/10.1103/PhysRevD.70.042002} {\bibfield  {journal} {\bibinfo
  {journal} {Physical Review D}\ }\textbf {\bibinfo {volume} {70}},\ \bibinfo
  {eid} {042002} (\bibinfo {year} {2004})},\ \Eprint
  {https://arxiv.org/abs/gr-qc/0406045} {arXiv:gr-qc/0406045 [gr-qc]}
  \BibitemShut {NoStop}%
\bibitem [{\citenamefont {{Knee}}\ \emph {et~al.}(2024)\citenamefont {{Knee}},
  \citenamefont {{Du}}, \citenamefont {{Goetz}}, \citenamefont {{McIver}},
  \citenamefont {{Carlin}}, \citenamefont {{Sun}}, \citenamefont {{Dunn}},
  \citenamefont {{Strang}}, \citenamefont {{Middleton}},\ and\ \citenamefont
  {{Melatos}}}]{KneeDu2024}%
  \BibitemOpen
  \bibfield  {author} {\bibinfo {author} {\bibfnamefont {A.~M.}\ \bibnamefont
  {{Knee}}}, \bibinfo {author} {\bibfnamefont {H.}~\bibnamefont {{Du}}},
  \bibinfo {author} {\bibfnamefont {E.}~\bibnamefont {{Goetz}}}, \bibinfo
  {author} {\bibfnamefont {J.}~\bibnamefont {{McIver}}}, \bibinfo {author}
  {\bibfnamefont {J.~B.}\ \bibnamefont {{Carlin}}}, \bibinfo {author}
  {\bibfnamefont {L.}~\bibnamefont {{Sun}}}, \bibinfo {author} {\bibfnamefont
  {L.}~\bibnamefont {{Dunn}}}, \bibinfo {author} {\bibfnamefont
  {L.}~\bibnamefont {{Strang}}}, \bibinfo {author} {\bibfnamefont
  {H.}~\bibnamefont {{Middleton}}},\ and\ \bibinfo {author} {\bibfnamefont
  {A.}~\bibnamefont {{Melatos}}},\ }\bibfield  {title} {\bibinfo {title}
  {{Search for continuous gravitational waves directed at subthreshold
  radiometer candidates in O3 LIGO data}},\ }\href
  {https://doi.org/10.1103/PhysRevD.109.062008} {\bibfield  {journal} {\bibinfo
   {journal} {Physical Review D}\ }\textbf {\bibinfo {volume} {109}},\ \bibinfo
  {eid} {062008} (\bibinfo {year} {2024})},\ \Eprint
  {https://arxiv.org/abs/2311.12138} {arXiv:2311.12138 [gr-qc]} \BibitemShut
  {NoStop}%
\bibitem [{\citenamefont {{Messenger}}\ \emph {et~al.}(2015)\citenamefont
  {{Messenger}}, \citenamefont {{Bulten}}, \citenamefont {{Crowder}},
  \citenamefont {{Dergachev}}, \citenamefont {{Galloway}}, \citenamefont
  {{Goetz}}, \citenamefont {{Jonker}}, \citenamefont {{Lasky}}, \citenamefont
  {{Meadors}}, \citenamefont {{Melatos}}, \citenamefont {{Premachandra}},
  \citenamefont {{Riles}}, \citenamefont {{Sammut}}, \citenamefont {{Thrane}},
  \citenamefont {{Whelan}},\ and\ \citenamefont
  {{Zhang}}}]{MessengerBulten2015}%
  \BibitemOpen
  \bibfield  {author} {\bibinfo {author} {\bibfnamefont {C.}~\bibnamefont
  {{Messenger}}}, \bibinfo {author} {\bibfnamefont {H.~J.}\ \bibnamefont
  {{Bulten}}}, \bibinfo {author} {\bibfnamefont {S.~G.}\ \bibnamefont
  {{Crowder}}}, \bibinfo {author} {\bibfnamefont {V.}~\bibnamefont
  {{Dergachev}}}, \bibinfo {author} {\bibfnamefont {D.~K.}\ \bibnamefont
  {{Galloway}}}, \bibinfo {author} {\bibfnamefont {E.}~\bibnamefont {{Goetz}}},
  \bibinfo {author} {\bibfnamefont {R.~J.~G.}\ \bibnamefont {{Jonker}}},
  \bibinfo {author} {\bibfnamefont {P.~D.}\ \bibnamefont {{Lasky}}}, \bibinfo
  {author} {\bibfnamefont {G.~D.}\ \bibnamefont {{Meadors}}}, \bibinfo {author}
  {\bibfnamefont {A.}~\bibnamefont {{Melatos}}}, \bibinfo {author}
  {\bibfnamefont {S.}~\bibnamefont {{Premachandra}}}, \bibinfo {author}
  {\bibfnamefont {K.}~\bibnamefont {{Riles}}}, \bibinfo {author} {\bibfnamefont
  {L.}~\bibnamefont {{Sammut}}}, \bibinfo {author} {\bibfnamefont {E.~H.}\
  \bibnamefont {{Thrane}}}, \bibinfo {author} {\bibfnamefont {J.~T.}\
  \bibnamefont {{Whelan}}},\ and\ \bibinfo {author} {\bibfnamefont
  {Y.}~\bibnamefont {{Zhang}}},\ }\bibfield  {title} {\bibinfo {title}
  {{Gravitational waves from Scorpius X-1: A comparison of search methods and
  prospects for detection with advanced detectors}},\ }\href
  {https://doi.org/10.1103/PhysRevD.92.023006} {\bibfield  {journal} {\bibinfo
  {journal} {Physical Review D}\ }\textbf {\bibinfo {volume} {92}},\ \bibinfo
  {eid} {023006} (\bibinfo {year} {2015})},\ \Eprint
  {https://arxiv.org/abs/1504.05889} {arXiv:1504.05889 [gr-qc]} \BibitemShut
  {NoStop}%
\bibitem [{\citenamefont {{Tan}}\ \emph {et~al.}(2018)\citenamefont {{Tan}},
  \citenamefont {{Bassa}}, \citenamefont {{Cooper}}, \citenamefont {{Dijkema}},
  \citenamefont {{Esposito}}, \citenamefont {{Hessels}}, \citenamefont
  {{Kondratiev}}, \citenamefont {{Kramer}}, \citenamefont {{Michilli}},
  \citenamefont {{Sanidas}}, \citenamefont {{Shimwell}}, \citenamefont
  {{Stappers}}, \citenamefont {{van Leeuwen}}, \citenamefont {{Cognard}},
  \citenamefont {{Grie{\ss}meier}}, \citenamefont {{Karastergiou}},
  \citenamefont {{Keane}}, \citenamefont {{Sobey}},\ and\ \citenamefont
  {{Weltevrede}}}]{TanBassa2018}%
  \BibitemOpen
  \bibfield  {author} {\bibinfo {author} {\bibfnamefont {C.~M.}\ \bibnamefont
  {{Tan}}}, \bibinfo {author} {\bibfnamefont {C.~G.}\ \bibnamefont {{Bassa}}},
  \bibinfo {author} {\bibfnamefont {S.}~\bibnamefont {{Cooper}}}, \bibinfo
  {author} {\bibfnamefont {T.~J.}\ \bibnamefont {{Dijkema}}}, \bibinfo {author}
  {\bibfnamefont {P.}~\bibnamefont {{Esposito}}}, \bibinfo {author}
  {\bibfnamefont {J.~W.~T.}\ \bibnamefont {{Hessels}}}, \bibinfo {author}
  {\bibfnamefont {V.~I.}\ \bibnamefont {{Kondratiev}}}, \bibinfo {author}
  {\bibfnamefont {M.}~\bibnamefont {{Kramer}}}, \bibinfo {author}
  {\bibfnamefont {D.}~\bibnamefont {{Michilli}}}, \bibinfo {author}
  {\bibfnamefont {S.}~\bibnamefont {{Sanidas}}}, \bibinfo {author}
  {\bibfnamefont {T.~W.}\ \bibnamefont {{Shimwell}}}, \bibinfo {author}
  {\bibfnamefont {B.~W.}\ \bibnamefont {{Stappers}}}, \bibinfo {author}
  {\bibfnamefont {J.}~\bibnamefont {{van Leeuwen}}}, \bibinfo {author}
  {\bibfnamefont {I.}~\bibnamefont {{Cognard}}}, \bibinfo {author}
  {\bibfnamefont {J.~M.}\ \bibnamefont {{Grie{\ss}meier}}}, \bibinfo {author}
  {\bibfnamefont {A.}~\bibnamefont {{Karastergiou}}}, \bibinfo {author}
  {\bibfnamefont {E.~F.}\ \bibnamefont {{Keane}}}, \bibinfo {author}
  {\bibfnamefont {C.}~\bibnamefont {{Sobey}}},\ and\ \bibinfo {author}
  {\bibfnamefont {P.}~\bibnamefont {{Weltevrede}}},\ }\bibfield  {title}
  {\bibinfo {title} {{LOFAR Discovery of a 23.5 s Radio Pulsar}},\ }\href
  {https://doi.org/10.3847/1538-4357/aade88} {\bibfield  {journal} {\bibinfo
  {journal} {The Astrophysical Journal}\ }\textbf {\bibinfo {volume} {866}},\
  \bibinfo {eid} {54} (\bibinfo {year} {2018})},\ \Eprint
  {https://arxiv.org/abs/1809.00965} {arXiv:1809.00965 [astro-ph.HE]}
  \BibitemShut {NoStop}%
\bibitem [{\citenamefont {{Reig}}(2011)}]{Reig2011}%
  \BibitemOpen
  \bibfield  {author} {\bibinfo {author} {\bibfnamefont {P.}~\bibnamefont
  {{Reig}}},\ }\bibfield  {title} {\bibinfo {title} {{Be/X-ray binaries}},\
  }\href {https://doi.org/10.1007/s10509-010-0575-8} {\bibfield  {journal}
  {\bibinfo  {journal} {Astrophysics and Space Science}\ }\textbf {\bibinfo
  {volume} {332}},\ \bibinfo {pages} {1} (\bibinfo {year} {2011})},\ \Eprint
  {https://arxiv.org/abs/1101.5036} {arXiv:1101.5036 [astro-ph.HE]}
  \BibitemShut {NoStop}%
\bibitem [{\citenamefont {{Hessels}}\ \emph {et~al.}(2006)\citenamefont
  {{Hessels}}, \citenamefont {{Ransom}}, \citenamefont {{Stairs}},
  \citenamefont {{Freire}}, \citenamefont {{Kaspi}},\ and\ \citenamefont
  {{Camilo}}}]{HesselsRansom2006}%
  \BibitemOpen
  \bibfield  {author} {\bibinfo {author} {\bibfnamefont {J.~W.~T.}\
  \bibnamefont {{Hessels}}}, \bibinfo {author} {\bibfnamefont {S.~M.}\
  \bibnamefont {{Ransom}}}, \bibinfo {author} {\bibfnamefont {I.~H.}\
  \bibnamefont {{Stairs}}}, \bibinfo {author} {\bibfnamefont {P.~C.~C.}\
  \bibnamefont {{Freire}}}, \bibinfo {author} {\bibfnamefont {V.~M.}\
  \bibnamefont {{Kaspi}}},\ and\ \bibinfo {author} {\bibfnamefont
  {F.}~\bibnamefont {{Camilo}}},\ }\bibfield  {title} {\bibinfo {title} {{A
  Radio Pulsar Spinning at 716 Hz}},\ }\href
  {https://doi.org/10.1126/science.1123430} {\bibfield  {journal} {\bibinfo
  {journal} {Science}\ }\textbf {\bibinfo {volume} {311}},\ \bibinfo {pages}
  {1901} (\bibinfo {year} {2006})},\ \Eprint
  {https://arxiv.org/abs/astro-ph/0601337} {arXiv:astro-ph/0601337 [astro-ph]}
  \BibitemShut {NoStop}%
\bibitem [{\citenamefont {{Manchester}}\ \emph {et~al.}(2005)\citenamefont
  {{Manchester}}, \citenamefont {{Hobbs}}, \citenamefont {{Teoh}},\ and\
  \citenamefont {{Hobbs}}}]{ManchesterHobbs2005}%
  \BibitemOpen
  \bibfield  {author} {\bibinfo {author} {\bibfnamefont {R.~N.}\ \bibnamefont
  {{Manchester}}}, \bibinfo {author} {\bibfnamefont {G.~B.}\ \bibnamefont
  {{Hobbs}}}, \bibinfo {author} {\bibfnamefont {A.}~\bibnamefont {{Teoh}}},\
  and\ \bibinfo {author} {\bibfnamefont {M.}~\bibnamefont {{Hobbs}}},\
  }\bibfield  {title} {\bibinfo {title} {{The Australia Telescope National
  Facility Pulsar Catalogue}},\ }\href {https://doi.org/10.1086/428488}
  {\bibfield  {journal} {\bibinfo  {journal} {The Astronomical Journal}\
  }\textbf {\bibinfo {volume} {129}},\ \bibinfo {pages} {1993} (\bibinfo {year}
  {2005})},\ \Eprint {https://arxiv.org/abs/astro-ph/0412641}
  {arXiv:astro-ph/0412641 [astro-ph]} \BibitemShut {NoStop}%
\bibitem [{\citenamefont {{Covas}}\ \emph {et~al.}(2018)\citenamefont
  {{Covas}}, \citenamefont {{Effler}}, \citenamefont {{Goetz}}, \citenamefont
  {{Meyers}}, \citenamefont {{Neunzert}}, \citenamefont {{Oliver}},
  \citenamefont {{Pearlstone}}, \citenamefont {{Roma}}, \citenamefont
  {{Schofield}}, \citenamefont {{Adya}},\ and\ \citenamefont
  {et~al.}}]{CovasEffler2018}%
  \BibitemOpen
  \bibfield  {author} {\bibinfo {author} {\bibfnamefont {P.~B.}\ \bibnamefont
  {{Covas}}}, \bibinfo {author} {\bibfnamefont {A.}~\bibnamefont {{Effler}}},
  \bibinfo {author} {\bibfnamefont {E.}~\bibnamefont {{Goetz}}}, \bibinfo
  {author} {\bibfnamefont {P.~M.}\ \bibnamefont {{Meyers}}}, \bibinfo {author}
  {\bibfnamefont {A.}~\bibnamefont {{Neunzert}}}, \bibinfo {author}
  {\bibfnamefont {M.}~\bibnamefont {{Oliver}}}, \bibinfo {author}
  {\bibfnamefont {B.~L.}\ \bibnamefont {{Pearlstone}}}, \bibinfo {author}
  {\bibfnamefont {V.~J.}\ \bibnamefont {{Roma}}}, \bibinfo {author}
  {\bibfnamefont {R.~M.~S.}\ \bibnamefont {{Schofield}}}, \bibinfo {author}
  {\bibfnamefont {V.~B.}\ \bibnamefont {{Adya}}},\ and\ \bibinfo {author}
  {\bibnamefont {et~al.}},\ }\bibfield  {title} {\bibinfo {title}
  {{Identification and mitigation of narrow spectral artifacts that degrade
  searches for persistent gravitational waves in the first two observing runs
  of Advanced LIGO}},\ }\href {https://doi.org/10.1103/PhysRevD.97.082002}
  {\bibfield  {journal} {\bibinfo  {journal} {Physical Review D}\ }\textbf
  {\bibinfo {volume} {97}},\ \bibinfo {eid} {082002} (\bibinfo {year}
  {2018})},\ \Eprint {https://arxiv.org/abs/1801.07204} {arXiv:1801.07204
  [astro-ph.IM]} \BibitemShut {NoStop}%
\bibitem [{\citenamefont {{Gravitational Wave Open Science
  Center}}()}]{O3LinesSite}%
  \BibitemOpen
  \bibfield  {author} {\bibinfo {author} {\bibnamefont {{Gravitational Wave
  Open Science Center}}},\ }\href@noop {} {\bibinfo {title} {{O3 Instrumental
  Lines}}},\ \bibinfo {howpublished}
  {\url{https://gwosc.org/O3/o3speclines/}}\BibitemShut {NoStop}%
\bibitem [{\citenamefont {{Sun}}\ \emph {et~al.}(2018)\citenamefont {{Sun}},
  \citenamefont {{Melatos}}, \citenamefont {{Suvorova}}, \citenamefont
  {{Moran}},\ and\ \citenamefont {{Evans}}}]{SunMelatos2018}%
  \BibitemOpen
  \bibfield  {author} {\bibinfo {author} {\bibfnamefont {L.}~\bibnamefont
  {{Sun}}}, \bibinfo {author} {\bibfnamefont {A.}~\bibnamefont {{Melatos}}},
  \bibinfo {author} {\bibfnamefont {S.}~\bibnamefont {{Suvorova}}}, \bibinfo
  {author} {\bibfnamefont {W.}~\bibnamefont {{Moran}}},\ and\ \bibinfo {author}
  {\bibfnamefont {R.~J.}\ \bibnamefont {{Evans}}},\ }\bibfield  {title}
  {\bibinfo {title} {{Hidden Markov model tracking of continuous gravitational
  waves from young supernova remnants}},\ }\href
  {https://doi.org/10.1103/PhysRevD.97.043013} {\bibfield  {journal} {\bibinfo
  {journal} {Physical Review D}\ }\textbf {\bibinfo {volume} {97}},\ \bibinfo
  {eid} {043013} (\bibinfo {year} {2018})},\ \Eprint
  {https://arxiv.org/abs/1710.00460} {arXiv:1710.00460 [astro-ph.IM]}
  \BibitemShut {NoStop}%
\bibitem [{\citenamefont {{Konar}}\ and\ \citenamefont
  {{Bhattacharya}}(1997)}]{KonarBhattacharya1997}%
  \BibitemOpen
  \bibfield  {author} {\bibinfo {author} {\bibfnamefont {S.}~\bibnamefont
  {{Konar}}}\ and\ \bibinfo {author} {\bibfnamefont {D.}~\bibnamefont
  {{Bhattacharya}}},\ }\bibfield  {title} {\bibinfo {title} {{Magnetic field
  evolution of accreting neutron stars}},\ }\href
  {https://doi.org/10.1093/mnras/284.2.311} {\bibfield  {journal} {\bibinfo
  {journal} {Monthly Notices of the Royal Astronomical Society}\ }\textbf
  {\bibinfo {volume} {284}},\ \bibinfo {pages} {311} (\bibinfo {year}
  {1997})}\BibitemShut {NoStop}%
\bibitem [{\citenamefont {{Melatos}}\ and\ \citenamefont
  {{Phinney}}(2001)}]{MelatosPhinney2001}%
  \BibitemOpen
  \bibfield  {author} {\bibinfo {author} {\bibfnamefont {A.}~\bibnamefont
  {{Melatos}}}\ and\ \bibinfo {author} {\bibfnamefont {E.~S.}\ \bibnamefont
  {{Phinney}}},\ }\bibfield  {title} {\bibinfo {title} {{Hydromagnetic
  Structure of a Neutron Star Accreting at Its Polar Caps}},\ }\href
  {https://doi.org/10.1071/AS01056} {\bibfield  {journal} {\bibinfo  {journal}
  {Publications of the Astronomical Society of Australia}\ }\textbf {\bibinfo
  {volume} {18}},\ \bibinfo {pages} {421} (\bibinfo {year} {2001})}\BibitemShut
  {NoStop}%
\bibitem [{\citenamefont {{Priymak}}\ \emph {et~al.}(2011)\citenamefont
  {{Priymak}}, \citenamefont {{Melatos}},\ and\ \citenamefont
  {{Payne}}}]{PriymakMelatos2011}%
  \BibitemOpen
  \bibfield  {author} {\bibinfo {author} {\bibfnamefont {M.}~\bibnamefont
  {{Priymak}}}, \bibinfo {author} {\bibfnamefont {A.}~\bibnamefont
  {{Melatos}}},\ and\ \bibinfo {author} {\bibfnamefont {D.~J.~B.}\ \bibnamefont
  {{Payne}}},\ }\bibfield  {title} {\bibinfo {title} {{Quadrupole moment of a
  magnetically confined mountain on an accreting neutron star: effect of the
  equation of state}},\ }\href
  {https://doi.org/10.1111/j.1365-2966.2011.19431.x} {\bibfield  {journal}
  {\bibinfo  {journal} {Monthly Notices of the Royal Astronomical Society}\
  }\textbf {\bibinfo {volume} {417}},\ \bibinfo {pages} {2696} (\bibinfo {year}
  {2011})},\ \Eprint {https://arxiv.org/abs/1109.1040} {arXiv:1109.1040
  [astro-ph.HE]} \BibitemShut {NoStop}%
\bibitem [{\citenamefont {{Mukherjee}}(2017)}]{Mukherjee2017}%
  \BibitemOpen
  \bibfield  {author} {\bibinfo {author} {\bibfnamefont {D.}~\bibnamefont
  {{Mukherjee}}},\ }\bibfield  {title} {\bibinfo {title} {{Revisiting Field
  Burial by Accretion onto Neutron Stars}},\ }\href
  {https://doi.org/10.1007/s12036-017-9465-6} {\bibfield  {journal} {\bibinfo
  {journal} {Journal of Astrophysics and Astronomy}\ }\textbf {\bibinfo
  {volume} {38}},\ \bibinfo {eid} {48} (\bibinfo {year} {2017})},\ \Eprint
  {https://arxiv.org/abs/1709.07332} {arXiv:1709.07332 [astro-ph.HE]}
  \BibitemShut {NoStop}%
\bibitem [{\citenamefont {{Palomba}}(2005)}]{Palomba2005}%
  \BibitemOpen
  \bibfield  {author} {\bibinfo {author} {\bibfnamefont {C.}~\bibnamefont
  {{Palomba}}},\ }\bibfield  {title} {\bibinfo {title} {{Simulation of a
  population of isolated neutron stars evolving through the emission of
  gravitational waves}},\ }\href
  {https://doi.org/10.1111/j.1365-2966.2005.08975.x} {\bibfield  {journal}
  {\bibinfo  {journal} {Monthly Notices of the Royal Astronomical Society}\
  }\textbf {\bibinfo {volume} {359}},\ \bibinfo {pages} {1150} (\bibinfo {year}
  {2005})},\ \Eprint {https://arxiv.org/abs/astro-ph/0503046}
  {arXiv:astro-ph/0503046 [astro-ph]} \BibitemShut {NoStop}%
\bibitem [{\citenamefont {{Prix}}(2007{\natexlab{a}})}]{Prix2007a}%
  \BibitemOpen
  \bibfield  {author} {\bibinfo {author} {\bibfnamefont {R.}~\bibnamefont
  {{Prix}}},\ }\bibfield  {title} {\bibinfo {title} {{Template-based searches
  for gravitational waves: efficient lattice covering of flat parameter
  spaces}},\ }\href {https://doi.org/10.1088/0264-9381/24/19/S11} {\bibfield
  {journal} {\bibinfo  {journal} {Classical and Quantum Gravity}\ }\textbf
  {\bibinfo {volume} {24}},\ \bibinfo {pages} {S481} (\bibinfo {year}
  {2007}{\natexlab{a}})},\ \Eprint {https://arxiv.org/abs/0707.0428}
  {arXiv:0707.0428 [gr-qc]} \BibitemShut {NoStop}%
\bibitem [{\citenamefont {{Prix}}(2007{\natexlab{b}})}]{Prix2007b}%
  \BibitemOpen
  \bibfield  {author} {\bibinfo {author} {\bibfnamefont {R.}~\bibnamefont
  {{Prix}}},\ }\bibfield  {title} {\bibinfo {title} {{Search for continuous
  gravitational waves: Metric of the multidetector F-statistic}},\ }\href
  {https://doi.org/10.1103/PhysRevD.75.023004} {\bibfield  {journal} {\bibinfo
  {journal} {Physical Review D}\ }\textbf {\bibinfo {volume} {75}},\ \bibinfo
  {eid} {023004} (\bibinfo {year} {2007}{\natexlab{b}})},\ \Eprint
  {https://arxiv.org/abs/gr-qc/0606088} {arXiv:gr-qc/0606088 [gr-qc]}
  \BibitemShut {NoStop}%
\bibitem [{\citenamefont {{Pletsch}}(2010)}]{Pletsch2010}%
  \BibitemOpen
  \bibfield  {author} {\bibinfo {author} {\bibfnamefont {H.~J.}\ \bibnamefont
  {{Pletsch}}},\ }\bibfield  {title} {\bibinfo {title} {{Parameter-space metric
  of semicoherent searches for continuous gravitational waves}},\ }\href
  {https://doi.org/10.1103/PhysRevD.82.042002} {\bibfield  {journal} {\bibinfo
  {journal} {Physical Review D}\ }\textbf {\bibinfo {volume} {82}},\ \bibinfo
  {eid} {042002} (\bibinfo {year} {2010})},\ \Eprint
  {https://arxiv.org/abs/1005.0395} {arXiv:1005.0395 [gr-qc]} \BibitemShut
  {NoStop}%
\bibitem [{\citenamefont {{Wette}}(2015)}]{Wette2015}%
  \BibitemOpen
  \bibfield  {author} {\bibinfo {author} {\bibfnamefont {K.}~\bibnamefont
  {{Wette}}},\ }\bibfield  {title} {\bibinfo {title} {{Parameter-space metric
  for all-sky semicoherent searches for gravitational-wave pulsars}},\ }\href
  {https://doi.org/10.1103/PhysRevD.92.082003} {\bibfield  {journal} {\bibinfo
  {journal} {Physical Review D}\ }\textbf {\bibinfo {volume} {92}},\ \bibinfo
  {eid} {082003} (\bibinfo {year} {2015})},\ \Eprint
  {https://arxiv.org/abs/1508.02372} {arXiv:1508.02372 [gr-qc]} \BibitemShut
  {NoStop}%
\bibitem [{\citenamefont {{Newville}}\ \emph {et~al.}(2016)\citenamefont
  {{Newville}}, \citenamefont {{Stensitzki}}, \citenamefont {{Allen}},
  \citenamefont {{Rawlik}}, \citenamefont {{Ingargiola}},\ and\ \citenamefont
  {{Nelson}}}]{NewvilleStensitzki2016}%
  \BibitemOpen
  \bibfield  {author} {\bibinfo {author} {\bibfnamefont {M.}~\bibnamefont
  {{Newville}}}, \bibinfo {author} {\bibfnamefont {T.}~\bibnamefont
  {{Stensitzki}}}, \bibinfo {author} {\bibfnamefont {D.~B.}\ \bibnamefont
  {{Allen}}}, \bibinfo {author} {\bibfnamefont {M.}~\bibnamefont {{Rawlik}}},
  \bibinfo {author} {\bibfnamefont {A.}~\bibnamefont {{Ingargiola}}},\ and\
  \bibinfo {author} {\bibfnamefont {A.}~\bibnamefont {{Nelson}}},\ }\href@noop
  {} {\bibinfo {title} {{Lmfit: Non-Linear Least-Square Minimization and
  Curve-Fitting for Python}}},\ \bibinfo {howpublished} {Astrophysics Source
  Code Library, record ascl:1606.014} (\bibinfo {year} {2016})\BibitemShut
  {NoStop}%
\bibitem [{\citenamefont {{Tiwari}}\ \emph {et~al.}(2015)\citenamefont
  {{Tiwari}}, \citenamefont {{Drago}}, \citenamefont {{Frolov}}, \citenamefont
  {{Klimenko}}, \citenamefont {{Mitselmakher}}, \citenamefont {{Necula}},
  \citenamefont {{Prodi}}, \citenamefont {{Re}}, \citenamefont {{Salemi}},
  \citenamefont {{Vedovato}},\ and\ \citenamefont
  {{Yakushin}}}]{TiwariDrago2015}%
  \BibitemOpen
  \bibfield  {author} {\bibinfo {author} {\bibfnamefont {V.}~\bibnamefont
  {{Tiwari}}}, \bibinfo {author} {\bibfnamefont {M.}~\bibnamefont {{Drago}}},
  \bibinfo {author} {\bibfnamefont {V.}~\bibnamefont {{Frolov}}}, \bibinfo
  {author} {\bibfnamefont {S.}~\bibnamefont {{Klimenko}}}, \bibinfo {author}
  {\bibfnamefont {G.}~\bibnamefont {{Mitselmakher}}}, \bibinfo {author}
  {\bibfnamefont {V.}~\bibnamefont {{Necula}}}, \bibinfo {author}
  {\bibfnamefont {G.}~\bibnamefont {{Prodi}}}, \bibinfo {author} {\bibfnamefont
  {V.}~\bibnamefont {{Re}}}, \bibinfo {author} {\bibfnamefont {F.}~\bibnamefont
  {{Salemi}}}, \bibinfo {author} {\bibfnamefont {G.}~\bibnamefont
  {{Vedovato}}},\ and\ \bibinfo {author} {\bibfnamefont {I.}~\bibnamefont
  {{Yakushin}}},\ }\bibfield  {title} {\bibinfo {title} {{Regression of
  environmental noise in LIGO data}},\ }\href
  {https://doi.org/10.1088/0264-9381/32/16/165014} {\bibfield  {journal}
  {\bibinfo  {journal} {Classical and Quantum Gravity}\ }\textbf {\bibinfo
  {volume} {32}},\ \bibinfo {eid} {165014} (\bibinfo {year} {2015})},\ \Eprint
  {https://arxiv.org/abs/1503.07476} {arXiv:1503.07476 [gr-qc]} \BibitemShut
  {NoStop}%
\bibitem [{\citenamefont {{Davis}}\ \emph {et~al.}(2019)\citenamefont
  {{Davis}}, \citenamefont {{Massinger}}, \citenamefont {{Lundgren}},
  \citenamefont {{Driggers}}, \citenamefont {{Urban}},\ and\ \citenamefont
  {{Nuttall}}}]{DavisMassinger2019}%
  \BibitemOpen
  \bibfield  {author} {\bibinfo {author} {\bibfnamefont {D.}~\bibnamefont
  {{Davis}}}, \bibinfo {author} {\bibfnamefont {T.}~\bibnamefont
  {{Massinger}}}, \bibinfo {author} {\bibfnamefont {A.}~\bibnamefont
  {{Lundgren}}}, \bibinfo {author} {\bibfnamefont {J.~C.}\ \bibnamefont
  {{Driggers}}}, \bibinfo {author} {\bibfnamefont {A.~L.}\ \bibnamefont
  {{Urban}}},\ and\ \bibinfo {author} {\bibfnamefont {L.}~\bibnamefont
  {{Nuttall}}},\ }\bibfield  {title} {\bibinfo {title} {{Improving the
  sensitivity of Advanced LIGO using noise subtraction}},\ }\href
  {https://doi.org/10.1088/1361-6382/ab01c5} {\bibfield  {journal} {\bibinfo
  {journal} {Classical and Quantum Gravity}\ }\textbf {\bibinfo {volume}
  {36}},\ \bibinfo {eid} {055011} (\bibinfo {year} {2019})},\ \Eprint
  {https://arxiv.org/abs/1809.05348} {arXiv:1809.05348 [astro-ph.IM]}
  \BibitemShut {NoStop}%
\bibitem [{\citenamefont {{Driggers}}\ \emph {et~al.}(2019)\citenamefont
  {{Driggers}}, \citenamefont {{Vitale}}, \citenamefont {{Lundgren}},
  \citenamefont {{Evans}}, \citenamefont {{Kawabe}}, \citenamefont {{Dwyer}},
  \citenamefont {{Izumi}}, \citenamefont {{Schofield}}, \citenamefont
  {{Effler}}, \citenamefont {{Sigg}},\ and\ \citenamefont
  {et~al.}}]{DriggersVitale2019}%
  \BibitemOpen
  \bibfield  {author} {\bibinfo {author} {\bibfnamefont {J.~C.}\ \bibnamefont
  {{Driggers}}}, \bibinfo {author} {\bibfnamefont {S.}~\bibnamefont
  {{Vitale}}}, \bibinfo {author} {\bibfnamefont {A.~P.}\ \bibnamefont
  {{Lundgren}}}, \bibinfo {author} {\bibfnamefont {M.}~\bibnamefont {{Evans}}},
  \bibinfo {author} {\bibfnamefont {K.}~\bibnamefont {{Kawabe}}}, \bibinfo
  {author} {\bibfnamefont {S.~E.}\ \bibnamefont {{Dwyer}}}, \bibinfo {author}
  {\bibfnamefont {K.}~\bibnamefont {{Izumi}}}, \bibinfo {author} {\bibfnamefont
  {R.~M.~S.}\ \bibnamefont {{Schofield}}}, \bibinfo {author} {\bibfnamefont
  {A.}~\bibnamefont {{Effler}}}, \bibinfo {author} {\bibfnamefont
  {D.}~\bibnamefont {{Sigg}}},\ and\ \bibinfo {author} {\bibnamefont
  {et~al.}},\ }\bibfield  {title} {\bibinfo {title} {{Improving astrophysical
  parameter estimation via offline noise subtraction for Advanced LIGO}},\
  }\href {https://doi.org/10.1103/PhysRevD.99.042001} {\bibfield  {journal}
  {\bibinfo  {journal} {Physical Review D}\ }\textbf {\bibinfo {volume} {99}},\
  \bibinfo {eid} {042001} (\bibinfo {year} {2019})},\ \Eprint
  {https://arxiv.org/abs/1806.00532} {arXiv:1806.00532 [astro-ph.IM]}
  \BibitemShut {NoStop}%
\bibitem [{\citenamefont {Kimpson}\ \emph {et~al.}(2024)\citenamefont {Kimpson}
  \emph {et~al.}}]{Kimpson2024}%
  \BibitemOpen
  \bibfield  {author} {\bibinfo {author} {\bibfnamefont {T.}~\bibnamefont
  {Kimpson}} \emph {et~al.},\ }\bibfield  {title} {\bibinfo {title} {{Adaptive
  cancellation of mains power interference in continuous gravitational wave
  searches with a hidden Markov model}},\ }\href@noop {} {\bibfield  {journal}
  {\bibinfo  {journal} {submitted}\ } (\bibinfo {year} {2024})}\BibitemShut
  {NoStop}%
\bibitem [{\citenamefont {{Abbott}}\ \emph
  {et~al.}(2021{\natexlab{b}})\citenamefont {{Abbott}}, \citenamefont
  {{Abbott}}, \citenamefont {{Abraham}}, \citenamefont {{Acernese}},
  \citenamefont {{Ackley}}, \citenamefont {{Adams}}, \citenamefont {{Adams}},
  \citenamefont {{Adhikari}}, \citenamefont {{Adya}}, \citenamefont
  {{Affeldt}},\ and\ \citenamefont {et~al.}}]{AbbottAbbott2021b}%
  \BibitemOpen
  \bibfield  {author} {\bibinfo {author} {\bibfnamefont {R.}~\bibnamefont
  {{Abbott}}}, \bibinfo {author} {\bibfnamefont {T.~D.}\ \bibnamefont
  {{Abbott}}}, \bibinfo {author} {\bibfnamefont {S.}~\bibnamefont {{Abraham}}},
  \bibinfo {author} {\bibfnamefont {F.}~\bibnamefont {{Acernese}}}, \bibinfo
  {author} {\bibfnamefont {K.}~\bibnamefont {{Ackley}}}, \bibinfo {author}
  {\bibfnamefont {A.}~\bibnamefont {{Adams}}}, \bibinfo {author} {\bibfnamefont
  {C.}~\bibnamefont {{Adams}}}, \bibinfo {author} {\bibfnamefont {R.~X.}\
  \bibnamefont {{Adhikari}}}, \bibinfo {author} {\bibfnamefont {V.~B.}\
  \bibnamefont {{Adya}}}, \bibinfo {author} {\bibfnamefont {C.}~\bibnamefont
  {{Affeldt}}},\ and\ \bibinfo {author} {\bibnamefont {et~al.}},\ }\bibfield
  {title} {\bibinfo {title} {{All-sky search for continuous gravitational waves
  from isolated neutron stars in the early O3 LIGO data}},\ }\href
  {https://doi.org/10.1103/PhysRevD.104.082004} {\bibfield  {journal} {\bibinfo
   {journal} {Physical Review D}\ }\textbf {\bibinfo {volume} {104}},\ \bibinfo
  {eid} {082004} (\bibinfo {year} {2021}{\natexlab{b}})},\ \Eprint
  {https://arxiv.org/abs/2107.00600} {arXiv:2107.00600 [gr-qc]} \BibitemShut
  {NoStop}%
\bibitem [{\citenamefont {Goetz}\ \emph {et~al.}()\citenamefont {Goetz},
  \citenamefont {Neunzert}, \citenamefont {Riles}, \citenamefont {Matas},
  \citenamefont {Kandhasamy} \emph {et~al.}}]{DCClines}%
  \BibitemOpen
  \bibfield  {author} {\bibinfo {author} {\bibfnamefont {E.}~\bibnamefont
  {Goetz}}, \bibinfo {author} {\bibfnamefont {A.}~\bibnamefont {Neunzert}},
  \bibinfo {author} {\bibfnamefont {K.}~\bibnamefont {Riles}}, \bibinfo
  {author} {\bibfnamefont {A.}~\bibnamefont {Matas}}, \bibinfo {author}
  {\bibfnamefont {S.}~\bibnamefont {Kandhasamy}}, \emph {et~al.},\ }\href
  {https://dcc.ligo.org/T2100200/public} {\emph {\bibinfo {title} {{O3 lines
  and combs found in self-gated C01 data}}}},\ \bibinfo {type} {LIGO Document}\
  \bibinfo {number} {T2100200}\BibitemShut {NoStop}%
\bibitem [{\citenamefont {{Jones}}\ \emph {et~al.}(2022)\citenamefont
  {{Jones}}, \citenamefont {{Sun}}, \citenamefont {{Carlin}}, \citenamefont
  {{Dunn}}, \citenamefont {{Millhouse}}, \citenamefont {{Middleton}},
  \citenamefont {{Meyers}}, \citenamefont {{Clearwater}}, \citenamefont
  {{Beniwal}}, \citenamefont {{Strang}}, \citenamefont {{Vargas}},\ and\
  \citenamefont {{Melatos}}}]{JonesSun2022}%
  \BibitemOpen
  \bibfield  {author} {\bibinfo {author} {\bibfnamefont {D.}~\bibnamefont
  {{Jones}}}, \bibinfo {author} {\bibfnamefont {L.}~\bibnamefont {{Sun}}},
  \bibinfo {author} {\bibfnamefont {J.}~\bibnamefont {{Carlin}}}, \bibinfo
  {author} {\bibfnamefont {L.}~\bibnamefont {{Dunn}}}, \bibinfo {author}
  {\bibfnamefont {M.}~\bibnamefont {{Millhouse}}}, \bibinfo {author}
  {\bibfnamefont {H.}~\bibnamefont {{Middleton}}}, \bibinfo {author}
  {\bibfnamefont {P.}~\bibnamefont {{Meyers}}}, \bibinfo {author}
  {\bibfnamefont {P.}~\bibnamefont {{Clearwater}}}, \bibinfo {author}
  {\bibfnamefont {D.}~\bibnamefont {{Beniwal}}}, \bibinfo {author}
  {\bibfnamefont {L.}~\bibnamefont {{Strang}}}, \bibinfo {author}
  {\bibfnamefont {A.}~\bibnamefont {{Vargas}}},\ and\ \bibinfo {author}
  {\bibfnamefont {A.}~\bibnamefont {{Melatos}}},\ }\bibfield  {title} {\bibinfo
  {title} {{Validating continuous gravitational-wave candidates from a
  semicoherent search using Doppler modulation and an effective point spread
  function}},\ }\href {https://doi.org/10.1103/PhysRevD.106.123011} {\bibfield
  {journal} {\bibinfo  {journal} {Physical Review D}\ }\textbf {\bibinfo
  {volume} {106}},\ \bibinfo {eid} {123011} (\bibinfo {year} {2022})},\ \Eprint
  {https://arxiv.org/abs/2203.14468} {arXiv:2203.14468 [gr-qc]} \BibitemShut
  {NoStop}%
\bibitem [{\citenamefont {{Jaume}}\ \emph {et~al.}(2024)\citenamefont
  {{Jaume}}, \citenamefont {{Tenorio}},\ and\ \citenamefont
  {{Sintes}}}]{JaumeTenorio2024}%
  \BibitemOpen
  \bibfield  {author} {\bibinfo {author} {\bibfnamefont {R.}~\bibnamefont
  {{Jaume}}}, \bibinfo {author} {\bibfnamefont {R.}~\bibnamefont {{Tenorio}}},\
  and\ \bibinfo {author} {\bibfnamefont {A.~M.}\ \bibnamefont {{Sintes}}},\
  }\bibfield  {title} {\bibinfo {title} {{Assessing the Similarity of
  Continuous Gravitational-Wave Signals to Narrow Instrumental Artifacts}},\
  }\href {https://doi.org/10.3390/universe10030121} {\bibfield  {journal}
  {\bibinfo  {journal} {Universe}\ }\textbf {\bibinfo {volume} {10}},\ \bibinfo
  {eid} {121} (\bibinfo {year} {2024})},\ \Eprint
  {https://arxiv.org/abs/2403.03027} {arXiv:2403.03027 [gr-qc]} \BibitemShut
  {NoStop}%
\bibitem [{\citenamefont {{Johnson-McDaniel}}\ and\ \citenamefont
  {{Owen}}(2013)}]{Johnson-McDanielOwen2013}%
  \BibitemOpen
  \bibfield  {author} {\bibinfo {author} {\bibfnamefont {N.~K.}\ \bibnamefont
  {{Johnson-McDaniel}}}\ and\ \bibinfo {author} {\bibfnamefont {B.~J.}\
  \bibnamefont {{Owen}}},\ }\bibfield  {title} {\bibinfo {title} {{Maximum
  elastic deformations of relativistic stars}},\ }\href
  {https://doi.org/10.1103/PhysRevD.88.044004} {\bibfield  {journal} {\bibinfo
  {journal} {Physical Review D}\ }\textbf {\bibinfo {volume} {88}},\ \bibinfo
  {eid} {044004} (\bibinfo {year} {2013})},\ \Eprint
  {https://arxiv.org/abs/1208.5227} {arXiv:1208.5227 [astro-ph.SR]}
  \BibitemShut {NoStop}%
\bibitem [{\citenamefont {{Morales}}\ and\ \citenamefont
  {{Horowitz}}(2022)}]{MoralesHorowitz2022}%
  \BibitemOpen
  \bibfield  {author} {\bibinfo {author} {\bibfnamefont {J.~A.}\ \bibnamefont
  {{Morales}}}\ and\ \bibinfo {author} {\bibfnamefont {C.~J.}\ \bibnamefont
  {{Horowitz}}},\ }\bibfield  {title} {\bibinfo {title} {{Neutron star crust
  can support a large ellipticity}},\ }\href
  {https://doi.org/10.1093/mnras/stac3058} {\bibfield  {journal} {\bibinfo
  {journal} {Monthly Notices of the Royal Astronomical Society}\ }\textbf
  {\bibinfo {volume} {517}},\ \bibinfo {pages} {5610} (\bibinfo {year}
  {2022})},\ \Eprint {https://arxiv.org/abs/2209.03222} {arXiv:2209.03222
  [gr-qc]} \BibitemShut {NoStop}%
\end{thebibliography}%

\appendix
\section{Transition probabilities}
\label{apdx:trans_fpe_calc}
The transition probabilities used in this work are essentially the same as those derived by \citet{MelatosClearwater2021}.
They are are derived from the system of equations (\ref{eqn:trans_f_evo}) and (\ref{eqn:trans_phi_evo}) except that we discard the damping term $-\gamma f$ included in Eq. (\ref{eqn:trans_f_evo}) by \citet{MelatosClearwater2021}.
In this appendix we reproduce the form for $A^{\mathrm{b}}(q_i, q_j) = \Pr[q(t_n) = q_i \mid q(t_{n+1}) = q_j] = p^{\mathrm{B}}(t_n, q_i)$ which appears in \citet{MelatosClearwater2021}, and subsequently justify the choice to drop the $-\gamma f$ term.
We then write down the transition probabilities in the $\gamma \to 0$ limit, which we adopt in the search.

With $\Delta q = (\Delta f, \Delta \Phi)$ specifying the difference between the final and initial states, the distribution $p^{\mathrm{B}}(t_n, \Delta q)$ is a wrapped Gaussian \cite{SuvorovaMelatos2018,MelatosClearwater2021}, \begin{align} p^{\mathrm{B}}(t_n, \Delta q) = &(2\pi)^{-1}(\det\mathbf{\Sigma})^{-1/2} \nonumber \\&\times\sum_{m=-\infty}^{\infty}\exp\left[-(\Delta q-Q_m)\mathbf{\Sigma}^{-1}(\Delta q-Q_m)^{\mathrm{T}}\right], \label{eqn:backwards_trans_probs} \end{align}
where we have
\begin{align}
(Q_m)_1 &= 0 \\
(Q_m)_2 &= -m, \\
\mathbf{\Sigma}_{11} &= \frac{\sigma^2}{2\gamma}\left[1 - \exp(-2\gamma \tau)\right], \label{eqn:apdx_sigma_11} \\
\mathbf{\Sigma}_{12} = \mathbf{\Sigma}_{21} &= \frac{\sigma^2}{2\gamma^2}\left[1-\exp(-\gamma \tau)\right]^2, \\
\mathbf{\Sigma}_{22} &= \frac{\sigma^2}{2\gamma^3}\left\{1 + 2\gamma\tau - \left[2 - \exp(-\gamma\tau)\right]^2\right\} \label{eqn:apdx_sigma_22},
\end{align}
with $\tau = t_{n+1} - t_n = T_{\text{coh}}$.
The condition $\gamma < (2fT_{\text{coh}}^2)^{-1}$ in Ref. \cite{MelatosClearwater2021} ensures that the argument of the exponent, $\gamma\tau \sim (2fT_{\text{coh}})^{-1}$, is $O(10^{-8})$.

Taylor expanding the exponentials in Eqs. (\ref{eqn:apdx_sigma_11})--(\ref{eqn:apdx_sigma_22}) about $\gamma\tau = 0$ allows us to write these expressions as a prefactor multiplied by a power series in $\gamma\tau$ with a constant term that is $O(1)$, e.g. \begin{equation} \mathbf{\Sigma}_{12} = \frac{\sigma^2\tau^2}{2}\left(1 - \gamma\tau + \ldots\right). \label{eqn:trans_gamma_small_eg} \end{equation}
Given $\gamma\tau \sim 10^{-8}$ we conclude by inspecting (\ref{eqn:trans_gamma_small_eg}) and its counterparts that the effect of the damping term on the moments of the transition probability distribution is negligible in this case.
We therefore take the $\gamma \to 0$ limit and write
\begin{align}
    \mathbf{\Sigma}_{11} &= \sigma^2\tau, \label{eqn:trans_sigma_11}\\
    \mathbf{\Sigma}_{12} = \mathbf{\Sigma}_{21} &= \frac{\sigma^2\tau^2}{2}, \\
    \mathbf{\Sigma}_{22} &= \frac{\sigma^2\tau^3}{2}. \label{eqn:trans_sigma_22}
\end{align} 
The transition probabilities are obtained by substituting (\ref{eqn:trans_sigma_11})--(\ref{eqn:trans_sigma_22}) into (\ref{eqn:backwards_trans_probs}).

\section{Detection thresholds}
\label{apdx:thresholds}
In order to reduce the number of candidates from a search to a manageable level, we set thresholds on the log likelihood of the paths returned from the search and do not retain any paths which fall below the threshold for followup analysis.
The thresholds are set to target a desired false alarm probability $P_{\text{fa}}$, i.e. one in every $P_{\text{fa}}^{-1}$ paths has $\mathcal{L} > \mathcal{L}_{\text{th}}$ in sub-bands where the data are not significantly polluted by non-Gaussian features.

Since in general we want $P_{\text{fa}}$ to be small, setting $\mathcal{L}_{\text{th}}$ requires knowledge of the behaviour of the distribution of path log likelihoods in the tail.
In this work we assume that the tail of the $\mathcal{L}$ distribution is exponential, as confirmed empirically in previous work \cite{AbbottAbbott2022c, KneeDu2024}.
We use off-target searches in the real data to infer the parameters of the exponential distribution.
We follow the method described in Appendix A 1 of \citet{AbbottAbbott2022c}, which we briefly recapitulate here.

We assume that the likelihood distribution $p(\mathcal{L})$ has the form \begin{align} p(\mathcal{L}) = A\lambda\exp[-\lambda(\mathcal{L}-\mathcal{L}_{\text{tail}})]&& \text{for } \mathcal{L} > \mathcal{L}_{\text{tail}}, \end{align} where $A$ and $\lambda$ are constants and $\mathcal{L}_{\text{tail}}$ is an empirically determined cut-off above which the distribution is approximately exponential.
We have a sample of $N$ paths, of which $N_{\text{tail}}$ have $\mathcal{L} > \mathcal{L}_{\text{tail}}$.
The sample is generated by searching random sky positions and $\dot{f}$ values in $0.5\,\mathrm{Hz}$ sub-bands that appear free from disturbance based on the detector noise ASDs.
Then $A$ is simply $N_{\text{tail}}/N$, and the maximum-likelihood estimate of $\lambda$ is \begin{equation} \hat{\lambda} = \frac{N_{\text{tail}}}{\sum_{i=1}^{N_{\text{tail}}}\mathcal{L}_i - \mathcal{L}_{\text{tail}}}.\end{equation} 
For a desired $P_{\text{fa}}$ we then estimate $\mathcal{L}_{\text{th}}$ to be \begin{equation} \mathcal{L}_{\text{th}}(P_{\text{fa}}) = -\frac{1}{\hat{\lambda}}\ln\left(\frac{N P_{\text{fa}}}{N_{\text{tail}}}\right) + \mathcal{L}_{\text{tail}}.\end{equation}

\section{Single-interferometer veto}
\label{apdx:1ifo_veto}
This appendix collects the plots of the single-detector log likelihood values $\mathcal{L}_X$ in the vicinity of the ten candidate groups which pass through the single-interferometer veto described in Section \ref{subsubsec:veto_single_ifo}.
The plots are displayed in Figure \ref{fig:single_ifo}.
The log likelihoods are computed using the template of the loudest candidate in each group.
The frequency of each candidate is indicated by a vertical dashed line, and the joint-detector log likelihood $\mathcal{L}_\cup$ is indicated by a horizontal dashed line.
A candidate group is vetoed if $\mathcal{L}_X > \mathcal{L}_\cup$ for either detector.
Nine of the ten candidate groups are vetoed this way.

\begin{figure*}
    \centering
    \includegraphics[width=0.9\columnwidth]{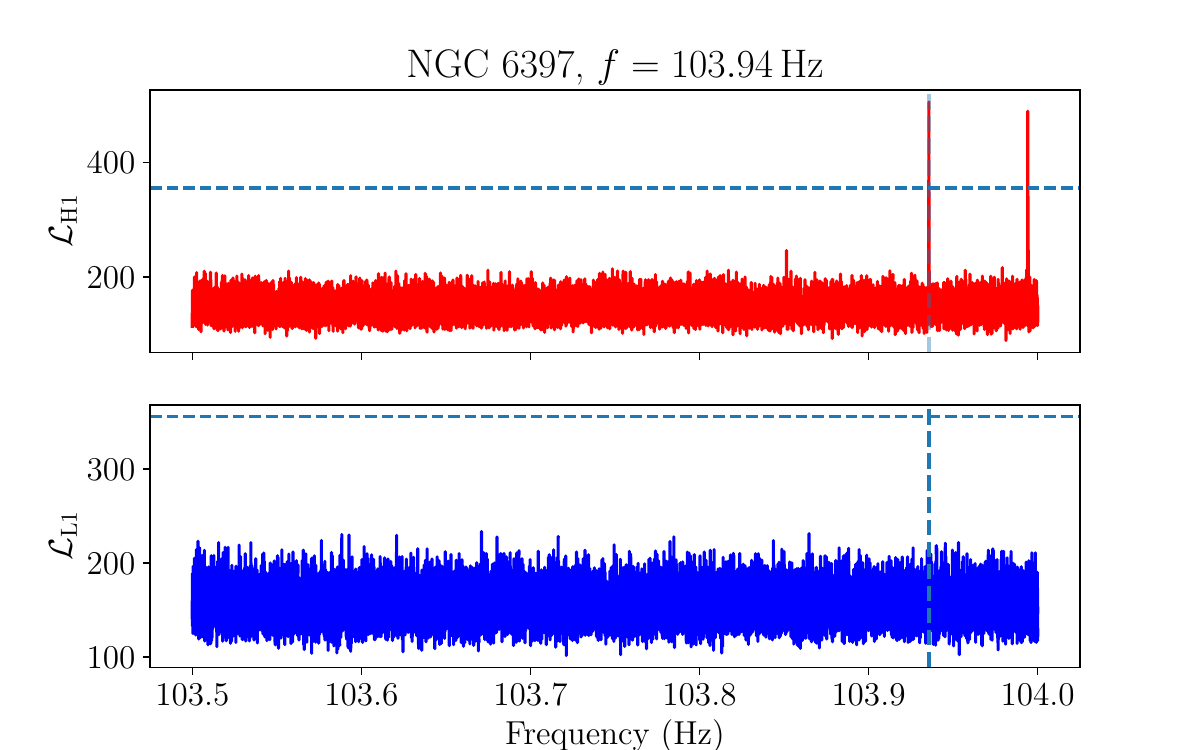}
    \includegraphics[width=0.9\columnwidth]{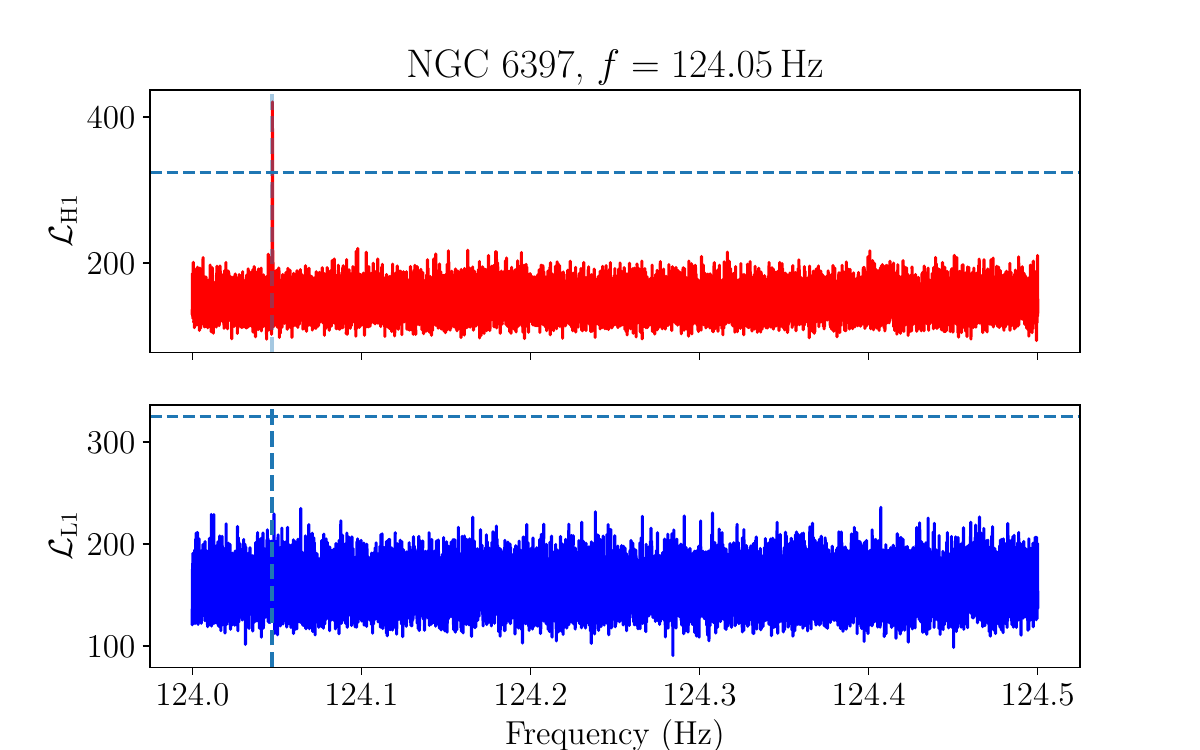}\\
    \includegraphics[width=0.9\columnwidth]{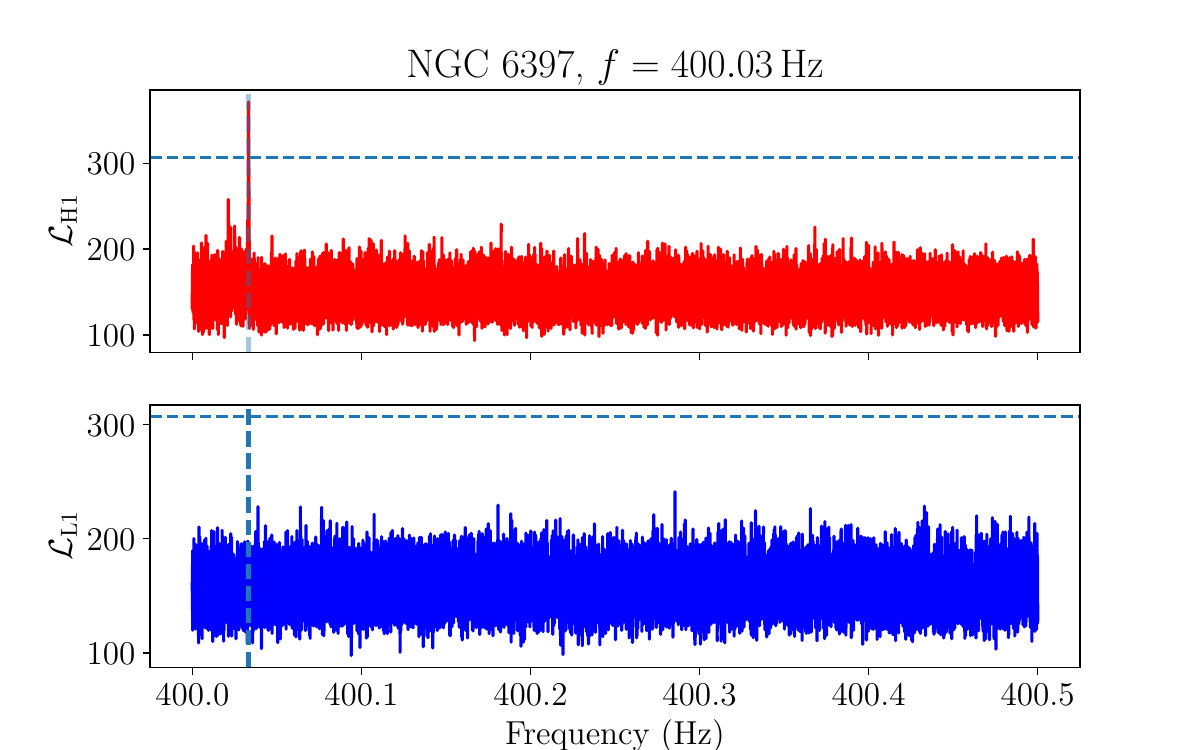}
    \includegraphics[width=0.9\columnwidth]{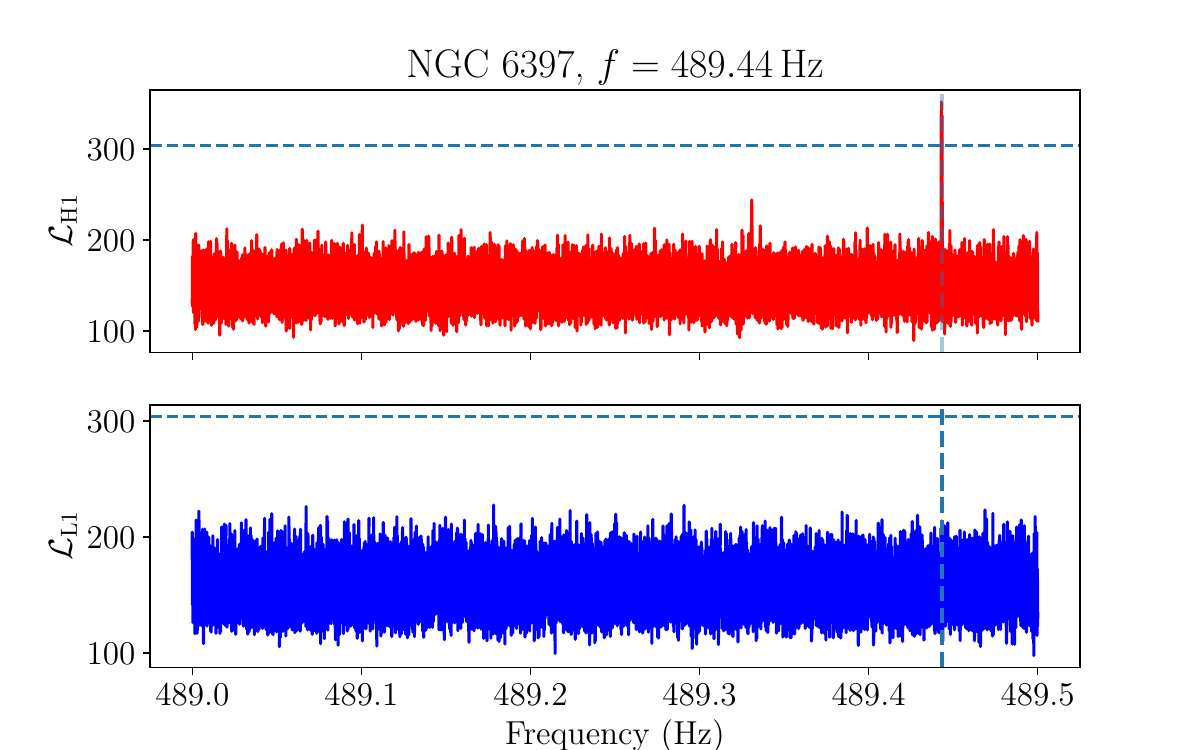}\\
    \includegraphics[width=0.9\columnwidth]{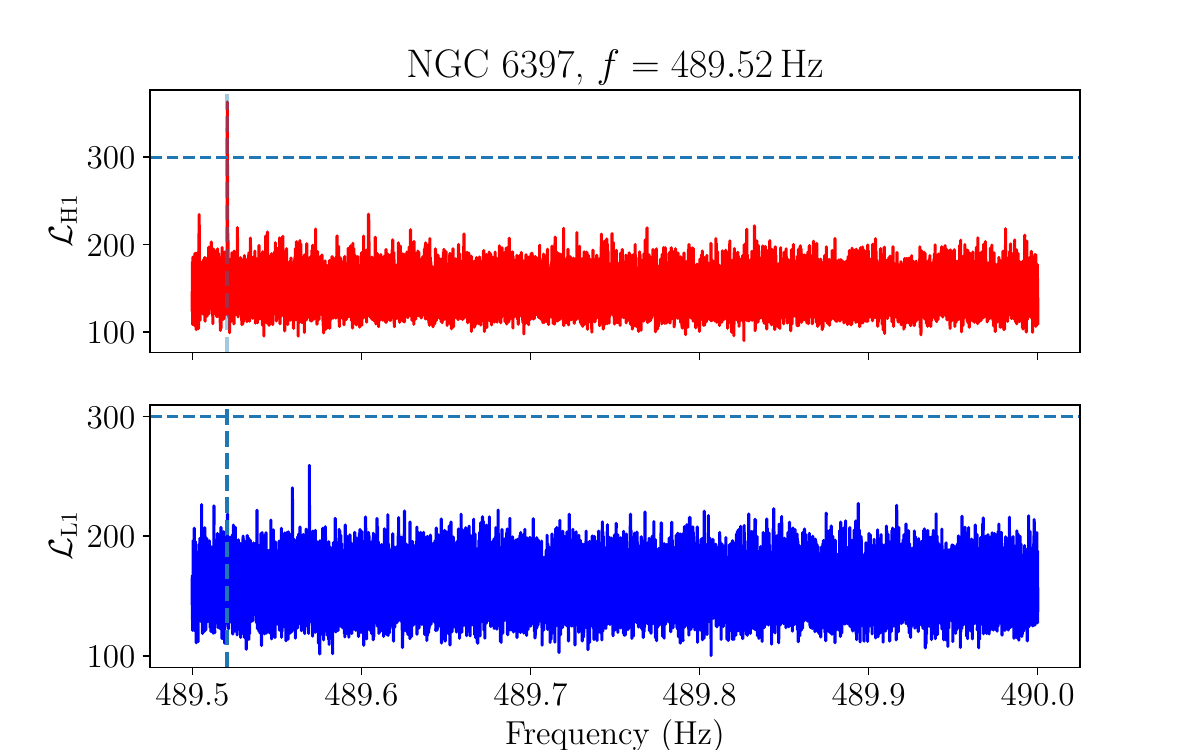}
    \includegraphics[width=0.9\columnwidth]{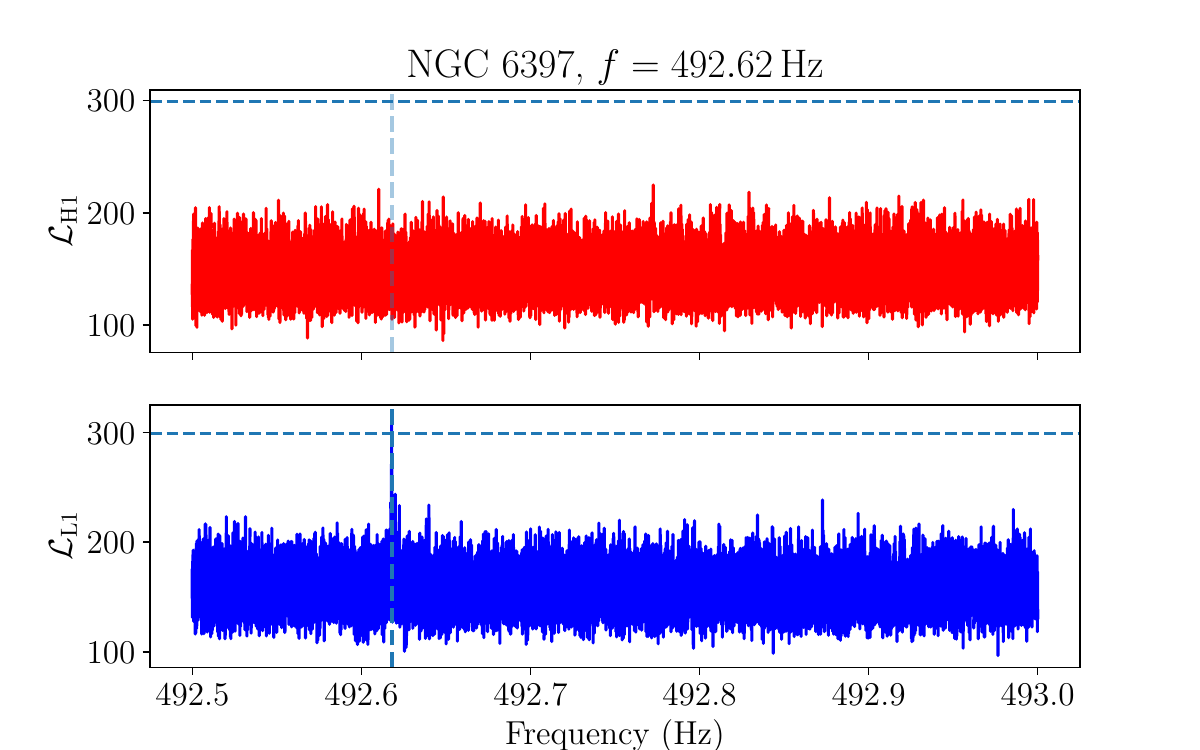}\\
    \includegraphics[width=0.9\columnwidth]{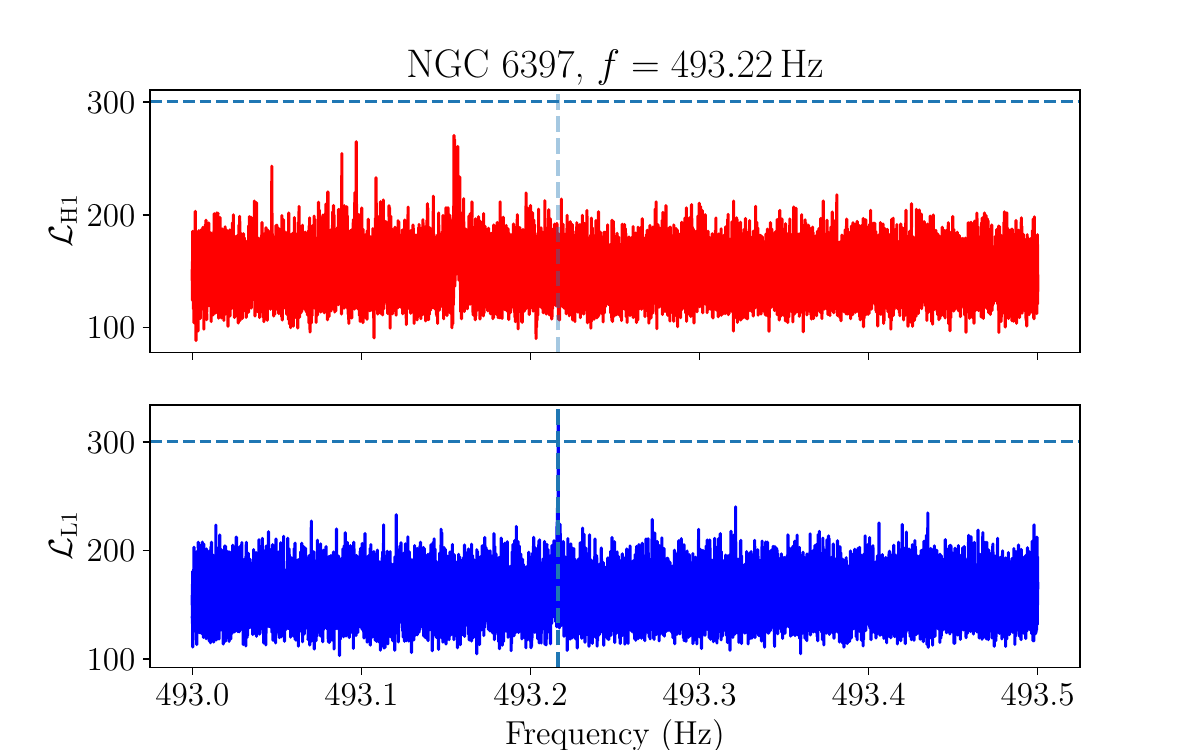}
    \includegraphics[width=0.9\columnwidth]{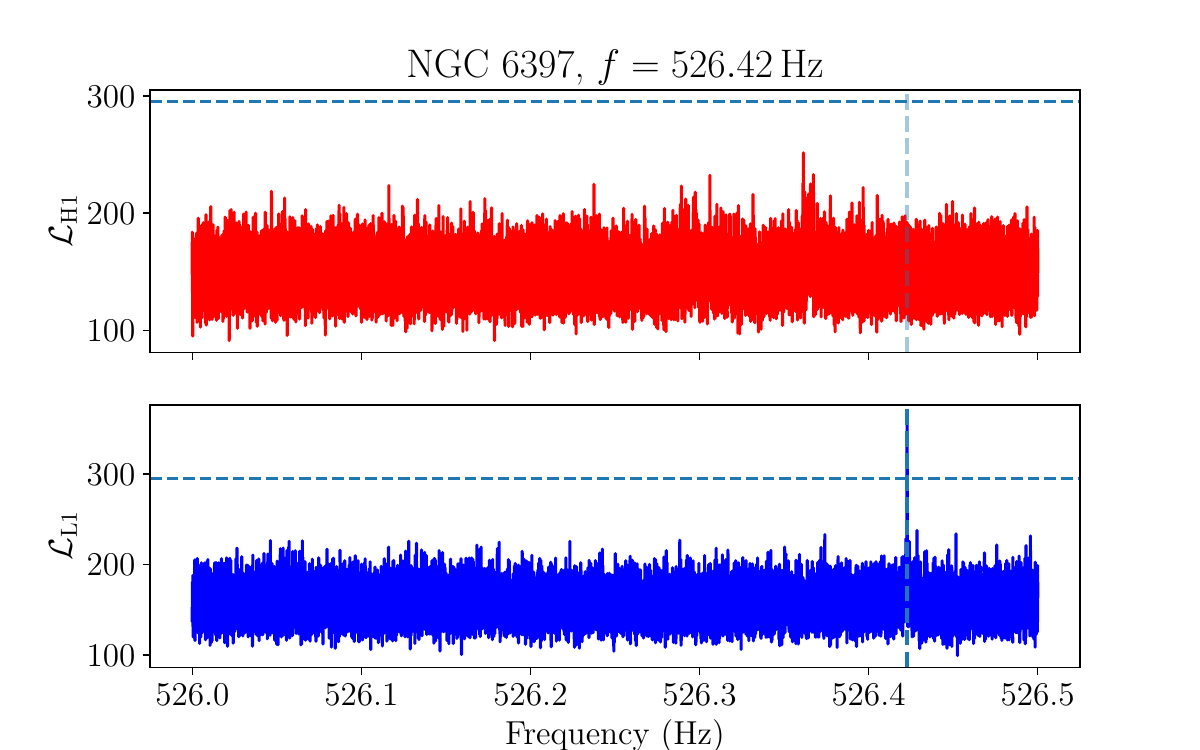}\\
    \caption{Single-detector log likelihoods $\mathcal{L}_X$ versus frequency for the ten candidate groups surviving the cross-cluster veto. The vertical dashed lines indicate the frequency of the loudest candidate in each group. The horizontal dashed lines indicate the loudest joint-detector log likelihood in each group, $\mathcal{L}_\cup$. Nine of the ten candidate groups satisfy $\mathcal{L}_X > \mathcal{L}_\cup$ and are therefore vetoed. The cluster name and approximate terminal frequency of the loudest candidate in each group are recorded above each panel.}
    \label{fig:single_ifo}
    \end{figure*}
\begin{figure*}\ContinuedFloat

    \includegraphics[width=0.9\columnwidth]{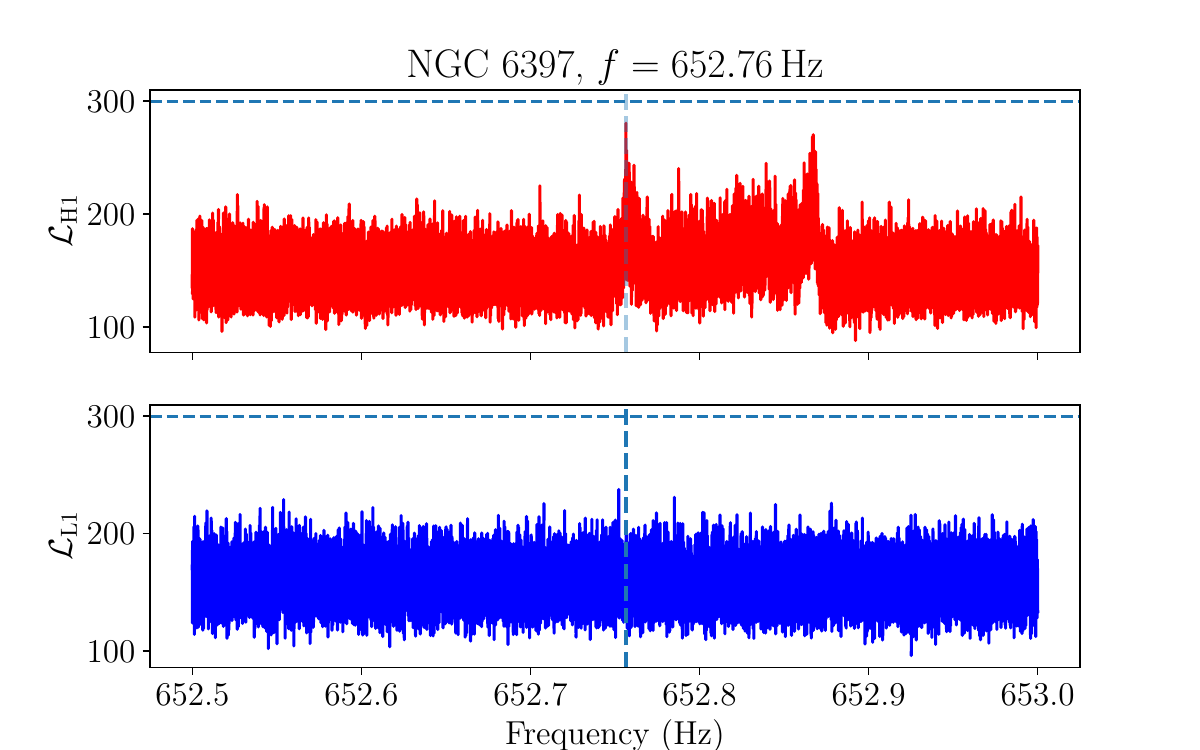}
    \includegraphics[width=0.9\columnwidth]{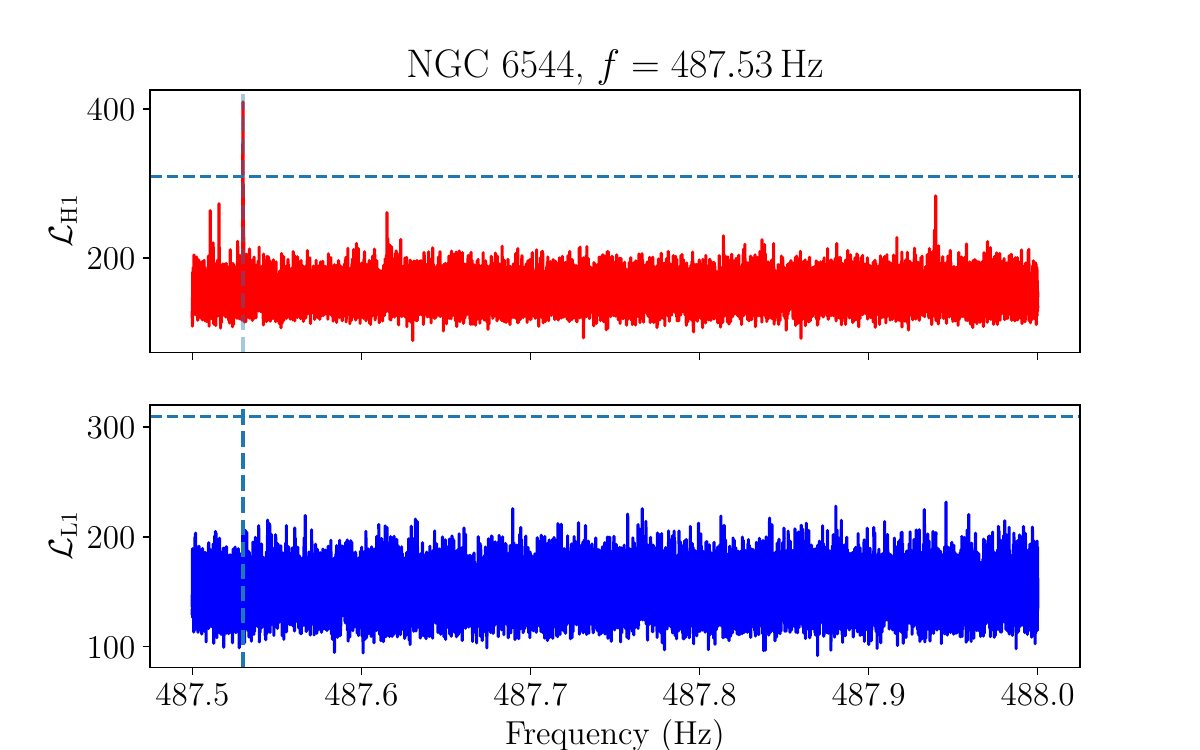}
    \caption{Single-detector log likelihoods $\mathcal{L}_X$ versus frequency for the ten candidate groups surviving the cross-cluster veto. The vertical dashed lines indicate the frequency of the loudest candidate in each group. The horizontal dashed lines indicate the loudest joint-detector log likelihood in each group, $\mathcal{L}_\cup$. Nine of the ten candidate groups satisfy $\mathcal{L}_X > \mathcal{L}_\cup$ and are therefore vetoed. The cluster name and approximate eding frequency of the loudest candidate in each group are recorded above each panel. (cont.)}
\end{figure*}

\section{Unknown lines}
\label{apdx:unknown_lines}
\begin{figure*}
    \centering
    \includegraphics[width=0.9\columnwidth]{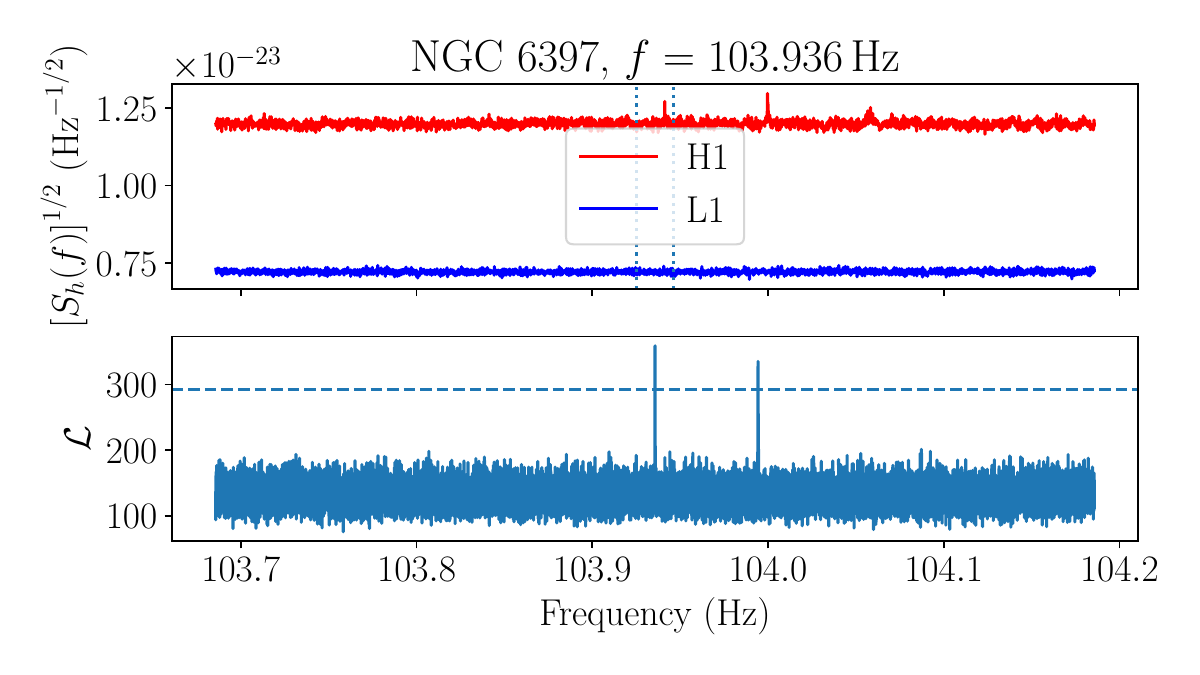}
    \includegraphics[width=0.9\columnwidth]{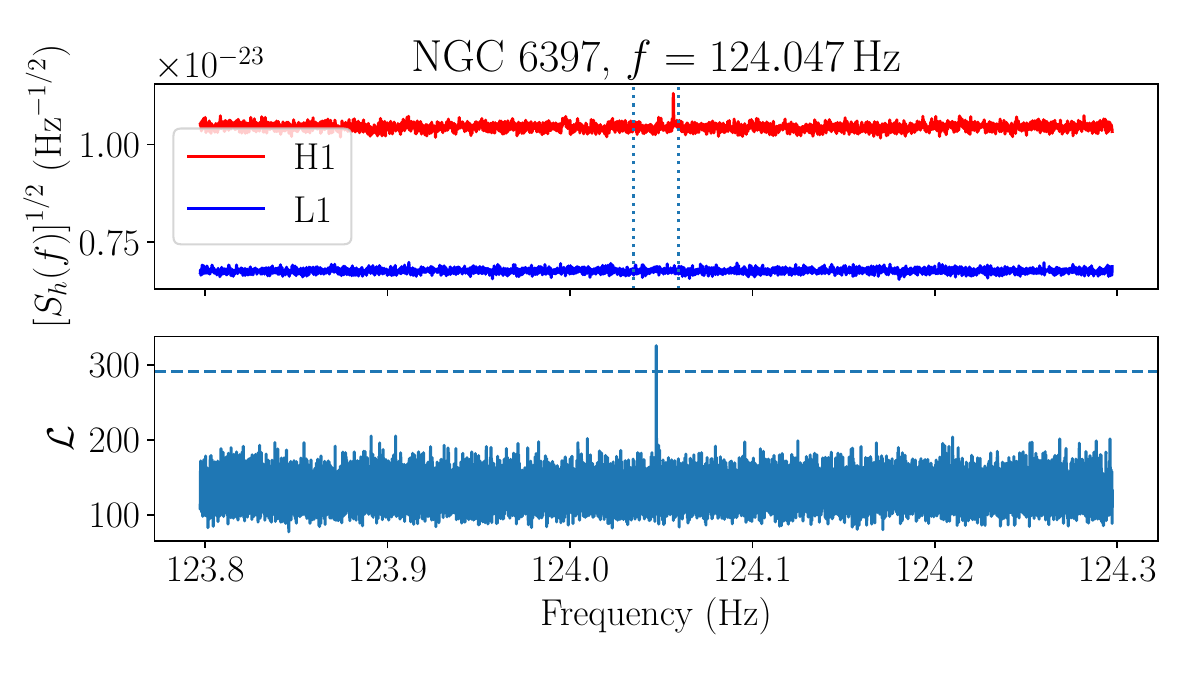}\\
    \includegraphics[width=0.9\columnwidth]{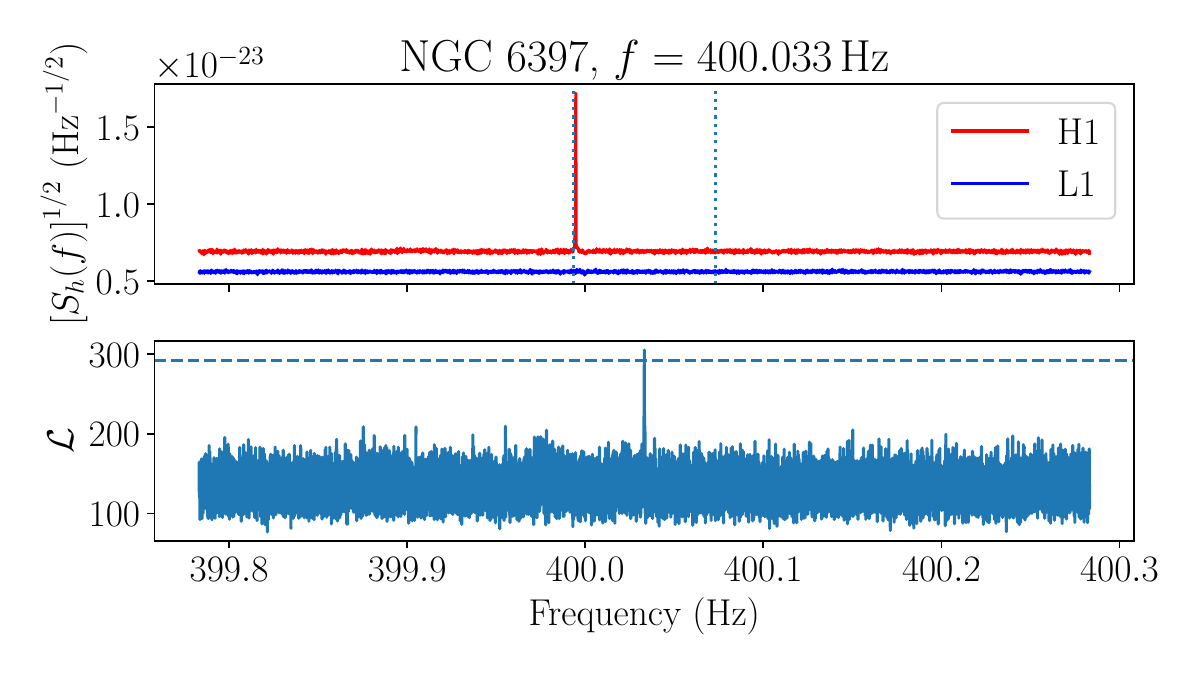}
    \includegraphics[width=0.9\columnwidth]{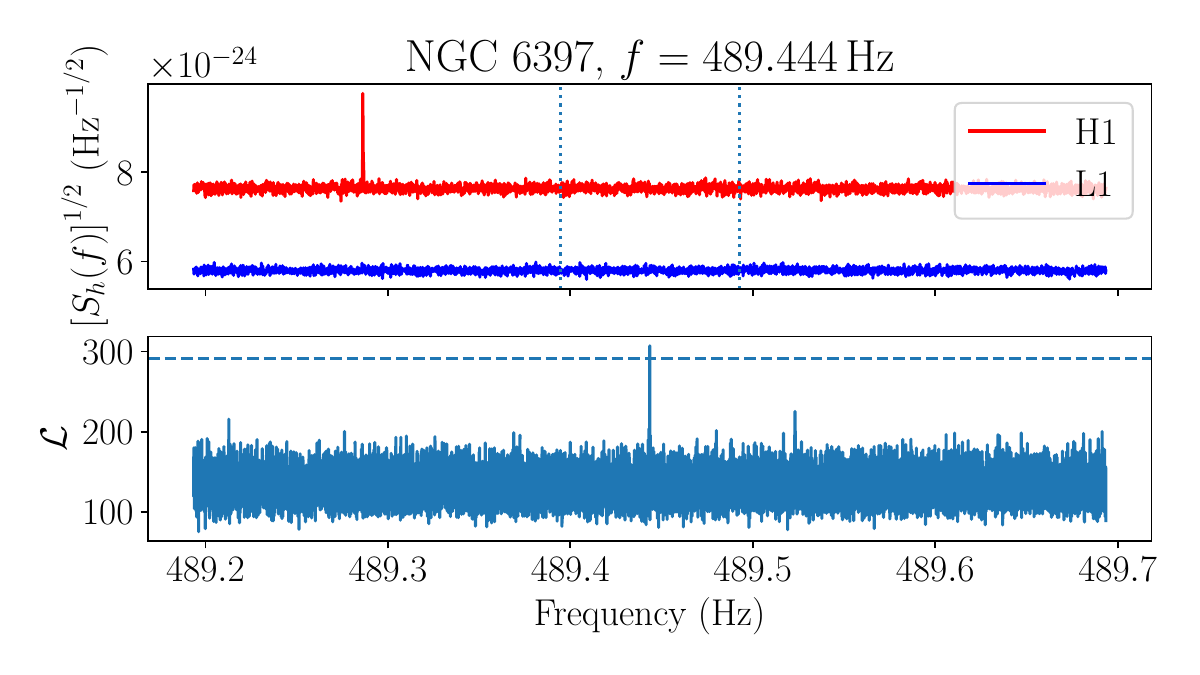}\\
    \includegraphics[width=0.9\columnwidth]{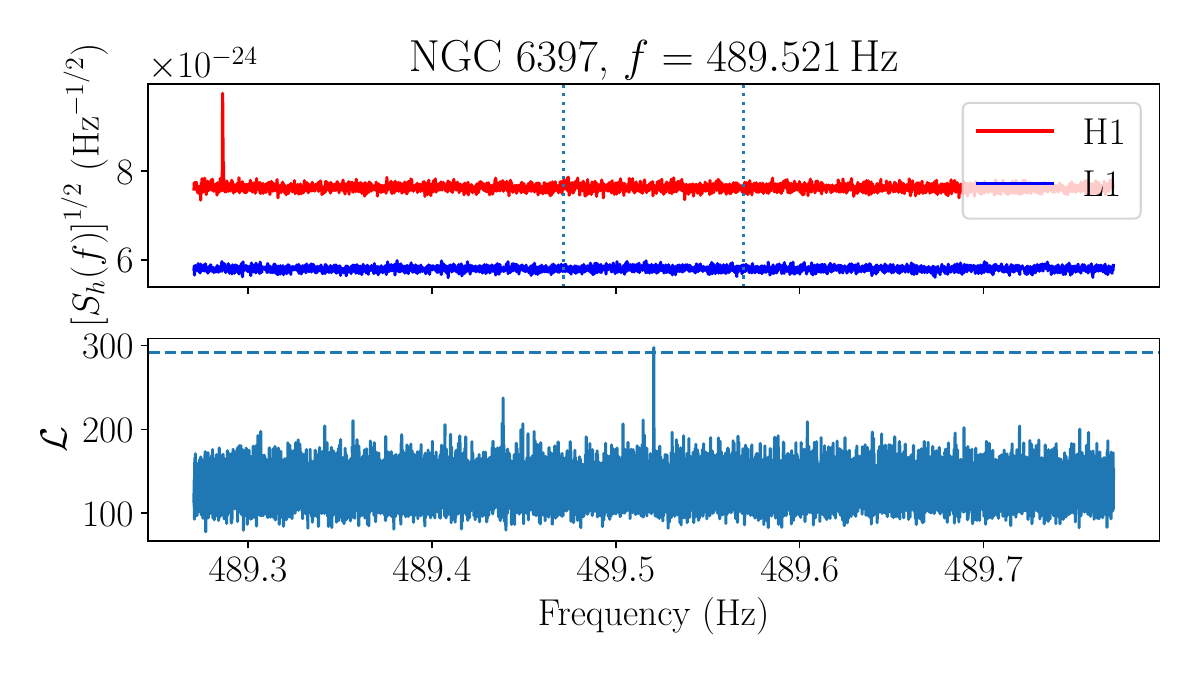}
    \includegraphics[width=0.9\columnwidth]{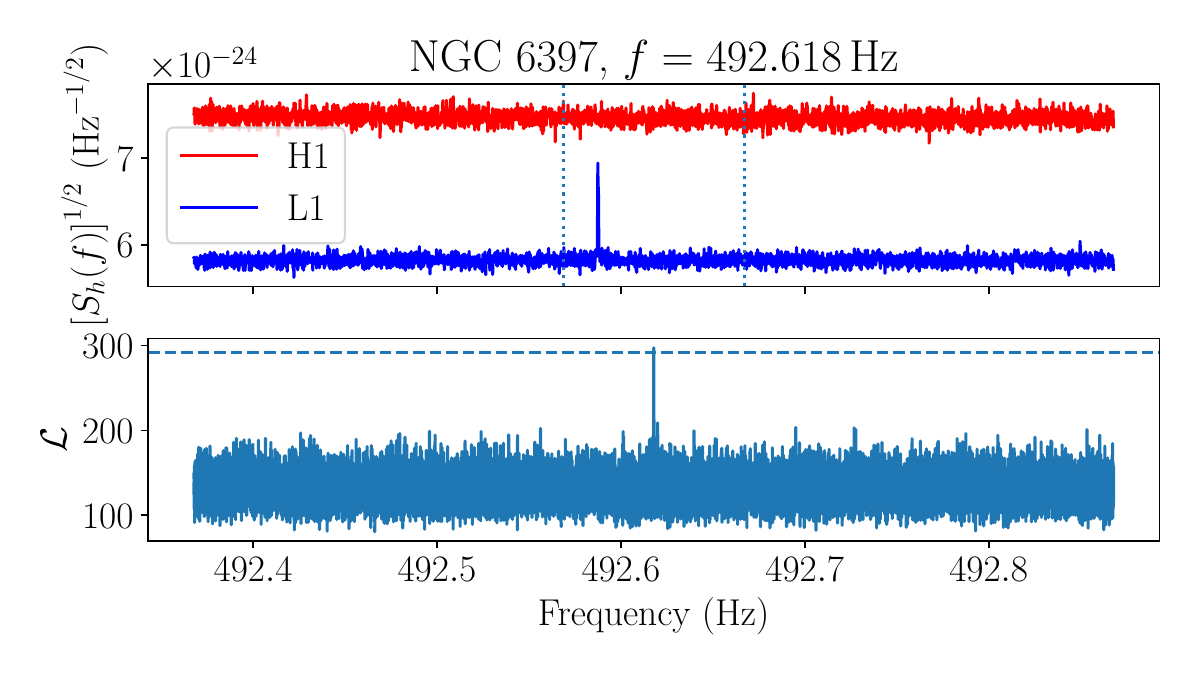}\\
    \includegraphics[width=0.9\columnwidth]{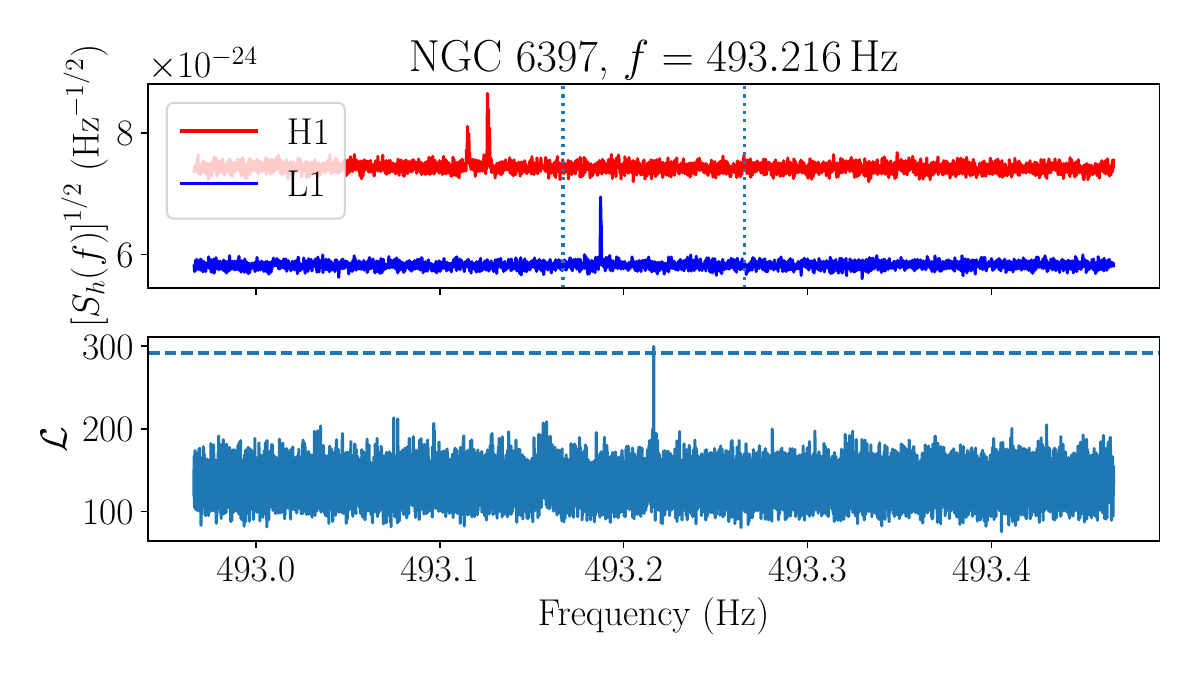}
    \includegraphics[width=0.9\columnwidth]{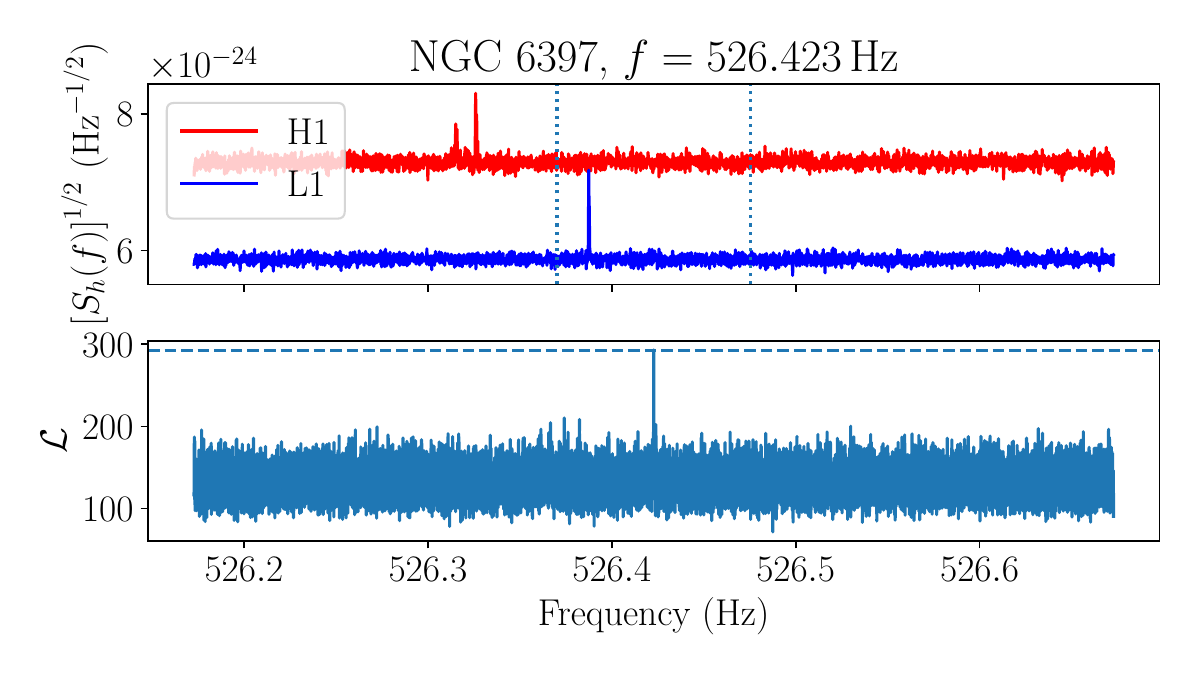}\\
    \includegraphics[width=0.9\columnwidth]{NGC6397_652.76_unknown.pdf}
    \includegraphics[width=0.9\columnwidth]{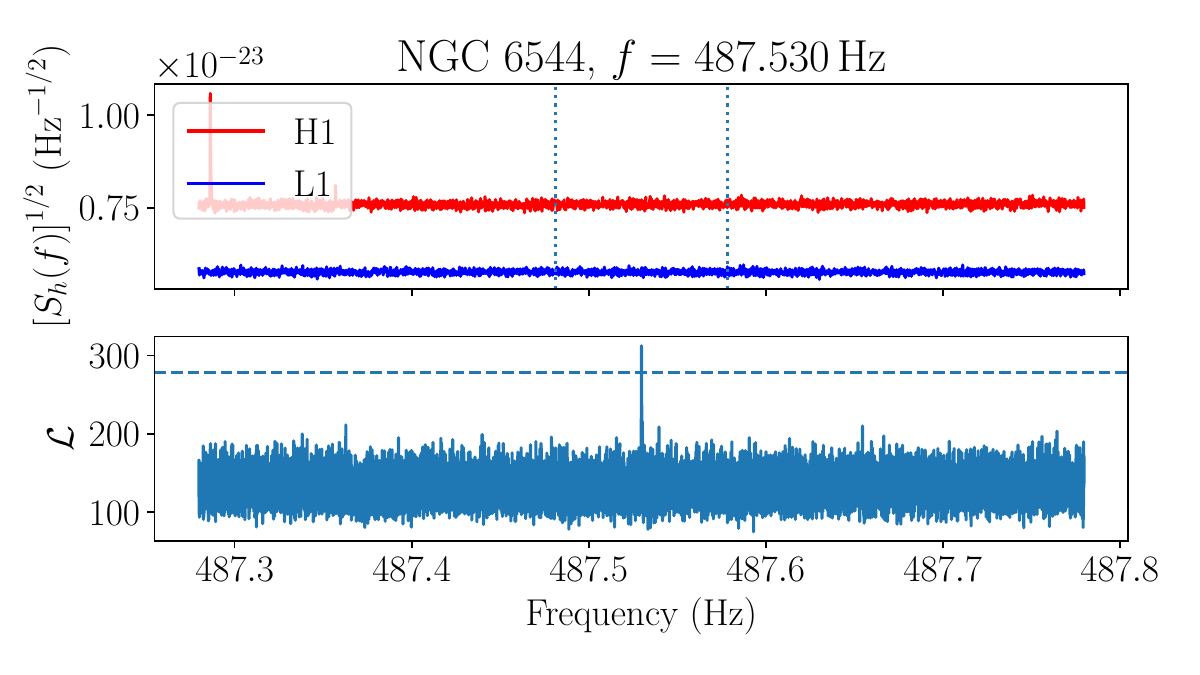}
    \caption{Single-detector ASDs (units: $\mathrm{Hz}^{-1/2}$) and log likelihoods versus frequency (units: Hz) for the ten candidate groups surviving the cross-cluster veto. The vertical, dotted lines in the ASD plots indicate the extent of the Doppler modulation of the candidate group. The horizontal, dashed lines in the log likelihood plots indicate the value of $\mathcal{L}_{\mathrm{th}}$ for the candidate's cluster. Seven of the ten candidate groups show a clear disturbance in one of the detector ASDs. The cluster name and approximate ending frequency of the loudest path in each candidate group are recorded above each panel.}
    \label{fig:unknown_lines_all}
    \end{figure*}

This appendix collects plots of the single-detector ASDs in the vicinity of the ten candidate groups which survive the cross-cluster veto (Section \ref{subsubsec:veto_cross_cluster}).
The plots are displayed in Figure \ref{fig:unknown_lines_all}.
Although all but one of these candidate groups are vetoed by the single-interferometer veto (Section \ref{subsubsec:veto_single_ifo}), we present the single-detector ASDs for completeness. 
Seven of the ten candidate groups show a disturbance in one ASD within the Doppler modulation window of the candidate group.
\end{document}